\documentclass[useAMS,usenatbib]{mn2e}

\usepackage{natbib} %for references
\usepackage{multirow} %for multicolumns in tables
\usepackage[caption=false]{subfig} %for figures over multiple pages
\usepackage{enumerate} %for lists
\usepackage[T1]{fontenc} %for emphasis in titles
\usepackage{amsmath} %for equations
\usepackage{amssymb} %for equations
\usepackage{rotating} %for sideways figures
\usepackage{chngcntr} %for appendix figure numbers
\usepackage{appendix} %for appendices
\usepackage{book tabs} %for horizontal lines in tables
\usepackage{pdflscape} %for landscape table
\usepackage{ae,aecompl} %for on-screen viewing zoom

\def\reff@jnl#1{{\rm#1\/}}

\def\aj{\reff@jnl{AJ}}                  % Astronomical Journal
\def\araa{\reff@jnl{ARA\&A}}            % Annual Review of Astron and Astrophys
\def\apj{\reff@jnl{ApJ}}                % Astrophysical Journal
\def\apjl{\reff@jnl{ApJ}}               % Astrophysical Journal, Letters
\def\apjs{\reff@jnl{ApJS}}              % Astrophysical Journal, Supplement
\def\ao{\reff@jnl{Appl.Optics}}         % Applied Optics
\def\apss{\reff@jnl{Ap\&SS}}            % Astrophysics and Space Science
\def\aap{\reff@jnl{A\&A}}               % Astronomy and Astrophysics
\def\aapr{\reff@jnl{A\&A~Rev.}}         % Astronomy and Astrophysics Reviews
\def\aaps{\reff@jnl{A\&AS}}             % Astronomy and Astrophysics, Supplement
\def\azh{\reff@jnl{AZh}}                        % Astronomicheskii Zhurnal
\def\baas{\reff@jnl{BAAS}}              % Bulletin of the AAS
\def\jrasc{\reff@jnl{JRASC}}            % Journal of the RAS of Canada
\def\memras{\reff@jnl{MmRAS}}           % Memoirs of the RAS
\def\mnras{\reff@jnl{MNRAS}}            % Monthly Notices of the RAS
\def\pra{\reff@jnl{Phys.Rev.A}}         % Physical Review A: General Physics
\def\prb{\reff@jnl{Phys.Rev.B}}         % Physical Review B: Solid State
\def\prc{\reff@jnl{Phys.Rev.C}}         % Physical Review C
\def\prd{\reff@jnl{Phys.Rev.D}}         % Physical Review D
\def\prl{\reff@jnl{Phys.Rev.Lett}}      % Physical Review Letters
\def\pasp{\reff@jnl{PASP}}              % Publications of the ASP
\def\pasj{\reff@jnl{PASJ}}              % Publications of the ASJ
\def\qjras{\reff@jnl{QJRAS}}            % Quarterly Journal of the RAS
\def\skytel{\reff@jnl{S\&T}}            % Sky and Telescope
\def\solphys{\reff@jnl{Solar~Phys.}}    % Solar Physics
\def\sovast{\reff@jnl{Soviet~Ast.}}     % Soviet Astronomy
\def\ssr{\reff@jnl{Space~Sci.Rev.}}     % Space Science Reviews
\def\zap{\reff@jnl{ZAp}}                        % Zeitschrift fuer Astrophysik
\def\nat{\reff@jnl{Nature}}             % Nature 

\title[CARMA Observations Of Galactic Cold Cores]{CARMA Observations of Galactic Cold Cores: Searching for Spinning Dust Emission}

\author[Tibbs et al.]{C.T.~Tibbs,$^{1,2}$\thanks{ESA Research Fellow}\thanks{E-mail: ctibbs@cosmos.esa.int}
R.~Paladini,$^{2}$
K.~Cleary,$^{3}$
S.J.C.~Muchovej,$^{3}$
A.M.M.~Scaife,$^{4}$
\and M.A.~Stevenson,$^{3}$
R.J.~Laureijs,$^{1}$
N.~Ysard,$^{5}$
K.J.B.~Grainge,$^{4}$
Y.C.~Perrott,$^{6}$
\and C.~Rumsey,$^{6}$
J.~Villadsen$^{3}$ 
\\
$^{1}$Scientific Support Office, Directorate of Science and Robotic Exploration, European Space Research and Technology Centre (ESA/ESTEC), \\ Keplerlaan 1, 2201 AZ, Noordwijk, The Netherlands \\
$^{2}$Infrared Processing Analysis Center, California Institute of Technology, Pasadena, CA 91125, USA \\
$^{3}$Cahill Center for Astronomy and Astrophysics, California Institute of Technology, Pasadena, CA 91125, USA \\
$^{4}$Jodrell Bank Centre for Astrophysics, The University of Manchester, Manchester, M13 9PL, UK \\
$^{5}$IAS, Universit\'{e} Paris-Sud, 91405 Orsay cedex, France \\
$^{6}$Astrophysics Group, Cavendish Laboratory, University of Cambridge, Cambridge, CB3 0HE, UK 
}

\date{Accepted 2015 July 30. Received 2015 July 30; in original form 2015 March 30}
\pubyear{2015}

\begin{document}
\label{first page}
\pagerange{\pageref{firstpage}--\pageref{lastpage}} \pubyear{}
\maketitle

%%%%%%% Abstract %%%%%%%%%%%%%%%%%%%%%%%%%%%%%%%%%%%%%%%%%%%%

\begin{abstract}
We present the first search for spinning dust emission from a sample of 34 Galactic cold cores, performed using the CARMA interferometer. For each of our cores we use photometric data from the \textit{Herschel Space Observatory} to constrain $\bar{N}$$_{\mathrm{H}}$, $\bar{T}$$_{\mathrm{d}}$, $\bar{n}$$_{\mathrm{H}}$, and $\bar{G}$$_{\mathrm{0}}$. By computing the mass of the cores and comparing it to the Bonnor-Ebert mass, we determined that 29 of the 34 cores are gravitationally unstable and undergoing collapse. In fact, we found that 6 cores are associated with at least one young stellar object, suggestive of their proto-stellar nature. By investigating the physical conditions within each core, we can shed light on the cm emission revealed (or not) by our CARMA observations. Indeed, we find that only 3 of our cores have any significant detectable cm emission. Using a spinning dust model, we predict the expected level of spinning dust emission in each core and find that for all 34 cores, the predicted level of emission is larger than the observed cm emission constrained by the CARMA observations. Moreover, even in the cores for which we do detect cm emission, we cannot, at this stage, discriminate between free-free emission from young stellar objects and spinning dust emission. We emphasise that, although the CARMA observations described in this analysis place important constraints on the presence of spinning dust in cold, dense environments, the source sample targeted by these observations is not statistically representative of the entire population of Galactic cores. 
\end{abstract}

%%%%%%% Key Words %%%%%%%%%%%%%%%%%%%%%%%%%%%%%%%%%%%%%%%%%%%%


%%%%%%% Introduction %%%%%%%%%%%%%%%%%%%%%%%%%%%%%%%%%%%%%%%%%

\section{Introduction}
\label{Sec:Intro}

In recent years, there has been mounting evidence for the existence of a new emission component of the Galactic interstellar medium~(ISM). This new component was first identified in the diffuse ISM, and due to its unexpected discovery, was simply referred to as anomalous microwave emission~\citep{Leitch:97}. We now believe that this anomalous microwave emission is due to electric dipole emission from small interstellar dust grains, commonly known as spinning dust emission~\citep{DaL:98}.

\begin{table*}
\begin{center}
\caption{Summary of the 15 cold clumps targeted in this analysis.}
\begin{tabular}{cccccccc}
\hline
 & & & & & \multicolumn{2}{c}{CARMA rms} \\ 
\cmidrule[0.4pt](l){6-7} 
Target & R.A. & Decl. & Distance & $T_{\mathrm{gas}}$ & $\sigma_{\mathrm{Short}}$ & $\sigma_{\mathrm{Long}}$ \\
 & (J2000) & (J2000) & (kpc) & (K) & (mJy~beam$^{-1}$) & (mJy~beam$^{-1}$) \\
\hline 
\hline 
 
ECC181 G102.19+15.24		& 20:41:10.74 		& +67:21:44.3 		& 0.33$^{a}$ 	& 9.6		& 0.58 	& 0.55 \\ 
ECC189 G103.71+14.88 		& 20:53:30.29 		& +68:19:32.9 		& 0.29$^{a}$ 	& 9.7   	& 0.37 	& 0.37 \\ 
ECC190 G103.77+13.90 		& 21:02:09.19 		& +67:45:51.8 		& 0.29$^{a}$ 	& 11.1 	& 0.45 	& 0.48 \\ 
ECC191 G103.90+13.97 		& 21:02:23.24 		& +67:54:43.3 		& 0.29$^{a}$ 	& 11.1   	& 0.90	& 0.86 \\ 
ECC223 G113.42+16.97 		& 21:59:59.03 		& +76:34:08.7 		& 0.99$^{b}$ 	& 8.9		& 0.49 	& 0.47 \\ 
ECC224 G113.62+15.01 		& 22:21:37.34 		& +75:06:33.5 		& 0.86$^{b}$ 	& 8.0		& 0.75	& 0.78 \\ 
ECC225 G113.75+14.90 		& 22:24:16.23 		& +75:05:01.8 		& 0.88$^{b}$ 	& 8.6 	& 0.67	& 0.61 \\ 
ECC229 G114.67+14.47 		& 22:39:35.57 		& +75:11:34.0 		& 0.77$^{b}$ 	& 10.3 	& 0.45	& 0.38 \\ 
ECC276 G127.88+02.66 		& 01:38:39.14 		& +65:05:06.5 		& 1.16$^{b}$ 	& 12.6 	& 0.33	& 0.37 \\ 
ECC332 G149.41+03.37 		& 04:17:09.10 		& +55:17:39.4 		& 0.18$^{b}$ 	& 8.7		& 0.52	& 0.59 \\ 
ECC334 G149.58+03.45 		& 04:18:23.96 		& +55:13:30.6 		& 0.20$^{b}$ 	& 8.7 	& 0.52	& 0.59 \\ 
ECC335 G149.65+03.54 		& 04:19:11.28 		& +55:14:44.4 		& 0.17$^{b}$ 	& 8.1 	& 0.57	& 0.65 \\ 
ECC340 G151.45+03.95 		& 04:29:56.29 		& +54:14:51.7 		& 0.19$^{a}$ 	& 10.1 	& 0.36	& 0.39 \\ 
ECC345 G154.07+05.09 		& 04:47:23.41 		& +53:03:31.4 		& 0.34$^{b}$ 	& 7.4 	& 0.59	& 0.63 \\ 
ECC346 G154.07+05.21 		& 04:47:57.83 		& +53:07:51.2 		& 0.23$^{b}$ 	& 10.0 	& 0.67	& 0.64 \\ 

\hline
\label{Table:Sources}
\end{tabular}
\end{center}
\vspace{-0.6cm}
$^{a}$Distances are based on association with known Lynds dark nebulae~\citep{Hilton:95}. $^{b}$Distances are kinematic distances based on $^{13}$CO observations~\citep{Wu:12}. $T_{\mathrm{gas}}$ values taken from \citet{Wu:12}.
\end{table*}

Spinning dust emission occurs in the wavelength range~$\sim$3~--~0.3~cm and produces a very distinctive peaked spectrum, peaking at wavelengths of~$\sim$1~cm, with the exact wavelength depending on the physical environmental conditions. Although this microwave emission from small spinning dust grains has been found in a wide variety of environments, the best examples of spinning dust detections occur in dense environments: the Perseus molecular cloud~\citep[][]{Watson:05, Tibbs:10, Tibbs:13a, Planck_Dickinson:11, Genova-Santos:15}; and the dark cloud LDN1622~\citep[][]{Finkbeiner:04, Casassus:06, Harper:15}. In this analysis we attempt, for the first time, to search for spinning dust emission in even more dense environments: Galactic cores. Galactic cores represent one of the earliest phases of star formation, created by the gravitational collapse of an over density within a molecular cloud~\citep[see][for a complete review]{Bergin:07}. Although no such observations of spinning dust emission from Galactic cores have been performed,~\citet{Ysard:11} conducted radiative transfer modelling of such dense cores and predicted that spinning dust emission should be detectable at microwave frequencies if there are small dust grains present. Additionally,~\citet{Ysard:11} showed that as the density of the cores increases, an anti-correlation between the microwave emission and the mid-infrared~(IR) emission occurs. The mid-IR emission is concentrated around the edge of the dense cores, while the microwave emission is contained within the centre of the cores. \citet{Ysard:11} proposed that this anti-correlation could be explained in two ways. Either the small grains that produce both the mid-IR emission and the microwave emission are present throughout the core, but due to the high densities, the stellar photons cannot penetrate deep enough into the core to cause the small grains to emit at mid-IR wavelengths, or that there is a deficit of small grains in the centre. Therefore, the motivation for this current work was to search for spinning dust emission in dense cores, and determine which of these two hypotheses is correct. To this end we observed a sample of Galactic cores with the CARMA interferometer at a wavelength of 1~cm, which based on the work by~\citet{Ysard:11}, is the expected peak wavelength of the spinning dust emission in these dense environments.

This paper is organized as follows. In Section~\ref{Sec:Source_Selection} we define our sample, and in Section~\ref{Sec:Observations} we describe the data that we use to perform this analysis. In Section~\ref{Sec:Dust_Properties} we characterize the physical properties of our cores, and in Section~\ref{Sec:Discussion} we compare our observations with the expected level of spinning dust emission. Finally, we present our conclusions in Section~\ref{Sec:Conclusions}.

%%%%%%% Source Selection %%%%%%%%%%%%%%%%%%%%%%%%%%%%%%%%%%%%%%%%%

\section{Source Selection}
\label{Sec:Source_Selection}

For the dust temperatures characteristic of Galactic cold cores~($\sim$8~--~14~K), the peak of the thermal dust emission occurs around~$\sim$200~$\mu$m, implying that the best wavelengths to detect these objects is in the far-IR and sub-mm bands. \textit{Planck}, with its all-sky coverage at sub-mm wavelengths, is an ideal instrument with which to detect cold cores. In fact, as part of the \textit{Planck} Early Release Compact Source Catalogue~\citep{Planck_Chary:11}, the Early Compact Core~(ECC) catalogue was published. The ECC catalogue contains 915 cold clumps\footnote{Since the angular resolution of \textit{Planck} is~$\sim$5~arcmin in the far-IR/sub-mm~bands, it can not detect individual cores, and therefore we refer to the \textit{Planck} sources as clumps.} identified by the \textit{Planck} team to be the coldest~(T~<~14~K) and most reliable~(SNR~>~15) sources in the Cold Core Catalogue of Planck Objects~(C3PO).

We selected a sample of 15 sources from the ECC catalogue. The sources were selected based on the following criteria: (a) the source must have been observed with the \textit{Herschel Space Observatory} PACS and SPIRE photometers~\citep{Pilbratt:10}; (b) the source must be visible from the site of the CARMA interferometer; (c) the source must have an angular size less than 12~arcmin to ensure that we do not resolve out any flux; (d) the source must have a known distance estimate; and (e) the sources must cover a range of densities, $n_{\mathrm{H}}$. The 15 sources included in our sample are listed in Table~\ref{Table:Sources}. 

Also listed in Table~\ref{Table:Sources} are the distance estimates for each source. \citet{Wu:12} observed the $J=1-0$ transition of $^{12}$CO, $^{13}$CO, and C$^{18}$O for 674 of the 915 ECC sources using the 13.7~m telescope of the Purple Mountain Observatory. Using the $^{13}$CO $V_{\mathrm{LSR}}$ measurements with the~\citet{Clemens:85} rotation curve, and adopting values of $R_{0}$~=~8.5~kpc and $\Theta_{0}$~=~220~km~s$^{-1}$, \citet{Wu:12} computed the kinematic distance to each source. For sources within the Solar circle, which suffer from the kinematic distance ambiguity, they selected the near distance. Since \citet{Wu:12} was unable to assign a distance estimate to all of their ECC sources, there were five sources in our sample ~(ECC181, ECC189, ECC190, ECC191, and ECC340) for which we did not have a kinematic distance estimate, and for these sources we estimated the distance based on associations with known Lynds dark nebulae~\citep{Hilton:95}. 

In addition to a distance estimate, \citet{Wu:12} also derived the temperature of the gas, $T_{\mathrm{gas}}$, for all of their ECC sources. The $T_{\mathrm{gas}}$ values for our 15 targets are also listed in Table~\ref{Table:Sources}, and these values are adopted throughout this analysis.

%%%%%%% Observations %%%%%%%%%%%%%%%%%%%%%%%%%%%%%%%%%%%%%%%%%

\section{Observations}
\label{Sec:Observations}

\begin{figure*}
\begin{center}
\includegraphics[angle=0,scale=0.50]{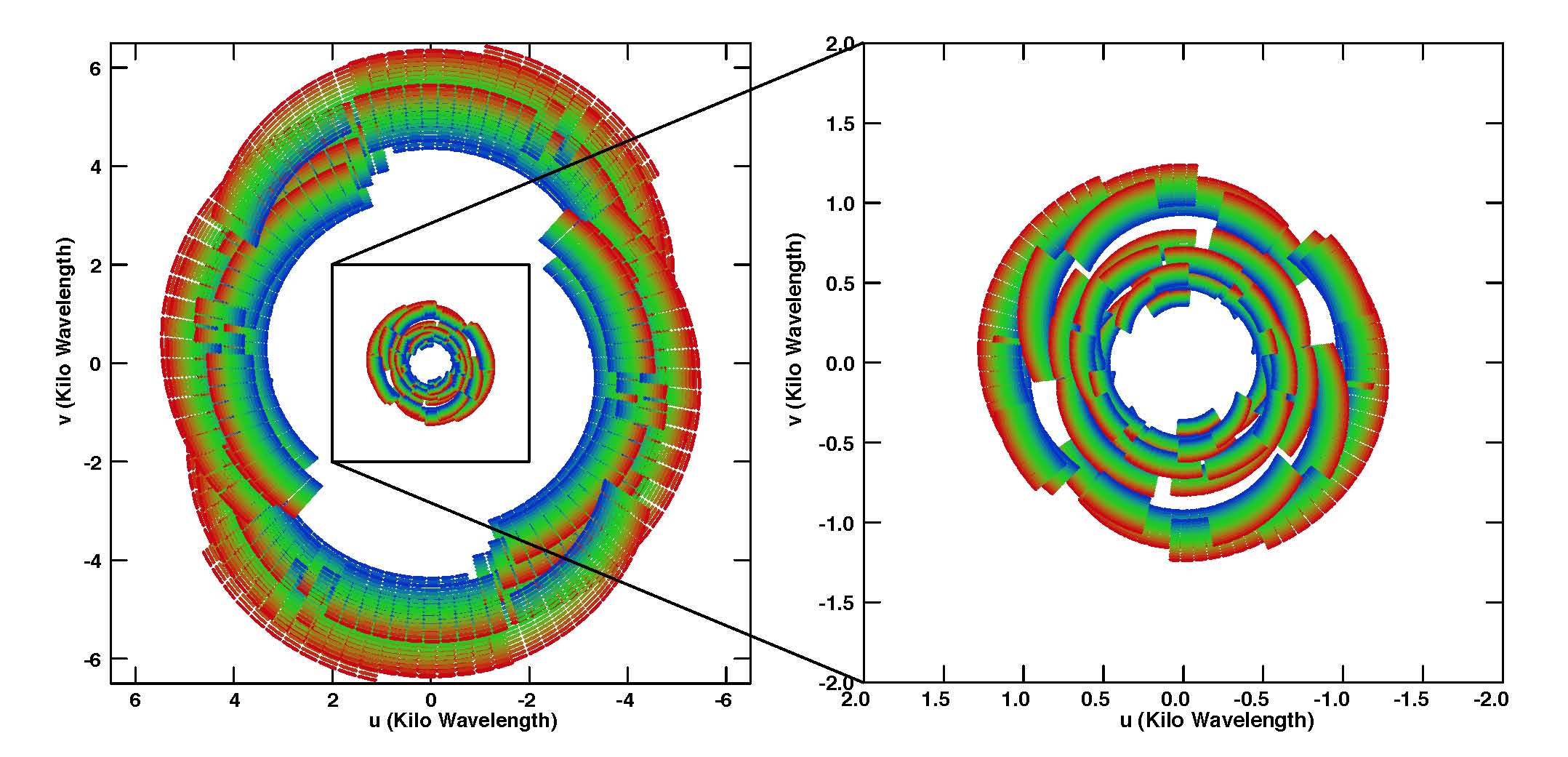}
\end{center}
\vspace{-0.3cm}
\caption{Complete \textit{u-v} coverage (\textit{left}) and the \textit{u-v} coverage for only the short-baselines (\textit{right}) for our observations of ECC181. The colour scale represents the 15 channels with red corresponding to the highest frequency channel and blue corresponding to the lowest frequency channel. The \textit{u,v} coverage for the rest of our targets is very similar.}
\label{Fig:uv_coverage}
\end{figure*}

%%%%%%% CARMA Data %%%%%%%%%%%%%%%%%%%%%%%%%%%%%%%%%%%%%%%%%

\subsection{CARMA Data}
\label{Subsec:Carma}

CARMA is an interferometer consisting of a total of 23 antennas located at a high-altitude site in eastern California. CARMA operates at three bands: 1~cm~(27~--~35~GHz), 3~mm~(85~--~116~GHz), and 1~mm~(215~--~270~GHz), and consists of six 10.4~m, nine 6.1~m, and eight 3.5~m antennas that are combined to form the full CARMA array. The data presented in this analysis were observed in the 1~cm band with fifteen 500~MHz channels between 27.5 and 35~GHz using the eight 3.5~m antennas. Six of the eight antennas formed a compact configuration with baseline lengths of~$\sim$0.3~--~2.0~k$\lambda$, while the remaining two antennas provided much longer baselines of~$\sim$2.0~--~8.0~k$\lambda$. We refer to data from the baselines between the six compact antennas as short-baseline data, and data from the baselines between the six compact antennas and the two outlying antennas as long-baseline data. In Figure~\ref{Fig:uv_coverage} we plot the Fourier space \textit{u,v} coverage for our observations of one of our targets, ECC181. The \textit{u,v} coverage of the rest of our fifteen targets is very similar. In the left panel of Figure~\ref{Fig:uv_coverage} we plot the complete \textit{u,v} coverage for all baselines, which clearly illustrates how we separate the data into short- and long-baseline data, while in the right panel we plot a zoomed in view of the short baselines.

\begin{table*}
\begin{center}
\caption{Flux densities of the four point sources identified in the CARMA maps.}
\begin{tabular}{ccccccc}
\hline
 & \multicolumn{3}{c}{Short Baselines} & \multicolumn{3}{c}{Long Baselines} \\ 
 \cmidrule[0.4pt](l){2-4} \cmidrule[0.4pt](l){5-7}
Source & R.A. & Decl. & $S_{1~\mathrm{cm}}$ & R.A. & Decl. & $S_{1~\mathrm{cm}}$ \\
 & (J2000) & (J2000) & (mJy) & (J2000) & (J2000) & (mJy) \\
\hline
\hline

NVSS~205303+682200	&	20:53:04.00	&	+68:22:06.4	&	2.85~$\pm$~0.64	&	20:53:04.44	&	+68:21:58.0	&	3.45~$\pm$~0.65 \\
NVSS~210243+675819	& 	21:02:43.91	&	+67:58:16.7	&	101.77~$\pm$~5.30	&	21:02:43.70	&	+67:58:17.1	&	102.52~$\pm$~5.27 \\
NVSS~043015+541429	&	04:30:15.40	&	+54:14:31.6	&	4.41~$\pm$~0.62	&	04:30:14.84	&	+54:14:28.1	&	4.32~$\pm$~0.69 \\
NVSS~044819+530830	&	04:48:18.88	&	+53:08:22.9	&	10.26~$\pm$~1.40	&	04:48:19.47	&	+53:08:30.4	&	9.98~$\pm$~1.07 \\

\hline
\label{Table:Radio_Sources}
\end{tabular}
\end{center}
\end{table*}

Our CARMA observations were performed during three separate observing runs in May 2012, August 2012, and January 2013. The data were reduced using a suite of \textsc{matlab} routines, which convert the data to physical units, correct for instrumental phase and amplitude variations, flag data that do not meet designed criteria at each stage of reduction, and perform the data calibration~\citep[see][for further details on the reduction pipeline]{Muchovej:07}. The absolute calibration is derived from observations of Mars, and based on long-term monitoring of flux density calibrators, we estimate that the calibration is accurate to 5\%. The reduction pipeline outputs the calibrated visibility data, which were then imported into \textsc{aips}, where the task \textsc{imagr} was used to perform the Fourier transform and deconvolution. This task uses a CLEAN-based algorithm to perform the deconvolution, and we implemented a Briggs robust parameter of 0, providing a compromise between uniform and natural weighting. We imaged the short-baseline data and the long-baseline data separately, using a 120~arcsec and 20~arcsec restoring beam~(FWHM), respectively. The final CARMA maps are displayed in Figure~\ref{Fig:CARMA_Maps}, and the rms noise in both the short-baseline~($\sigma_{\mathrm{Short}}$) and long-baseline~($\sigma_{\mathrm{Long}}$) images for each source is tabulated in Table~\ref{Table:Sources}. 

The long-baseline data are ideal for identifying compact radio sources~($\lesssim$~1 arcmin), while the short-baseline data are sensitive to the more diffuse emission on angular scales of~$\sim$2~--~12~arcmin. Looking at the maps in Figure~\ref{Fig:CARMA_Maps}, we detected four compact sources that are visible in both the short- and long-baseline data~(one source each in the ECC189, ECC191, ECC340, and ECC346 maps). Using the \textsc{aips} task \textsc{jmfit} we fitted 2-dimensional Gaussians to each source to derive positions and flux densities. We searched the literature to identify these sources, and the NVSS counterpart for each source is listed in Table~\ref{Table:Radio_Sources}. Also listed in Table~\ref{Table:Radio_Sources} are the measured CARMA 1~cm flux densities. The CARMA flux density uncertainties include a 5\% calibration uncertainty combined in quadrature with the uncertainty from the fit to the source. Throughout the rest of this analysis we ignore these four point sources.

In addition to the point sources, there is not much extended emission present in the short-baseline maps displayed in Figure~\ref{Fig:CARMA_Maps}. In fact, we find that only 1 of our 15 clumps, ECC229, has a significant detection of emission, which consists of a more compact component to the west and a more extended structure, likely consisting of two separate components, to the east.

%%%%%%% Herschel Data %%%%%%%%%%%%%%%%%%%%%%%%%%%%%%%%%%%%%%%%%

\subsection{\textit{Herschel} Data}
\label{Subsec:Herschel}

In order to characterize the physical conditions of the observed CARMA clumps, we used \textit{Herschel} PACS and SPIRE data. Eight of our targets are covered by the Guaranteed Time Key Programme ``Probing the origin of the stellar initial mass function: A wide-field \textit{Herschel} photometric survey of nearby star-forming cloud complexes''~\citep[KPGT\_pandre\_1;][]{Andre:10} while the remaining seven are covered by the Open Time Key Programme ``Galactic Cold Cores: A \textit{Herschel} survey of the source populations revealed by Planck''~\citep[KPOT\_mjuvela\_1;][]{Juvela:10}.

The \textit{Herschel} data were downloaded from the \textit{Herschel} Data Archive and re-processed using the \textit{Herschel} Interactive Processing Environment~(HIPE) version 10.0.2747. The PACS data were processed using the standard pipeline for extended sources. Starting with the level 1 PACS data, this processing performed a correction for the global correlated signal drift and produced the time ordered data, which were used to create a map using the Madmap map-maker. Finally, a point source artifact correction was applied, resulting in maps at 70, 100, and 160~$\mu$m with an angular resolution of~$\sim$6, 7, and 11~arcsec, respectively. The absolute calibration uncertainty of the PACS data was conservatively estimated to be 10~$\%$. The SPIRE data were processed from level 1 to level 2.5 using the SPIA reduction routines, with the Destriper map-maker used to produce maps at 250, 350, and 500~$\mu$m with an angular resolution of~$\sim$18, 25, and 36~arcsec, respectively. The absolute calibration uncertainty of the SPIRE data was conservatively estimated to be 7~$\%$. 

Although PACS has three channels (70, 100, and 160~$\mu$m), only two~(either 70 or 100~$\mu$m along with 160~$\mu$m) can be observed simultaneously for each observation. Therefore, to ensure that observations in all three PACS channels are obtained, a few of our sources had both PACS/SPIRE parallel observations and PACS photometry observations. However, this not only results in observations in all three PACS channels, it also results in a duplication of observations in the 160~$\mu$m channel. Not all of our sources had duplicate 160~$\mu$m observations, but for the ones that did, we used the mean of 
the two 160~$\mu$m maps. All fifteen sources did have at least one set of PACS and SPIRE observations, meaning that we have \textit{Herschel} observations at 160, 250, 350, and 500~$\mu$m for all of our sources which, as we will discuss in Section~\ref{Subsec:H_Column_Density}, is adequate for our analysis.

%%%%%%% Dust Properties %%%%%%%%%%%%%%%%%%%%%%%%%%%%%%%%%%%%%%%%%

\section{Clump Properties}
\label{Sec:Dust_Properties}

To understand why only 1 of our 15 clumps had detectable 1~cm emission, it is crucial to explore the physical conditions of our sample. Properties such as the density and radiation field are important parameters in spinning up the grains to produce spinning dust emission. Additionally, we need to determine the evolutionary stage of the clumps, specifically, whether these sources are gravitationally bound and, if so, whether they are pre-stellar or proto-stellar. If indeed some of the clumps are proto-stellar, we also need to identify the location of young stellar objects~(YSOs) as they can generate emission at cm wavelengths due to a variety of mechanisms such as stellar winds and/or shock-induced ionization, which can mimic the spinning dust emission spectral signal. Additionally, knowledge about the evolutionary stage of the clumps and, in particular, understanding if the observed sources are both simultaneously forming stars and harbouring spinning dust emission can help to understand the potential role of spinning dust emission in the star formation process.

Therefore, in Section~\ref{Subsec:H_Column_Density} we compute the hydrogen column density and dust temperature within our clumps, and in Section~\ref{Subsec:H_Number_Density} we estimate the hydrogen number density and the radiation field. Then in Section~\ref{Subsec:Mass} we estimate the mass and determine if the clump is stable or likely to undergo collapse, and in Section~\ref{Subsec:YSOs} we identify and classify candidate YSOs in the vicinity of our clumps.

%%%%%%% Column Density and Temperature %%%%%%%%%%%%%%%%%%%%%%%%%%%%%%%%%%%%%%%%%%%%

\subsection{Hydrogen Column Density and Dust Temperature}
\label{Subsec:H_Column_Density}

All of the \textit{Herschel} maps were convolved to the angular resolution of the 500~$\mu$m map using the convolution kernels produced by~\citet{Aniano:11} to allow for accurate comparisons between each of the \textit{Herschel} bands. To ensure we are probing the dense cores themselves and are not contaminated by foreground or background emission, we performed a background subtraction on the \textit{Herschel} maps. For each of the \textit{Herschel} maps, we computed the median value of the flux in a reference position that is devoid of emission, and subtracted this from the map. The emission in these background subtracted \textit{Herschel} maps was then modelled using

\begin{equation}
I_{\nu} = \mu m_{\mathrm{H}} N_{\mathrm{H}} \kappa_{\nu} B_{\nu}(T_{\mathrm{d}}) ,
\label{Equ:mod_BB}
\end{equation}

\noindent
where $I_{\nu}$ is the intensity at frequency $\nu$, $\mu$ is the molecular weight per hydrogen atom, which we assume to be 1.4, $m_{\mathrm{H}}$ is the mass of a H atom, $N_{\mathrm{H}}$ is the hydrogen column density, $\kappa_{\nu}$ is the dust opacity, and $B_{\nu}(T_{\mathrm{d}})$ is the Planck function for dust temperature $T_{\mathrm{d}}$. The dust opacity is the subject of much debate, with large variations observed between models, depending on the physical properties of the dust grains such as size, composition, and structure~\citep[e.g.,][]{Ossenkopf:94}. In this analysis we use the dust opacity parameterization defined by~\citet{Beckwith:90}

\begin{equation}
\kappa_{\nu} = 0.1 \left( \frac{\nu}{1000~\mathrm{GHz}} \right)^{\beta} \mathrm{cm^{2}~g^{-1}} ,
\label{Equ:kappa}
\end{equation}

\noindent
which is applicable to these cold, dense environments~\citep[e.g.,][]{Ward-Thompson:10, Planck_Ristorcelli:11, Planck_Montier:11, Juvela:12}. This normalization assumes a standard gas to dust mass ratio of 100.

We used the model defined by Equations~\eqref{Equ:mod_BB} and~\eqref{Equ:kappa} to fit the background subtracted \textit{Herschel} maps at 160, 250, 350, and 500~$\mu$m. We excluded the 70 and 100~$\mu$m \textit{Herschel} maps from the fit because our clumps are cold~(< 14~K; this was one of the selection criteria used to produce the \textit{Planck} ECC catalogue), and therefore they do not emit strongly at wavelengths $\le$ 100~$\mu$m. Furthermore, excluding the 70 and 100~$\mu$m maps ensures that we avoid possible contamination from stochastically heated small dust grains. We inspected the 70 and 100~$\mu$m maps for all of the clumps and confirmed that there is very little emission present at these wavelengths, justifying our decision to exclude them from the fit. 

We fitted the data for $N_{\mathrm{H}}$ and $T_{\mathrm{d}}$, fixing the dust opacity index $\beta$~=~2~\citep[e.g.,][]{Planck_Ristorcelli:11, Planck_Montier:11, Juvela:12}. This fit was performed on a pixel-by-pixel basis using MPFIT~\citep{Markwardt:09} resulting in maps of $N_{\mathrm{H}}$ and $T_{\mathrm{d}}$, along with their associated uncertainty maps, which were computed from the covariance matrix. The $N_{\mathrm{H}}$ and $T_{\mathrm{d}}$ maps for all 15 clumps are displayed in Figure~\ref{Fig:NH_Td_Maps}. For our 15 targets, we find values of $N_{\mathrm{H}}$ ranging from~$\sim$6$\times$10$^{21}$ to~$\sim$7$\times$10$^{22}$ H~cm$^{-2}$ and dust temperatures between $T_{\mathrm{d}}$~$\sim$10 to~$\sim$15~K. These values are well within the the range found for the entire C3PO catalogue by~\citet{Planck_Montier:11}, and are consistent with those estimated by~\citet{Juvela:12} using \textit{Herschel} SPIRE data for their sample of cores.

%%%%%%% Number Density and Radiation Field %%%%%%%%%%%%%%%%%%%%%%%%%%%%%%%%%%%%%%%%%%%%

\subsection{Hydrogen Number Density and Radiation Field}
\label{Subsec:H_Number_Density}

\textit{Herschel}, with its sub-arcmin angular resolution, is able to resolve the sources observed by \textit{Planck} into sub-structures. In the case of the ECC clumps, we can therefore identify individual cores within each clump. For this purpose, we used the \textsc{clumpfind} package~\citep{Williams:94}, which identifies local peaks in the $N_{\mathrm{H}}$ maps and follows them down to lower levels, resulting in a decomposition of the map into multiple cores. \textsc{clumpfind} identified 34 cores, the positions of which are listed in Table~\ref{Table:Clump_Props} and labelled in Figure~\ref{Fig:NH_Td_Maps}. In addition to the position, \textsc{clumpfind} also computes the angular size of each identified core. Since this angular size is the effective circular radius, $R$ ($ = (N_{\mathrm{pixels}} / \pi)^{1/2}$, where $N_{\mathrm{pixels}}$ is the total number of pixels), we assumed that the cores are spherical, and for each core we computed the mean column density, $\bar{N}$$_{\mathrm{H}}$, and the mean dust temperature, $\bar{T}$$_{\mathrm{d}}$. The values of $\bar{N}$$_{\mathrm{H}}$ and $\bar{T}$$_{\mathrm{d}}$ for each core are listed in Table~\ref{Table:Clump_Props}. 

Using the distance to each clump~(see Table~\ref{Table:Sources}) we converted the angular size of each core computed by \textsc{clump find} to a physical, linear size, $L$. Since we have performed a background/foreground subtraction, by combining $\bar{N}$$_{\mathrm{H}}$ and the linear size we can estimate the mean density, $\bar{n}$$_{\mathrm{H}}$. Based on the conservation of mass, the total mass in the $N_{\mathrm{H}}$ map within a given radius, $R$, is equivalent to the mass within a sphere of the same radius, with a mean density, $\bar{n}_{\mathrm{H}}$, i.e.,

\begin{equation}
\pi R^{2} \mu m_{\mathrm{H}} \bar{N}_{\mathrm{H}} = \frac{4}{3} \pi R^{3} \mu m_{\mathrm{H}} \bar{n}_{\mathrm{H}} .
\label{Equ:con_mass}
\end{equation}

\noindent
Since $R$ = $L$/2, this results in

\begin{equation}
\bar{n}_{\mathrm{H}} = \frac{3}{2} \frac{\bar{N}_{\mathrm{H}}} {L} .
\label{Equ:nh}
\end{equation}

\noindent
Using Equation~\eqref{Equ:nh} we estimated $\bar{n}$$_{\mathrm{H}}$ for each core. The uncertainty on $\bar{n}$$_{\mathrm{H}}$ was calculated assuming a 10$\%$ uncertainty on the distance. The derived $\bar{n}$$_{\mathrm{H}}$ values range from~$\sim$5$\times$10$^{3}$~H~cm$^{-3}$ in core 1 in ECC223 to~$\sim$125$\times$10$^{3}$~H~cm$^{-3}$ in core 2 in ECC229 (see Table~\ref{Table:Clump_Props}). Although these values are mean quantities, they are comparable to the densities estimated in the entire C3PO catalogue~\citep{Planck_Montier:11}. 

In addition to estimating $\bar{n}$$_{\mathrm{H}}$, we also estimated the mean radiation field, $\bar{G}$$_{\mathrm{0}}$, in each core by converting $\bar{T}$$_{\mathrm{d}}$ into $\bar{G}$$_{0}$ using

\begin{equation}
\bar{G}_{0}  = \left( \frac{\bar{T}_{\mathrm{d}}}{17.5~\mathrm{K}} \right) ^{4 + \beta} ,
\label{Equ:Td_G0}
\end{equation}

\noindent
where we fixed $\beta$ = 2 as before. The computed values of $\bar{G}$$_{\mathrm{0}}$ are also listed in Table~\ref{Table:Clump_Props} and we find values ranging from~$\sim$0.04 to 0.30, where a value of 1 corresponds to the~\citet{Mathis:83} solar neighbourhood radiation field. Such low values of $\bar{G}$$_{\mathrm{0}}$ indicate that the interior of these dense environments are shielded from the surrounding interstellar radiation field.

%%%%%%% Mass %%%%%%%%%%%%%%%%%%%%%%%%%%%%%%%%%%%%%%%%%%%%

\subsection{Mass}
\label{Subsec:Mass}

We estimated the total (dust $+$ gas) mass for each of our cores using Equation~\eqref{Equ:con_mass}. The computed mass estimates are listed in Table~\ref{Table:Clump_Props} and range between~$\sim$0.4~--~115~M$_{\odot}$.

To determine if our cores are gravitationally stable or in the process of collapsing, we also computed the Bonnor-Ebert mass, $M_{\mathrm{BE}}$, the mass of an isothermal sphere in hydrostatic and pressure equilibrium. Cores with $M > M_{\mathrm{BE}}$ are unstable, and will therefore collapse, while cores with $M < M_{\mathrm{BE}}$ are stable. We estimated the Bonnor-Ebert mass as 

\begin{equation}
M_{BE} \approx 1.82 \left( \frac{\bar{n}_{\mathrm{H}}}{10^{4}~\mathrm{H~cm^{-3}}} \right) ^{-0.5} \left( \frac{T_{\mathrm{gas}}}{10~\mathrm{K}} \right) ^{1.5} \mathrm{M_{\sun}} ,
\end{equation}

\noindent
from~\citet{Lada:08}. Both the values of $M$ and $M_{\mathrm{BE}}$ for all of the cores are listed in Table~\ref{Table:Clump_Props}. We find that 29 of our 34 cores are unstable, and therefore undergoing collapse.

%%%%%%% YSOs %%%%%%%%%%%%%%%%%%%%%%%%%%%%%%%%%%%%%%%%%%%%

\subsection{Young Stellar Objects}
\label{Subsec:YSOs}

Since the unstable cores will collapse and eventually form stars, we also investigated the ongoing star formation by searching for YSOs in the vicinity of our cores. Various IR colour-cuts have been used to identify YSOs, e.g., based on \textit{Spitzer}~\citep[e.g.,][]{Allen:04} and \textit{WISE}~\citep[e.g.,][]{Koenig:12} data. Since not all of our clumps have been observed with \textit{Spitzer}, we take advantage of the all-sky coverage of \textit{WISE}, which mapped the entire sky simultaneously at four wavelengths: 3.4, 4.6, 12, and 22~$\mu$m~\citep{Wright:10}. We apply the colour-colour selection criteria developed by~\citet{Koenig:12} to the \textit{AllWISE} all-sky source catalog to identify YSO candidates, excluding sources with a signal-to-noise ratio~$<$ 5 and contaminated sources identified by the catalog contamination flags ``D'' (diffraction spikes), ``P'' (persistent latent artefacts), ``H'' (halos from bright sources), and ``O'' (optical ghosts from bright sources). This selection criteria allows us to reject extragalactic contaminants and distinguish between class I and class II YSO candidates~\citep[for details of the specific selection criteria, see Appendices A.1, A.2, and A.3 in][]{Koenig:12}.

This analysis enabled us to identify YSO candidates associated with each clump. We found that only 5 of our 15 clumps have at least one YSO candidate within our 20~arcmin~$\times$~20~arcmin maps displayed in Figure~\ref{Fig:NH_Td_Maps}, and that only 4 of these~(ECC190, ECC191, ECC229, and ECC276) have a YSO candidate within the CARMA primary beam. The location of the YSO candidates, and their classification~(either class~\textsc{i} or class~\textsc{ii}), are marked on Figure~\ref{Fig:NH_Td_Maps}. It is widely accepted that no colour-colour identification criteria are 100\% accurate. Using the contamination rate of false identifications from~\citet{Koenig:12}, we estimate that there are $\sim$0.4~--~0.9 false identifications in each of our maps.

Looking at the location of these YSO candidates relative to our cores, we determined that 6 cores are possibly associated with a YSO (see Table~\ref{Table:Clump_Props}). Core 1 in ECC191 is coincident with a class \textsc{i} YSO candidate, while cores 2, 3, and 4 in ECC229 are coincident with two class \textsc{i} YSO candidates, a class \textsc{ii} YSO candidate, and a class \textsc{i} YSO candidate, respectively. In addition, there are 2 cores which have YSO candidates nearby: core 1 in ECC229 has two nearby class \textsc{ii} YSO candidates, and core 2 in ECC276 has a class \textsc{i} YSO candidate nearby.

YSOs have previously been detected at cm wavelengths~\citep[e.g.,][]{Scaife:11}, but of the 6 cores that are possibly associated with a YSO candidate, only three have detectable cm emission (ECC229 cores 1, 2, and 3). However, given the variety of mechanisms by which YSOs can produce cm emission, perhaps it is not surprising that we do not detect cm emission from all of our candidates. This also makes it more difficult to interpret the cm emission that we detect in ECC229, as we will discuss in Section~\ref{Subsec:ECC229}.

%%%%%%% Discussion %%%%%%%%%%%%%%%%%%%%%%%%%%%%%%%%%%%%%%%%%

\section{Discussion}
\label{Sec:Discussion}

%%%%%%% ECC229 %%%%%%%%%%%%%%%%%%%%%%%%%%%%%%%%%%%%%%%%%

\subsection{ECC229}
\label{Subsec:ECC229}

Out of all of our 15 clumps, ECC229 is the only clump for which we detect any extended cm emission, making it particularly interesting. In our CARMA map of ECC229 we detected three components of cm emission~(associated with cores 1, 2, and 3), although we identified four cores. Based on the values listed in Table~\ref{Table:Clump_Props}, the cores in ECC229 have the highest values of $\bar{N}$$_{\mathrm{H}}$. Moreover, as shown in Figure~\ref{Fig:NH_Td_Maps}, the cm emission appears to be spatially correlated with $N_{\mathrm{H}}$ and originating from the densest regions.

Core 2 in ECC229 has the highest density~($\bar{n}$$_{\mathrm{H}}$~=~(123.7 $\pm$ 17.3) $\times$~10$^{3}$~H~cm$^{-3}$), but also the warmest dust temperature~($\bar{T}_{\mathrm{d}}$~=~14.3~$\pm$~0.3~K) and highest radiation field value~($\bar{G}_{\mathrm{0}}$~=~0.298~$\pm$~0.041), potentially suggesting that the star formation process has already begun. Indeed two YSO candidates, likely heating the dust in the core, are found to be associated with core 2. In fact all 4 of the cores in ECC229 appear to be associated with a YSO candidate, and the fact that we only detect cm emission from 3 of the cores might suggest that the cm emission is not linked to the presence of the YSOs. However, given that YSOs can produce cm emission via a variety of different mechanisms~(accretion, stellar winds, shock ionization), it is far from clear. For example, the spectral index at cm wavelengths~(defined as $\alpha$, where $S_{\nu} \propto \nu^{\alpha}$) predicted from free-free emission from an ionized stellar wind can cover the range from $-0.1 \le \alpha \le+2.0$, with typical values of~$\alpha$~$\sim$0.6 in the case of a spherical, isothermal stellar wind~\citep{Reynolds:86}. A similar range of values for~$\alpha$ are produced for free-free emission from accretion and shock ionization. The spectral index from spinning dust emission at these wavelengths is dependent on the frequency of the peak of the emission, and so~$\alpha$ can be both positive or negative. This makes it difficult to distinguish between the cm emission arising from YSOs as opposed to spinning dust emission, especially in the absence of multi-wavelength cm observations, which can closely trace the microwave emission, combined with IR spectroscopic data, which can provide details on the YSO geometry (e.g., disk inclination, etc.). Therefore, lacking this ancillary information, to determine the presence of spinning dust emission~(or a lack thereof), we compared our CARMA observations to the predicted level of spinning dust emission based on the physical conditions in each core.

\begin{figure*}
\begin{center}
\includegraphics[angle=0,scale=0.450]{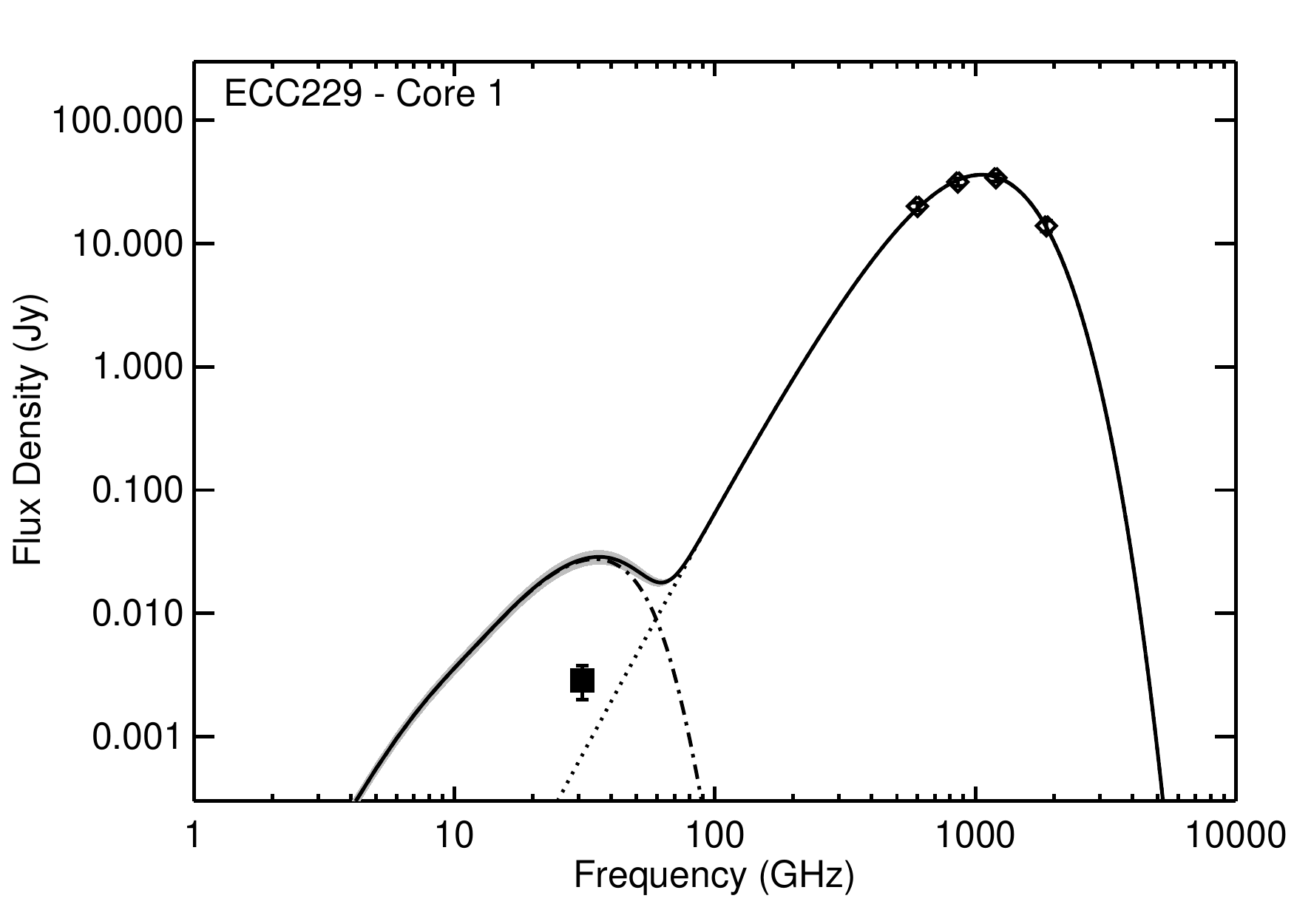}
\includegraphics[angle=0,scale=0.450]{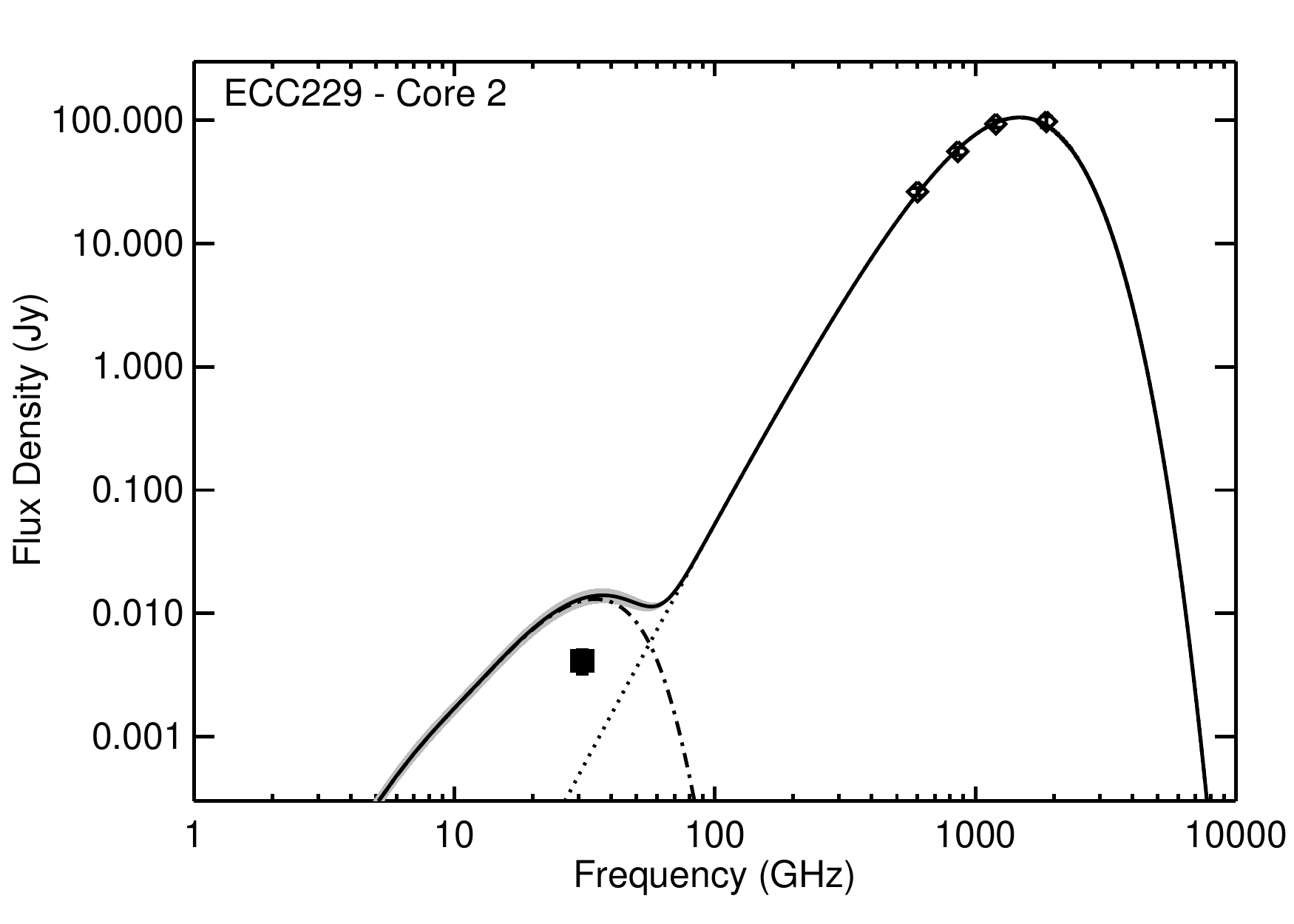} \\
\vspace{-0.35cm}
\includegraphics[angle=0,scale=0.450]{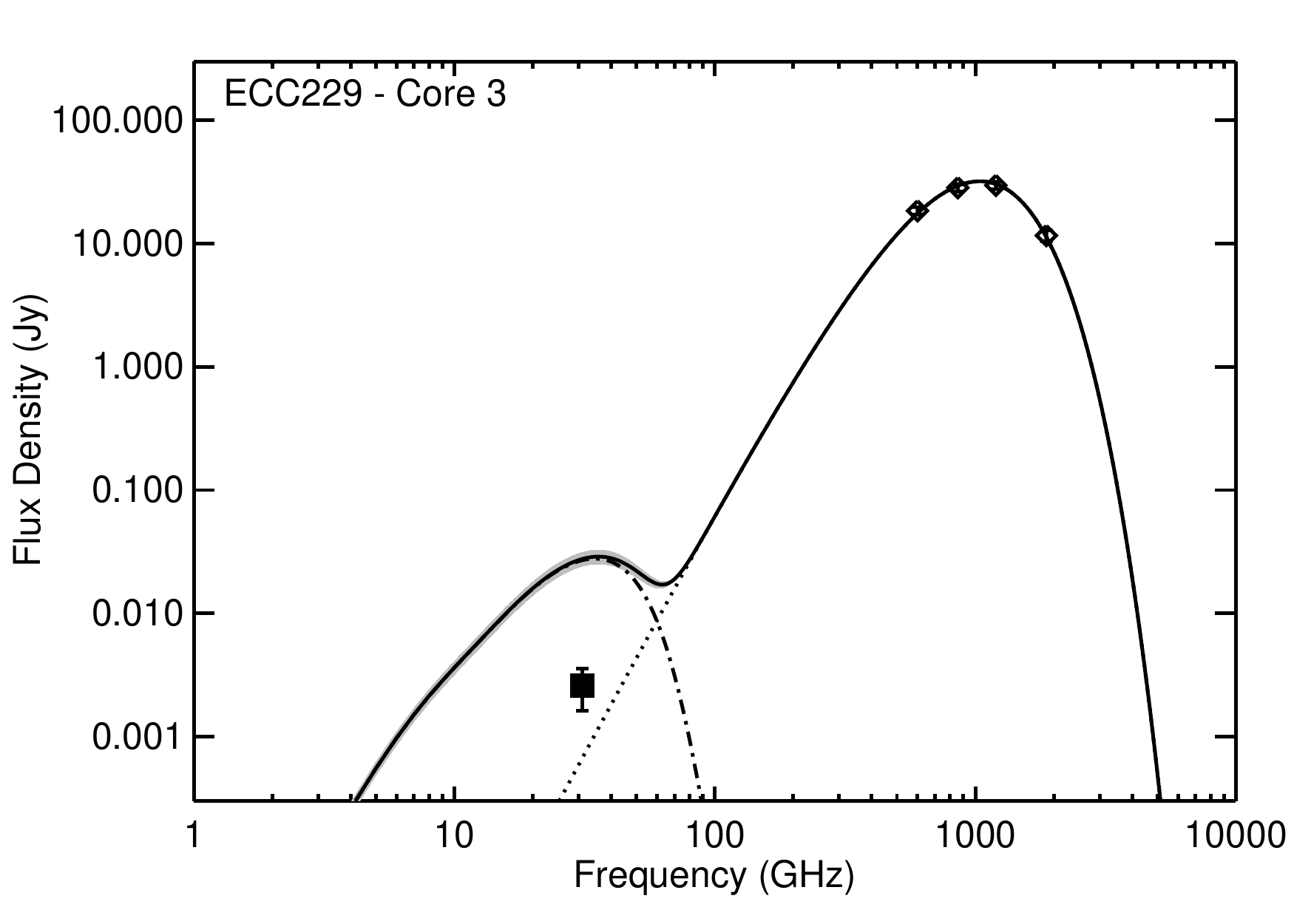} 
\includegraphics[angle=0,scale=0.450]{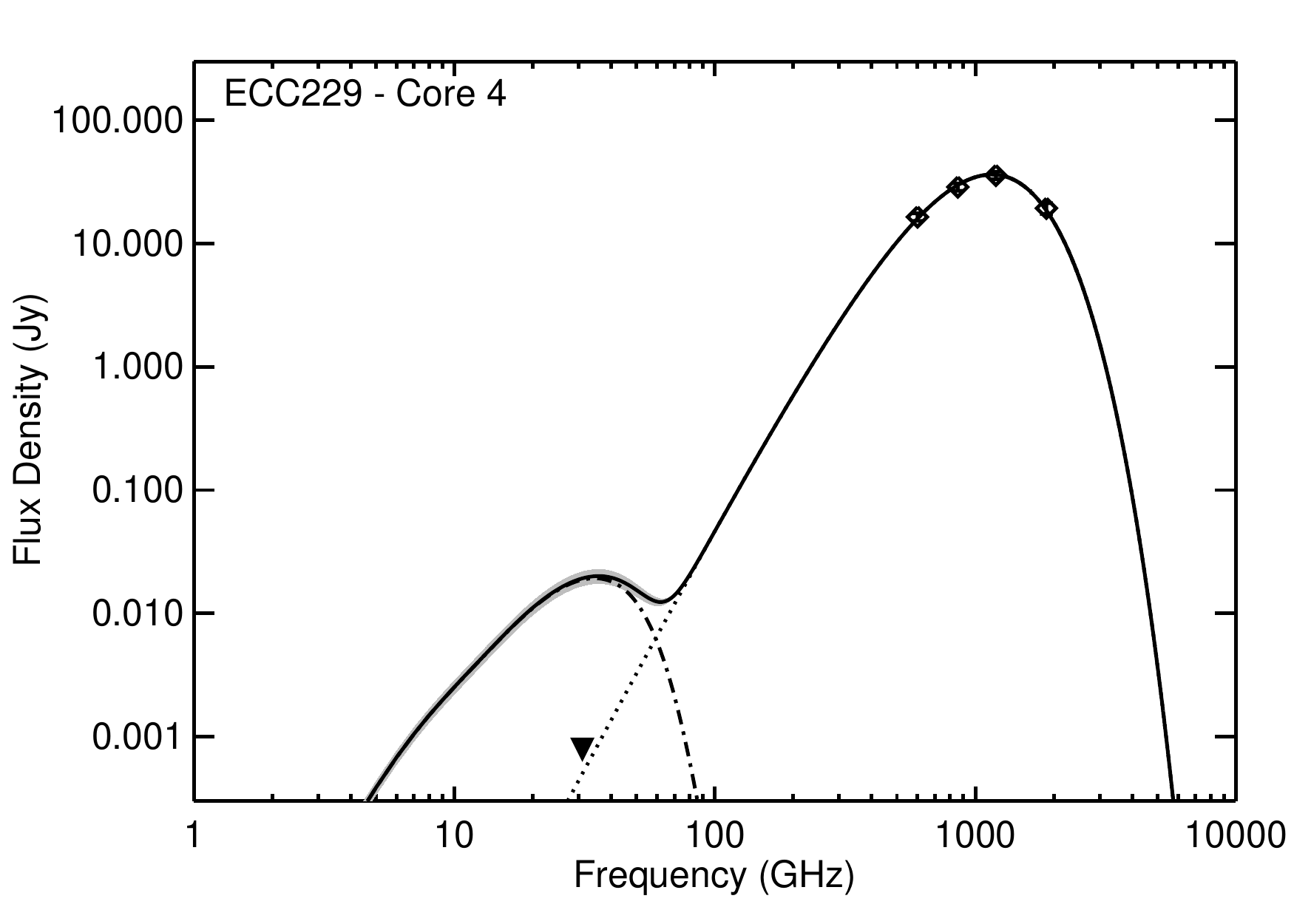} \\
\end{center}
\vspace{-0.30cm}
\caption{Spectrum for the four cores in ECC229, consisting of both a spinning dust~(\textit{dot-dashed line}) and thermal dust~(\textit{dotted line}) component. The thermal dust emission is modelled using a modified black body with $\beta$ fixed at 2, and is fitted to the integrated flux density at 160, 250, 350, and 500~$\mu$m~(\textit{diamonds}) in each core, while the spinning dust curve, and associated uncertainty~(\textit{shaded region}), is the predicted level of spinning dust emission based on the modelling analysis. Also plotted is the measured flux density at 1~cm~(for core 4 in which we do not detect any cm emission, we have plotted the 5$\sigma$ upper limit), illustrating how the predicted spinning dust emission is larger than what is observed with CARMA.}
\label{Fig:Spinning_Dust_Curves}
\end{figure*}

%%%%%%% Spinning Dust Emission %%%%%%%%%%%%%%%%%%%%%%%%%%%%%%%%%%%%%%%%%

\subsection{Spinning Dust Emission}
\label{Subsec:Spinning_Dust}

With the physical properties of the clumps characterized, as described in Section~\ref{Sec:Dust_Properties}, we can investigate the significance of our CARMA observations, namely: does the fact that we did not detect cm emission in 14 of our 15 clumps signify that there is no spinning dust emission present? In order to address this question, we computed the expected level of spinning dust emission, using the constraints provided by the analysis of the \textit{Herschel} data.

We characterized our cores by using the $\bar{n}$$_{\mathrm{H}}$ and $\bar{G}$$_{\mathrm{0}}$ values derived in Section~\ref{Subsec:H_Number_Density}, the gas temperatures, $T_{\mathrm{gas}}$, listed in Table~\ref{Table:Sources}, and the standard values for dense environments for the hydrogen ionization fraction~($x_{\mathrm{H}}$ = 0), the carbon ionization fraction ($x_{\mathrm{C}}$ = 10$^{-6}$), and the molecular hydrogen fraction ($y$ = 0.999) as defined by~\citet[][]{DaL:98}. In addition to these parameters, and since these are dense environments, we used the~\citet{Weingartner:01} grain size distribution for $R_{\mathrm{V}} = 5.5$ with $b_{\mathrm{C}} = 3\times10^{-5}$~\citep[corresponding to the parameters in][Table 1, Line 16]{Weingartner:01}. With these parameters as inputs, we used the spinning dust model, \textsc{spdust}~\citep{Silsbee:11} to predict the expected level of spinning dust emission at 1~cm.

\textsc{spdust} estimates the spinning dust emissivity in units of Jy~sr$^{-1}$~H$^{-1}$~cm$^{2}$ which we converted to a flux by using our computed value of $\bar{N}$$_{\mathrm{H}}$. To estimate the uncertainty on the predicted spinning dust emission, we repeated this modelling analysis 1000 times, randomising the input $\bar{n}$$_{\mathrm{H}}$ and $\bar{G}$$_{\mathrm{0}}$ values within their uncertainty, resulting in a range of predicted values for the spinning dust emission at 1~cm. We then compared the predicted spinning dust emission with our CARMA maps. For the cores for which we do not detect any cm emission we computed a conservative 5$\sigma$ upper limit by scaling the rms noise in the CARMA maps based on the ratio of the size of the beam to the size of the core, while for ECC229, we fitted the extended emission with three 2-dimensional Gaussian components using the \textsc{aips} task \textsc{jmfit}. The last two columns in Table~\ref{Table:Clump_Props} list our measured 1~cm flux densities~($S_{1~\mathrm{cm}}^{\mathrm{observed}}$) along with the predicted levels of the spinning dust emission~($S_{1~\mathrm{cm}}^{\mathrm{predicted}}$). For all of our cores we find that the predicted level of spinning dust emission is much larger than what we observe in our CARMA maps. We illustrate this result for the four cores in ECC229 in Figure~\ref{Fig:Spinning_Dust_Curves}, where we plot the predicted spinning dust emission, with its associated uncertainty, the thermal dust emission, modelled using a modified black body with $\beta$ fixed at 2, and our CARMA 1~cm flux density measurements.

We do note, however, that this result depends strongly on the spinning dust modelling analysis. Firstly, since we are only comparing our observed cm emission with the predicted spinning dust emission, we are actually being conservative as we are ignoring any possible contribution to the 1~cm emission from thermal dust emission or free-free emission, which would increase the expected level of the 1~cm emission. Secondly, it is possible that the discrepancy between the observed and the predicted level of spinning dust emission could be due to an incorrect assumption in our spinning dust modelling. For example, there is much uncertainty on the value of the electric dipole moment of dust grains. The value used in this analysis is consistent with the standard value estimated by~\citet{DaL:98}~--~an average rms dipole moment per atom of~0.38~D. If this value is overestimated, it would result in an overestimate of the level of the spinning dust emission. Likewise, we might have overestimated $\bar{n}$$_{\mathrm{H}}$, as a consequence of the uncertainty in the distance estimate or due to our assumption that the core geometry is spherical. Similarly, our estimate of $\bar{T}$$_{\mathrm{d}}$, and hence $\bar{G}$$_{0}$, may be overestimated due to the mixing of different line-of-sight temperature variations~\citep[e.g.][]{Nielbock:12, Roy:14}. Again, such overestimates would result in us overestimating the spinning dust emission. Finally, our adopted grain size distribution might not be appropriate for Galactic cores. Although we used the grain size distribution for dense environments with $R_{\mathrm{V}}$ = 5.5 (as opposed to values of $R_{\mathrm{V}}$ = 3.1 or $R_{\mathrm{V}}$ = 4.0), this could still be overestimating the abundance of the small grains responsible for the spinning dust emission. In a forthcoming paper~\citep{Tibbs:15b} we will investigate these scenarios in more detail and, in particular, use these observations to place a constraint on the abundance of the small dust grains in our sample of cores.

Finally, we stress that the sample of Galactic cores which was the subject of this study does not span the full parameter space (in terms of density, temperature, mass, etc.) relative to the Galactic cold core population. For this reason, the conclusions reached by our analysis cannot easily be generalized, and more cm observations like the ones presented are needed to allow confirmation of our findings.

%%%%%%% Conclusions %%%%%%%%%%%%%%%%%%%%%%%%%%%%%%%%%%%%%%%%%%%%%%%%

\section{Conclusions}
\label{Sec:Conclusions}

We have attempted, for the first time, to search for spinning dust emission in a sample of Galactic cold clumps. To do this we observed a sample of 15 cold clumps with CARMA at 1~cm, and found that only 1 of our 15 clumps exhibited significant extended emission. To determine if the lack of detection of 1~cm emission could rule out a spinning dust detection, we investigated the physical properties of our sample of clumps using ancillary \textit{Herschel} and \textit{WISE} data. Using \textit{Herschel} photometric data at 160, 250, 350, and 500~$\mu$m of our 15 cold clumps we produced maps of $N_{\mathrm{H}}$ and $T_{\mathrm{d}}$, from which we identified 34 cores, characterized by the densest and coldest regions within each of the 15 clumps. For each core we estimated $\bar{N}$$_{\mathrm{H}}$, $\bar{T}$$_{\mathrm{d}}$, $\bar{n}$$_{\mathrm{H}}$, and $\bar{G}$$_{\mathrm{0}}$. Making use of the all-sky coverage of \textit{WISE}, we used the \textit{AllWISE} source catalog to identify candidate YSOs and found that 6 of our 34 cores were associated with a YSO candidate. 

With the physical environments of each core constrained, we used \textsc{spdust} to model the spinning dust emission, which we compared to our CARMA observations, and we found that the observed cm emission in all of our cores was below the predicted level. This implies that we do not detect spinning dust emission from 14 of our 15 clumps, and in the 1 clump in which we do detect cm emission, it could be due to either spinning dust emission, but at a level much lower than predicted based on the modelling, or it could be due to free-free emission from YSOs. 

This analysis is the first attempt to detect spinning dust emission in Galactic cores, however, we emphasize that our sample is not statistically representative of the entire Galactic cold core population, and therefore we are cautious to extend this result to all of the cold cores in the Galaxy.

%%%%%%% Acknowledgments %%%%%%%%%%%%%%%%%%%%%%%%%%%%%%%%%%%%%%%%%%%%

\section*{Acknowledgments}

We thank the anonymous referee for providing detailed comments that have improved the content of this paper. 

This work has been performed within the framework of a NASA/ADP ROSES-2009 grant, no. 09-ADP09-0059. 

Support for CARMA construction was derived from the Gordon and Betty Moore Foundation, the Kenneth T. and Eileen L. Norris Foundation, the James S. McDonnell Foundation, the Associates of the California Institute of Technology, the University of Chicago, the states of California, Illinois, and Maryland, and the National Science Foundation. Ongoing CARMA development and operations are supported by the National Science Foundation under a cooperative agreement, and by the CARMA partner universities.

This research has made use of the NASA/IPAC Infrared Science Archive, which is operated by the Jet Propulsion Laboratory, California Institute of Technology, under contract with the National Aeronautics and Space Administration.

This publication makes use of data products from the Wide-field Infrared Survey Explorer, which is a joint project of the University of California, Los Angeles, and the Jet Propulsion Laboratory/California Institute of Technology, and NEOWISE, which is a project of the Jet Propulsion Laboratory/California Institute of Technology. WISE and NEOWISE are funded by the National Aeronautics and Space Administration.

%%%%%%%%%%%%%%%  BIBLIOGRAPHY  %%%%%%%%%%%%%%%

%%%%%%%%%%%%%%%%%%%%%%%%%%%%%%%%%%%%%%%%%%%%%%%%%%%%%%%%%%%%%%%%%%%%

\bsp % ``This paper has been produced using the ...''

\newpage
\begin{landscape}
\begin{table}
\begin{center}
\caption{Derived properties of the cores analysed in this work, including the core size, mean column density ($\bar{N}$$_{\mathrm{H}}$), mean density ($\bar{n}$$_{\mathrm{H}}$), mean dust temperature ($\bar{T}$$_{\mathrm{d}}$), mean radiation field ($\bar{G}$$_{\mathrm{0}}$), mass ($M$), Bonnor-Ebert mass ($M_{BE}$), associated YSOs (YSO), observed 1~cm flux density ($S_{1~\mathrm{cm}}^{\mathrm{observed}}$), and the predicted 1~cm spinning dust emission ($S_{1~\mathrm{cm}}^{\mathrm{predicted}}$).}
\scalebox{0.95}{%
\begin{tabular}{cccccccccccccc}
\hline
Clump & Core & R.A. & Decl. & Size & $\bar{N}$$_{\mathrm{H}}$ & $\bar{n}$$_{\mathrm{H}}$ & $\bar{T}$$_{\mathrm{d}}$ & $\bar{G}$$_{\mathrm{0}}$ & $M$ & $M_{BE}$ & YSO & $S_{1~\mathrm{cm}}^{\mathrm{observed}}$ & $S_{1~\mathrm{cm}}^{\mathrm{predicted}}$ \\
 & & (J2000) & (J2000) & (pc) & ($10^{20} \mathrm{H~cm^{-2}}$) & ($10^{3} \mathrm{H~cm}^{-3}$) & (K) & & (M$_{\odot}$) & (M$_{\odot}$) & & (mJy) & (mJy) \\
 \hline
 \hline

\multirow{3}{*}{ECC181} 	& 1 & 20:41:13.2 & 67:20:34.2 & 0.23 & 121.1 $\pm$ 10.5 & 25.6 $\pm$ 3.4 & 11.6 $\pm$ 0.2 & 0.086 $\pm$ 0.008 & 5.6 $\pm$ 0.9 & 1.1 & - & <~2.87 & 9.09 -- 12.16 \\
 					& 2 & 20:40:56.2 & 67:22:54.4 & 0.18 & 115.2 $\pm$ 10.0 & 32.0 $\pm$ 4.2 & 11.7 $\pm$ 0.2 & 0.090 $\pm$ 0.009 & 3.1 $\pm$ 0.5 & 1.0 & - & <~1.66 & 5.10 -- 6.79 \\
 					& 3 & 20:40:31.9 & 67:20:48.1 & 0.13 & 107.4 $\pm$ 9.3 & 41.3 $\pm$ 5.5 & 11.4 $\pm$ 0.2 & 0.078 $\pm$ 0.007 & 1.5 $\pm$ 0.2 & 0.8 & - & <~0.86 & 2.53 -- 3.34 \\
\\
\multirow{1}{*}{ECC189} 	& 1 & 20:53:35.4 & 68:19:19.9 & 0.17 & 65.0 $\pm$ 5.7 & 18.7 $\pm$ 2.5 & 12.3 $\pm$ 0.2 & 0.120 $\pm$ 0.012 & 1.6 $\pm$ 0.3 & 1.3 & - & <~1.30 & 3.47 -- 4.84 \\
\\
\multirow{2}{*}{ECC190} 	& 1 & 21:01:54.4 & 67:43:45.7 & 0.20 & 128.3 $\pm$ 11.1 & 31.6 $\pm$ 4.2 & 12.3 $\pm$ 0.2 & 0.122 $\pm$ 0.012 & 4.4 $\pm$ 0.7 & 1.2 & - & <~2.15 & 10.53 -- 14.04 \\
 					& 2 & 21:02:21.5 & 67:45:37.7 & 0.21 & 110.7 $\pm$ 9.7 & 25.4 $\pm$ 3.4 & 12.7 $\pm$ 0.2 & 0.146 $\pm$ 0.015 & 4.4 $\pm$ 0.7 & 1.3 & - & <~2.48 & 10.25 -- 13.72 \\
\\
\multirow{1}{*}{ECC191} 	& 1 & 21:02:23.2 & 67:54:15.2 & 0.21 & 274.9 $\pm$ 23.8 & 63.8 $\pm$ 8.4 & 11.6 $\pm$ 0.2 & 0.083 $\pm$ 0.008 & 10.6 $\pm$ 1.8 & 0.8 & Class~\textsc{i} & <~4.80 & 26.62 -- 34.99 \\
\\
\multirow{1}{*}{ECC223} 	& 1 & 21:59:38.8 & 76:33:13.1 & 1.06 & 115.2 $\pm$ 10.2 & 5.3 $\pm$ 0.7 & 12.1 $\pm$ 0.2 & 0.108 $\pm$ 0.011 & 113.6 $\pm$ 18.9 & 2.1 & - & <~5.69 & 25.15 -- 48.26 \\
\\
\multirow{3}{*}{ECC224} 	& 1 & 22:21:27.6 & 75:04:26.0 & 0.48 & 130.7 $\pm$ 11.6 & 13.4 $\pm$ 1.8 & 12.2 $\pm$ 0.2 & 0.112 $\pm$ 0.011 & 25.9 $\pm$ 4.3 & 1.1 & - & <~2.36 & 6.29 -- 12.85 \\
 					& 2 & 22:21:41.2 & 75:06:06.0 & 0.51 & 117.2 $\pm$ 10.4 & 11.2 $\pm$ 1.5 & 12.4 $\pm$ 0.2 & 0.129 $\pm$ 0.013 & 26.8 $\pm$ 4.5 & 1.2 & - & <~2.72 & 9.59 -- 13.79 \\
 					& 3 & 22:21:50.5 & 75:09:09.5 & 0.28 & 83.7 $\pm$ 7.4 & 14.6 $\pm$ 1.9 & 12.6 $\pm$ 0.2 & 0.138 $\pm$ 0.014 & 5.7 $\pm$ 1.0 & 1.1 & - & <~0.81 & 1.65 -- 3.13 \\
\\
\multirow{1}{*}{ECC225} 	& 1 & 22:24:09.1 & 75:04:33.1 & 0.71 & 110.5 $\pm$ 9.6 & 7.5 $\pm$ 1.0 & 12.0 $\pm$ 0.2 & 0.105 $\pm$ 0.010 & 49.4 $\pm$ 8.2 & 1.7 & - & <~4.49 & 14.97 -- 22.46 \\
\\
\multirow{4}{*}{ECC229} 	& 1 & 22:39:39.5 & 75:12:01.5 & 0.34 & 676.0 $\pm$ 57.6 & 97.0 $\pm$ 12.7 & 10.2 $\pm$ 0.1 & 0.040 $\pm$ 0.003 & 68.1 $\pm$ 11.3 & 0.6 & 2$\times$Class~\textsc{ii} & 2.87~$\pm$~0.88 & 23.16 -- 30.30 \\
 					& 2 & 22:38:47.9 & 75:11:27.2 & 0.24 & 619.5 $\pm$ 60.5 & 123.7 $\pm$ 17.3 & 14.3 $\pm$ 0.3 & 0.298 $\pm$ 0.041 & 32.3 $\pm$ 5.6 & 0.5 & 2$\times$Class~\textsc{i} & 4.12~$\pm$~0.91 & 10.83 -- 14.45 \\
 					& 3 & 22:39:31.6 & 75:11:06.6 & 0.35 & 642.6 $\pm$ 54.8 & 89.5 $\pm$ 11.8 & 10.1 $\pm$ 0.1 & 0.036 $\pm$ 0.003 & 68.8 $\pm$ 11.4 & 0.6 & Class~\textsc{ii} & 2.59~$\pm$~0.96 & 23.36 -- 30.59 \\
 					& 4 & 22:39:06.5 & 75:11:52.5 & 0.32 & 540.3 $\pm$ 47.4 & 82.7 $\pm$ 11.0 & 11.2 $\pm$ 0.2 & 0.068 $\pm$ 0.006 & 48.0 $\pm$ 8.0 & 0.7 & Class~\textsc{i} & <~0.78 & 16.18 -- 21.28 \\
\\
\multirow{2}{*}{ECC276} 	& 1 & 01:38:34.7 & 65:05:48.7 & 0.47 & 161.4 $\pm$ 14.4 & 16.7 $\pm$ 2.2 & 11.1 $\pm$ 0.2 & 0.065 $\pm$ 0.006 & 31.5 $\pm$ 5.3 & 2.0 & - & <~0.56 & 4.94 -- 6.70 \\
 					& 2 & 01:38:30.3 & 65:04:52.7 & 0.45 & 173.6 $\pm$ 15.9 & 18.6 $\pm$ 2.5 & 11.6 $\pm$ 0.2 & 0.086 $\pm$ 0.009 & 31.4 $\pm$ 5.3 & 1.9 & Class~\textsc{i} & <~0.52 & 4.97 -- 6.74 \\
\\
\multirow{4}{*}{ECC332} 	& 1 & 04:17:23.9 & 55:16:15.7 & 0.20 & 79.5 $\pm$ 7.3 & 19.8 $\pm$ 2.7 & 13.4 $\pm$ 0.3 & 0.203 $\pm$ 0.024 & 2.7 $\pm$ 0.5 & 1.1 & - & <~6.24 & 20.90 -- 37.23 \\
 					& 2 & 04:17:04.3 & 55:13:55.0 & 0.08 & 74.5 $\pm$ 7.0 & 47.2 $\pm$ 6.5 & 13.4 $\pm$ 0.3 & 0.204 $\pm$ 0.025 & 0.4 $\pm$ 0.1 & 0.7 & - & <~0.96 & 1.97 -- 2.64 \\
 					& 3 & 04:17:02.6 & 55:15:47.0 & 0.15 & 68.2 $\pm$ 6.4 & 22.1 $\pm$ 3.0 & 13.6 $\pm$ 0.3 & 0.218 $\pm$ 0.027 & 1.4 $\pm$ 0.2 & 1.0 & - & <~3.69 & 10.20 -- 19.20 \\
 					& 4 & 04:16:57.5 & 55:20:13.2 & 0.16 & 63.8 $\pm$ 5.9 & 19.2 $\pm$ 2.6 & 13.6 $\pm$ 0.3 & 0.222 $\pm$ 0.027 & 1.5 $\pm$ 0.2 & 1.1 & - & <~4.25 & 12.86 -- 20.27 \\
\\
\multirow{1}{*}{ECC334} 	& 1 & 04:18:25.6 & 55:12:48.6 & 0.29 & 89.4 $\pm$ 8.2 & 14.8 $\pm$ 2.0 & 13.3 $\pm$ 0.3 & 0.196 $\pm$ 0.023 & 6.7 $\pm$ 1.1 & 1.2 & - & <~11.35 & 50.94 -- 72.20 \\
\\
\multirow{3}{*}{ECC335} 	& 1 & 04:18:51.6 & 55:14:44.9 & 0.15 & 77.7 $\pm$ 7.0 & 24.4 $\pm$ 3.3 & 13.3 $\pm$ 0.3 & 0.191 $\pm$ 0.022 & 1.6 $\pm$ 0.3 & 0.8 & - & <~4.80 & 10.43 -- 21.45 \\
 					& 2 & 04:19:16.2 & 55:14:02.2 & 0.16 & 91.9 $\pm$ 8.2 & 28.1 $\pm$ 3.8 & 12.9 $\pm$ 0.2 & 0.163 $\pm$ 0.018 & 2.0 $\pm$ 0.3 & 0.8 & - & <~5.08 & 11.22 -- 16.06 \\
 					& 3 & 04:19:22.8 & 55:15:54.1 & 0.13 & 86.6 $\pm$ 7.6 & 33.2 $\pm$ 4.4 & 12.4 $\pm$ 0.2 & 0.126 $\pm$ 0.013 & 1.2 $\pm$ 0.2 & 0.7 & - & <~3.24 & 6.48 -- 8.63 \\
\\
\multirow{4}{*}{ECC340} 	& 1 & 04:29:27.5 & 54:14:37.5 & 0.07 & 95.3 $\pm$ 8.7 & 63.7 $\pm$ 8.6 & 12.9 $\pm$ 0.2 & 0.159 $\pm$ 0.018 & 0.4 $\pm$ 0.1 & 0.7 & - & <~0.54 & 2.37 -- 3.15 \\
 					& 2 & 04:29:48.3 & 54:16:01.8 & 0.08 & 94.7 $\pm$ 9.0 & 57.8 $\pm$ 8.0 & 13.8 $\pm$ 0.3 & 0.238 $\pm$ 0.030 & 0.5 $\pm$ 0.1 & 0.8 & - & <~0.65 & 2.79 -- 3.74 \\
 					& 3 & 04:29:41.9 & 54:14:09.6 & 0.08 & 90.6 $\pm$ 8.5 & 52.7 $\pm$ 7.2 & 13.4 $\pm$ 0.3 & 0.203 $\pm$ 0.025 & 0.6 $\pm$ 0.1 & 0.8 & - & <~0.72 & 2.93 -- 3.92 \\
 					& 4 & 04:29:51.5 & 54:14:09.7 & 0.09 & 93.2 $\pm$ 8.8 & 52.6 $\pm$ 7.2 & 13.5 $\pm$ 0.3 & 0.215 $\pm$ 0.026 & 0.6 $\pm$ 0.1 & 0.8 & - & <~0.76 & 3.20 -- 4.28 \\
\\
\multirow{2}{*}{ECC345} 	& 1 & 04:47:15.7 & 53:01:39.2 & 0.18 & 77.1 $\pm$ 7.3 & 20.8 $\pm$ 2.9 & 13.0 $\pm$ 0.3 & 0.171 $\pm$ 0.020 & 2.2 $\pm$ 0.4 & 0.8 & - & <~1.70 & 3.50 -- 7.52 \\
 					& 2 & 04:47:17.2 & 53:04:27.4 & 0.21 & 70.4 $\pm$ 6.7 & 16.0 $\pm$ 2.2 & 13.2 $\pm$ 0.3 & 0.183 $\pm$ 0.022 & 2.9 $\pm$ 0.5 & 0.9 & - & <~2.40 & 6.82 -- 9.67 \\
\\
\multirow{2}{*}{ECC346} 	& 1 & 04:48:08.7 & 53:07:23.0 & 0.13 & 208.0 $\pm$ 18.3 & 78.8 $\pm$ 10.5 & 11.9 $\pm$ 0.2 & 0.098 $\pm$ 0.010 & 3.0 $\pm$ 0.5 & 0.6 & - & <~2.13 & 11.06 -- 14.57 \\
 					& 2 & 04:48:01.0 & 53:09:15.3 & 0.12 & 205.7 $\pm$ 18.1 & 83.6 $\pm$ 11.1 & 11.6 $\pm$ 0.2 & 0.086 $\pm$ 0.008 & 2.6 $\pm$ 0.4 & 0.6 & - & <~1.85 & 9.52 -- 12.53 \\
\\

\hline
\label{Table:Clump_Props}
\end{tabular}}
\end{center}
\end{table}
\end{landscape}
\newpage

%%%%%%% Appendices %%%%%%%%%%%%%%%%%%%%%%%%%%%%%%%%%%%%%%%%%%%%

%\newpage
\appendix

\section{CARMA Maps}
\label{appendix1}

\vspace{-0.5cm}
\begin{figure*}
\begin{center}
\includegraphics[angle=0,scale=0.375]{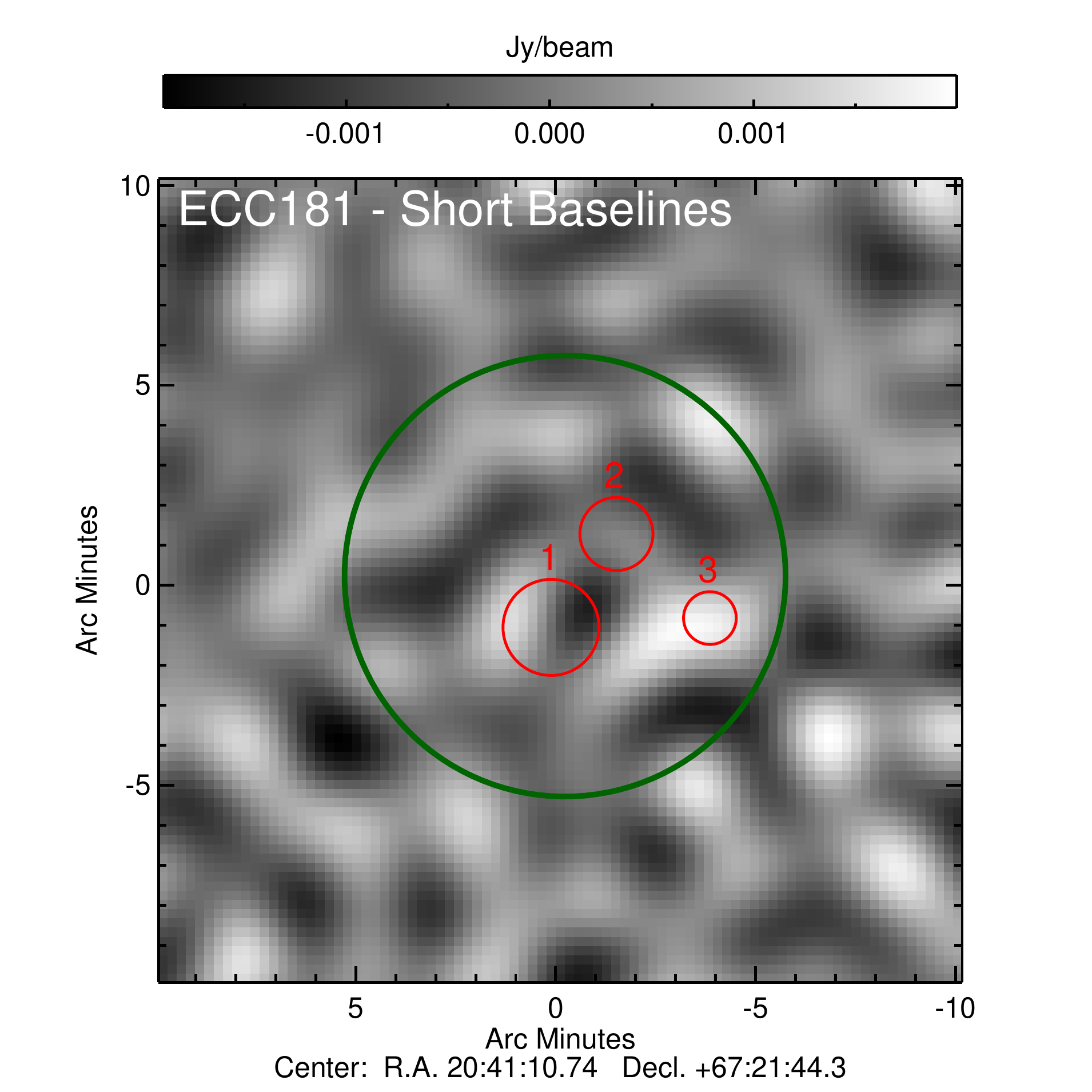}
\includegraphics[angle=0,scale=0.375]{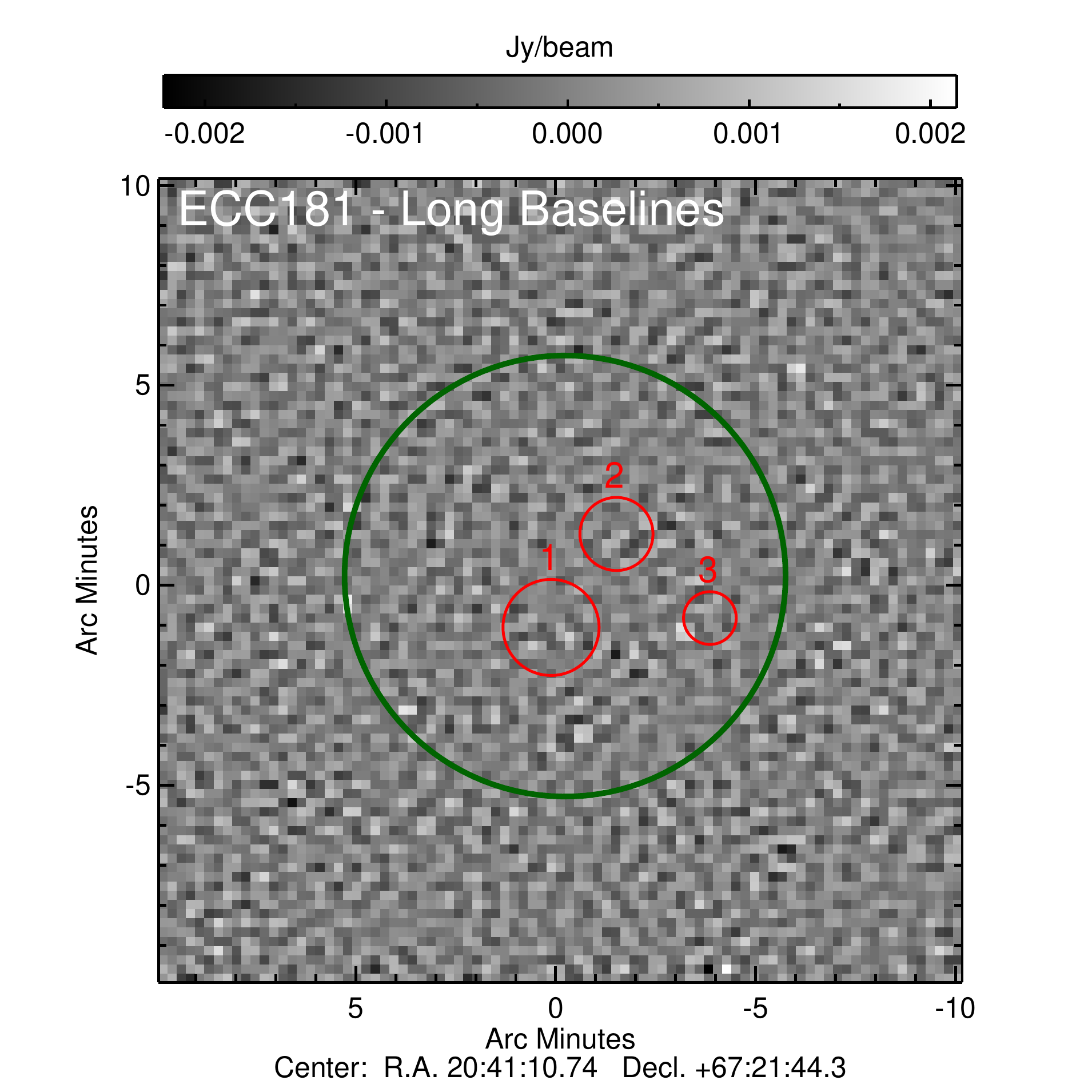} \\

\includegraphics[angle=0,scale=0.375]{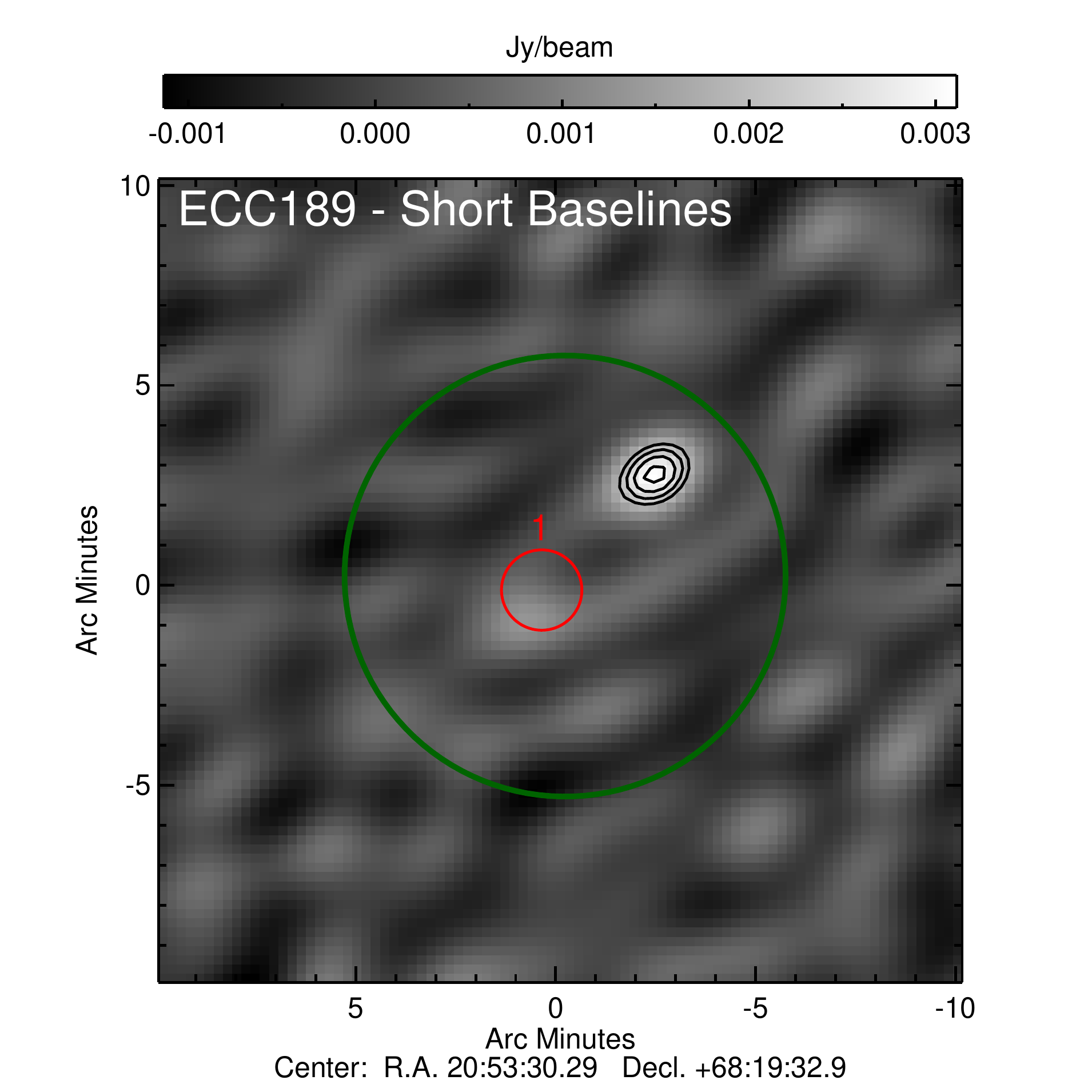}
\includegraphics[angle=0,scale=0.375]{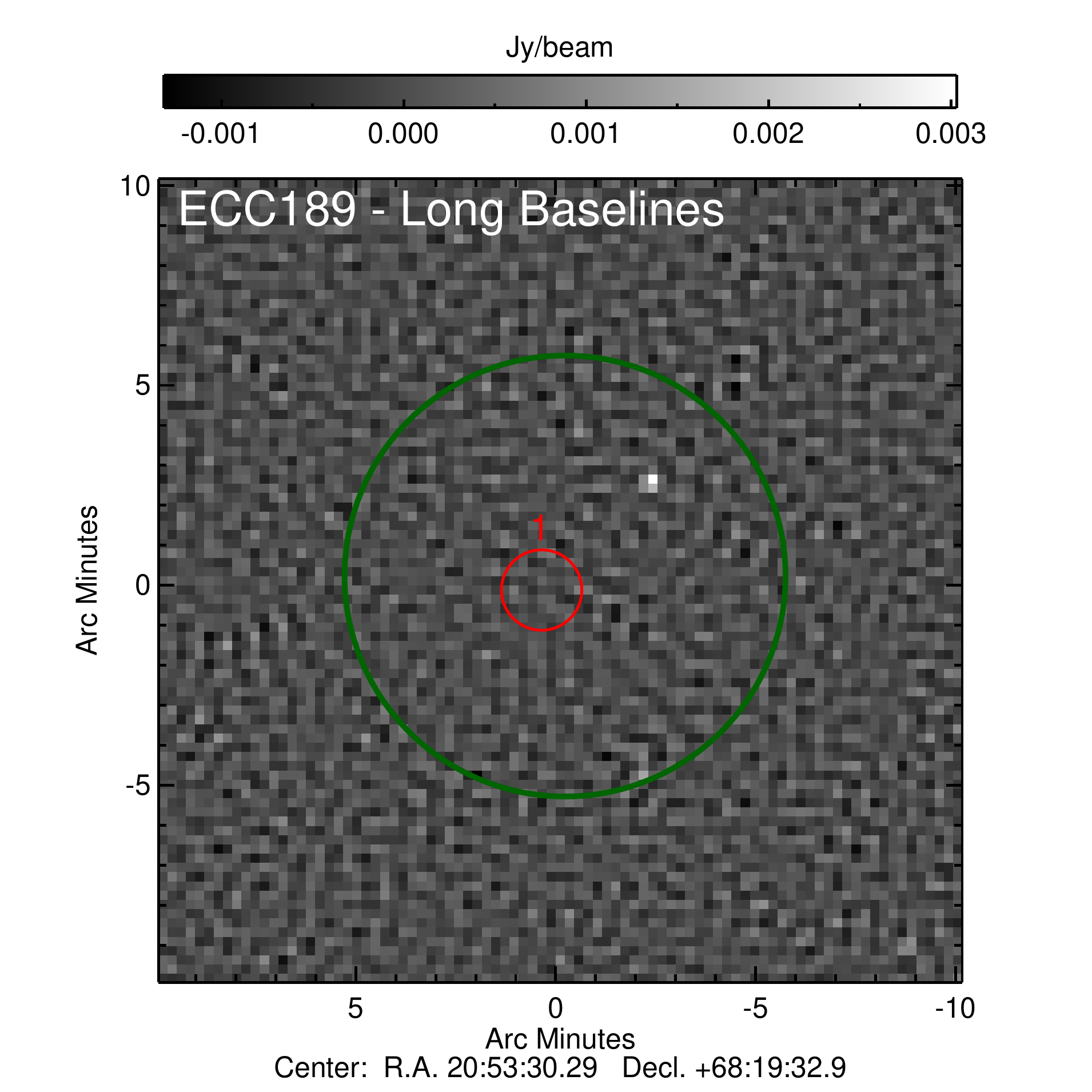} \\

\includegraphics[angle=0,scale=0.375]{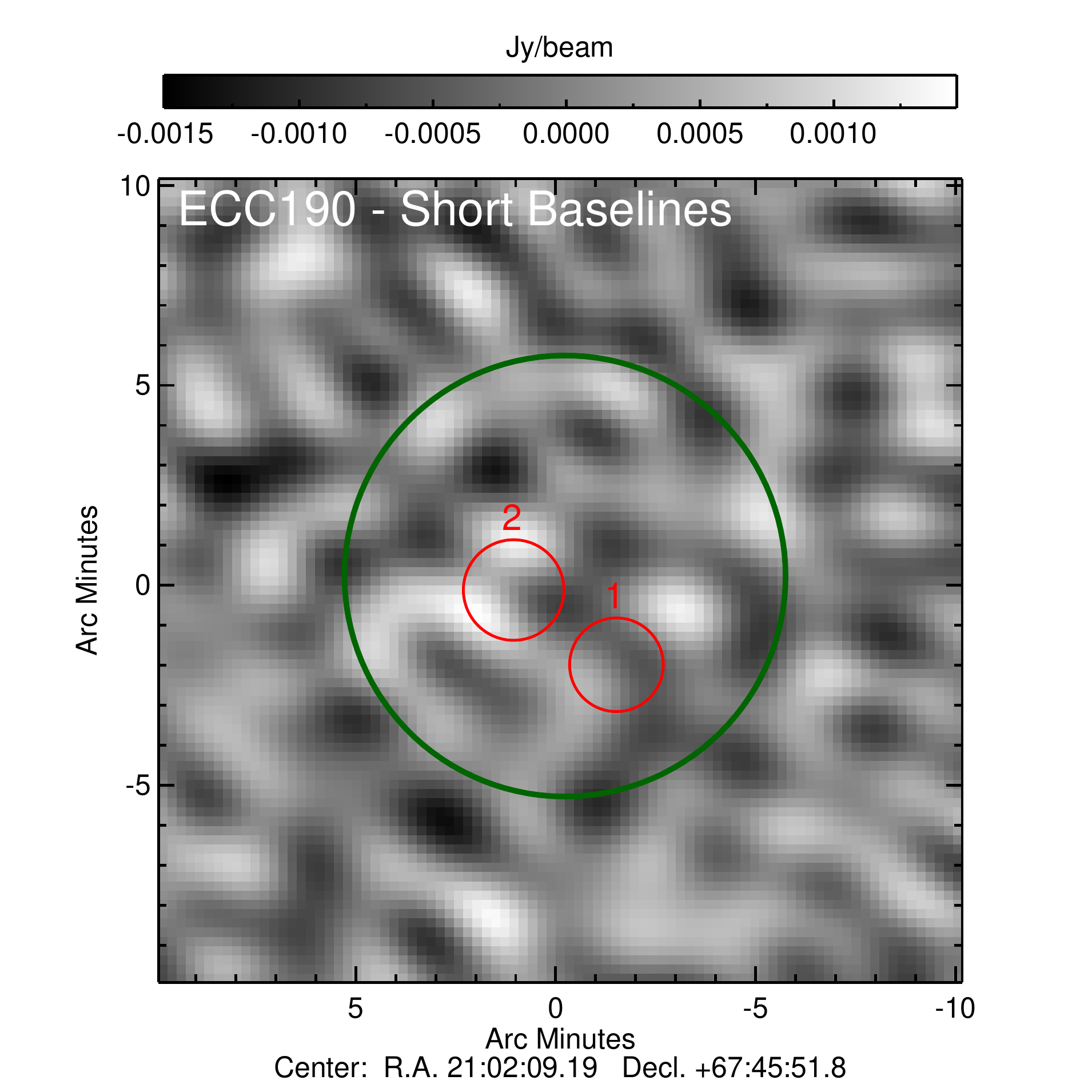}
\includegraphics[angle=0,scale=0.375]{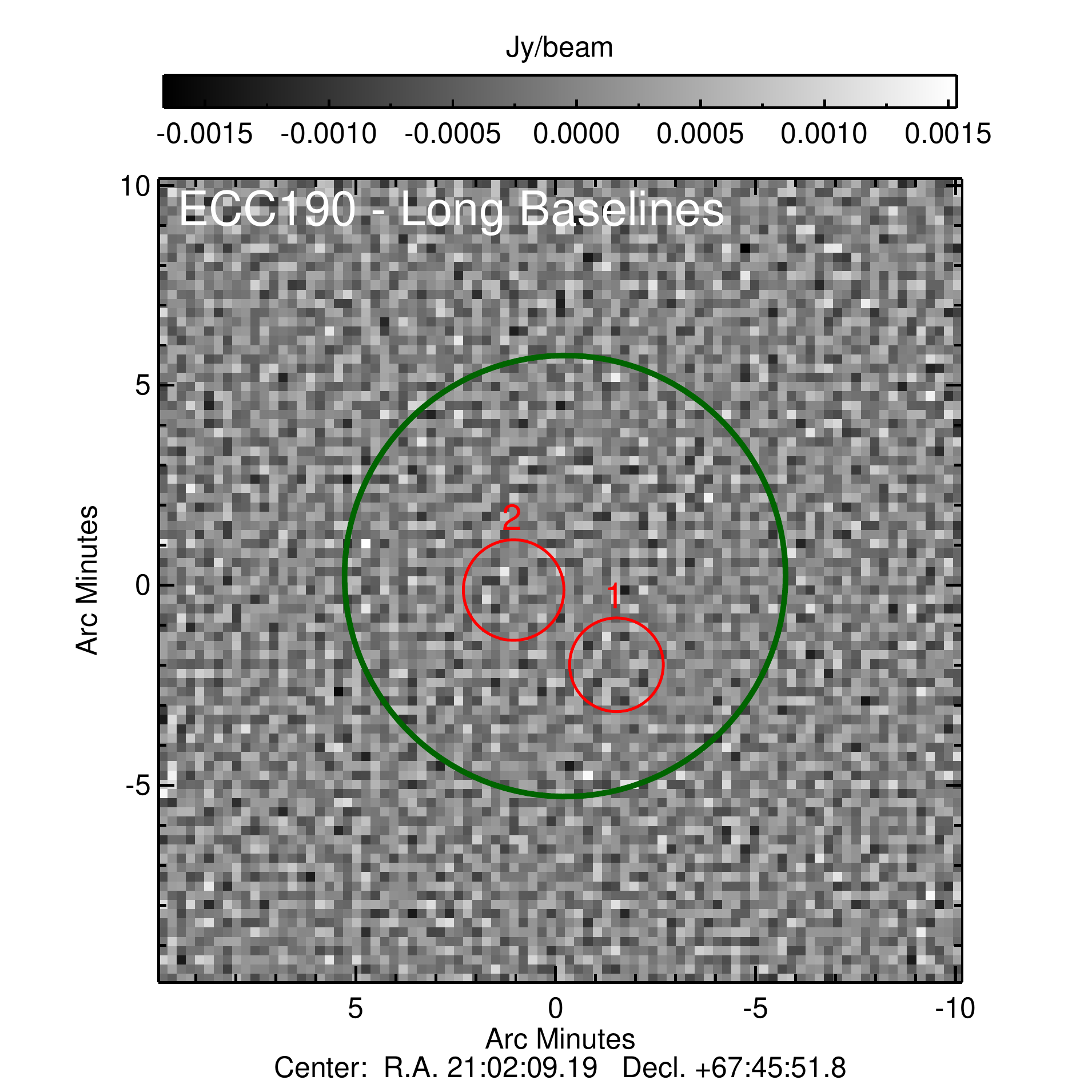} \\
\end{center}
\vspace{-0.4cm}¡
\caption{CARMA 1~cm maps for the short~(\textit{left}) and long~(\textit{right}) baseline data for all fifteen cold clumps. The location of the cores are numbered and identified by red circles. CARMA contours, starting at 5$\sigma$ and increasing linearly, are displayed for the short baseline data (except for ECC191 where the contours increase in steps of 10$\sigma$) along with the CARMA 11~arcmin primary beam~(\textit{green circle}).}
\label{Fig:CARMA_Maps}
\end{figure*}

\begin{figure*}
\ContinuedFloat
\begin{center}
\includegraphics[angle=0,scale=0.375]{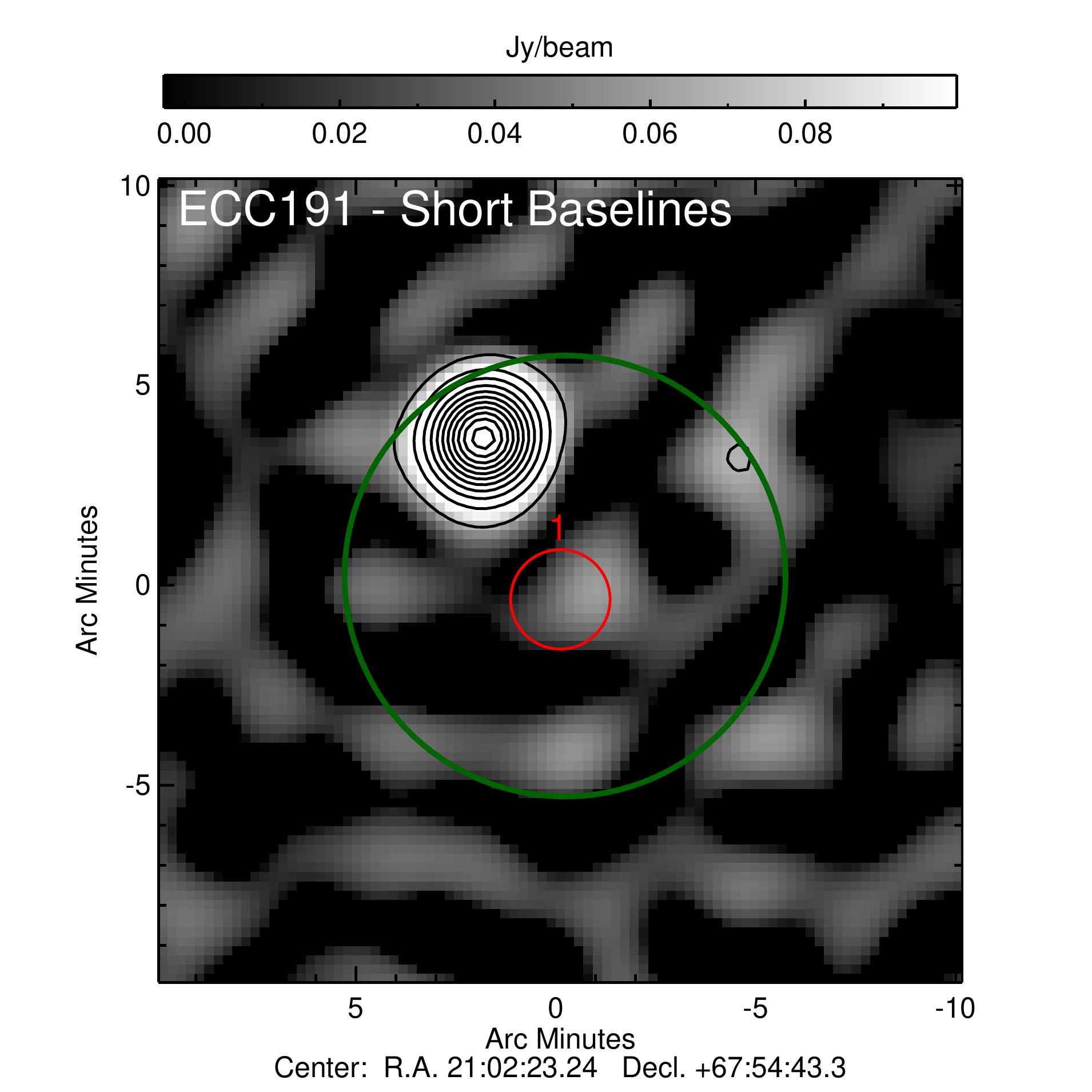}
\includegraphics[angle=0,scale=0.375]{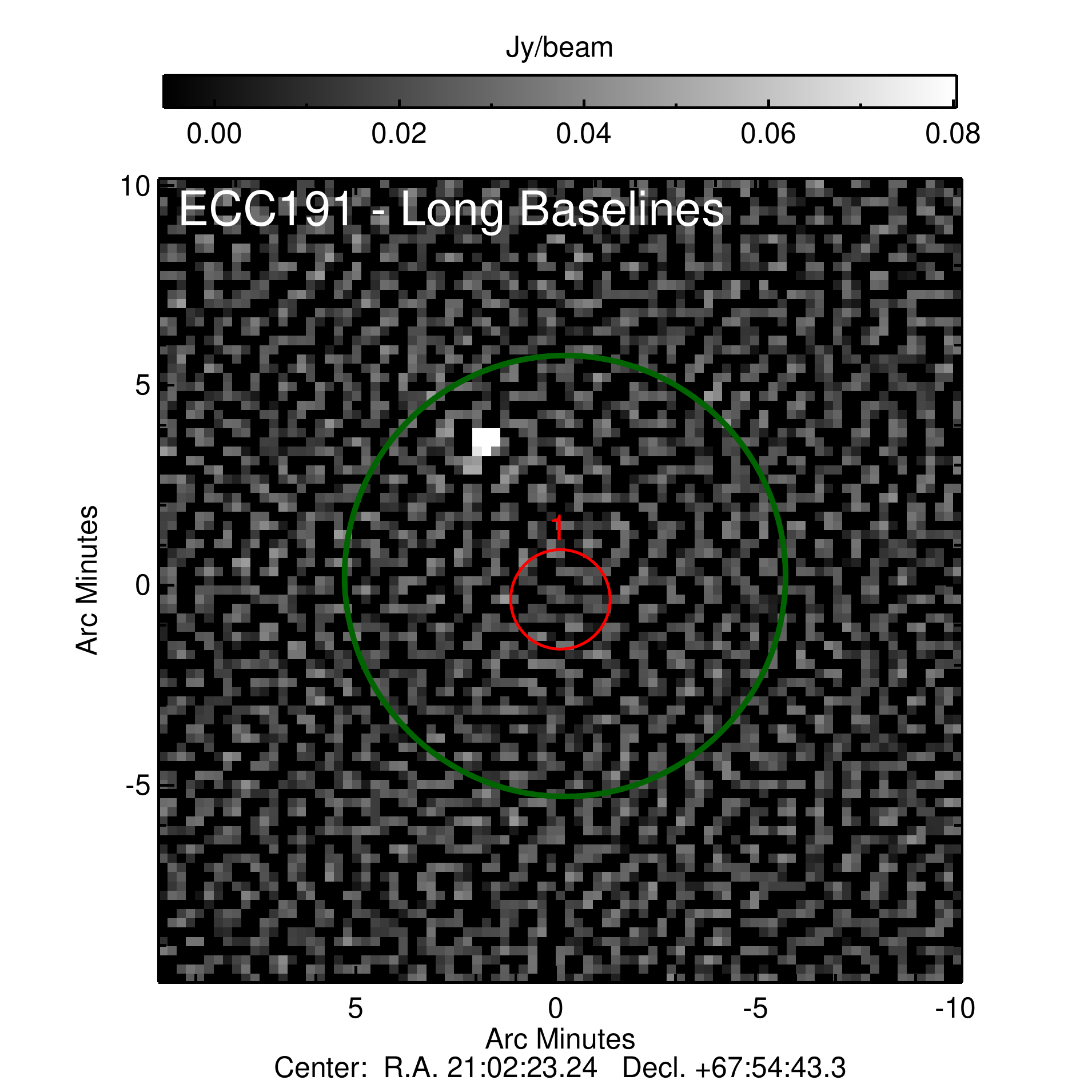} \\

\includegraphics[angle=0,scale=0.375]{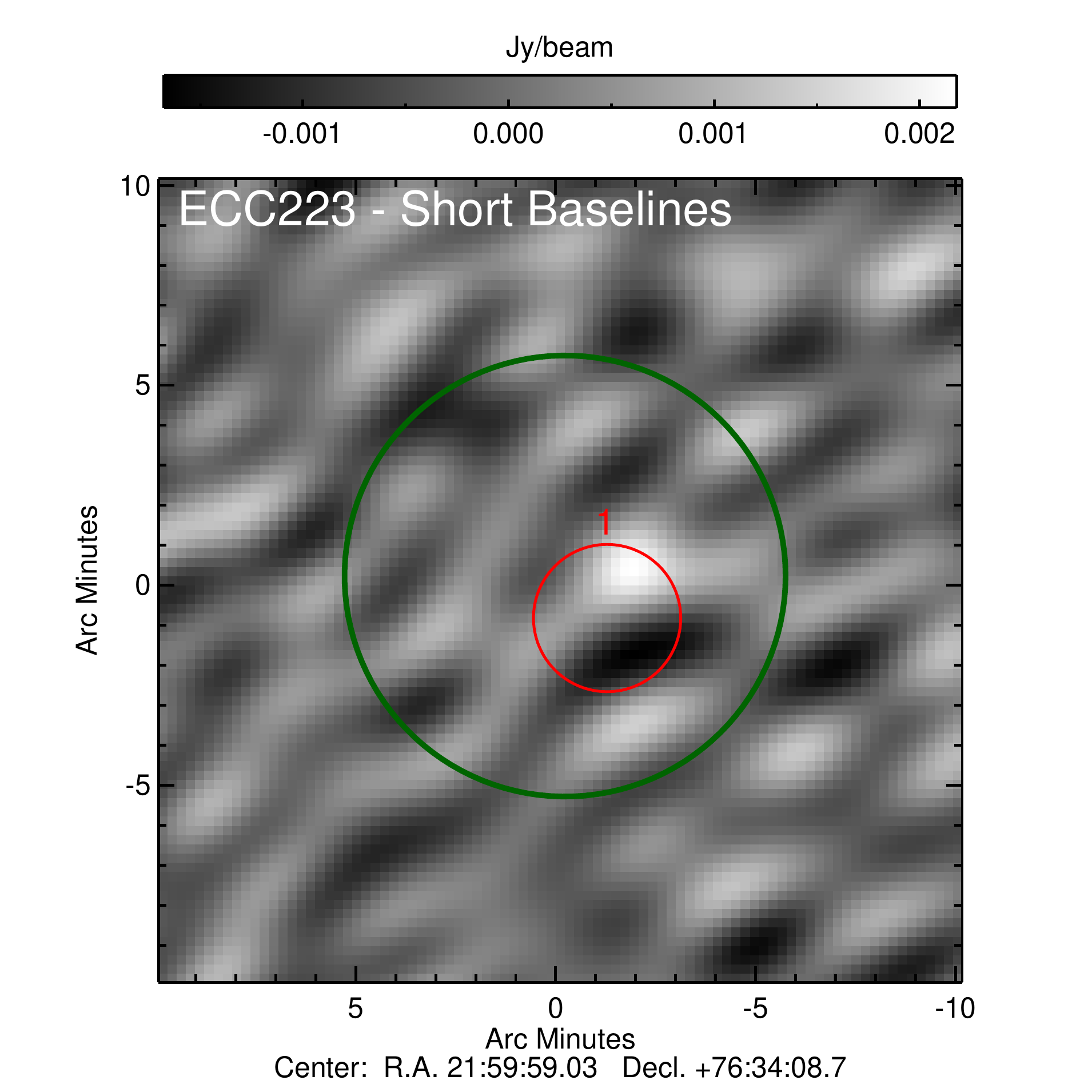}
\includegraphics[angle=0,scale=0.375]{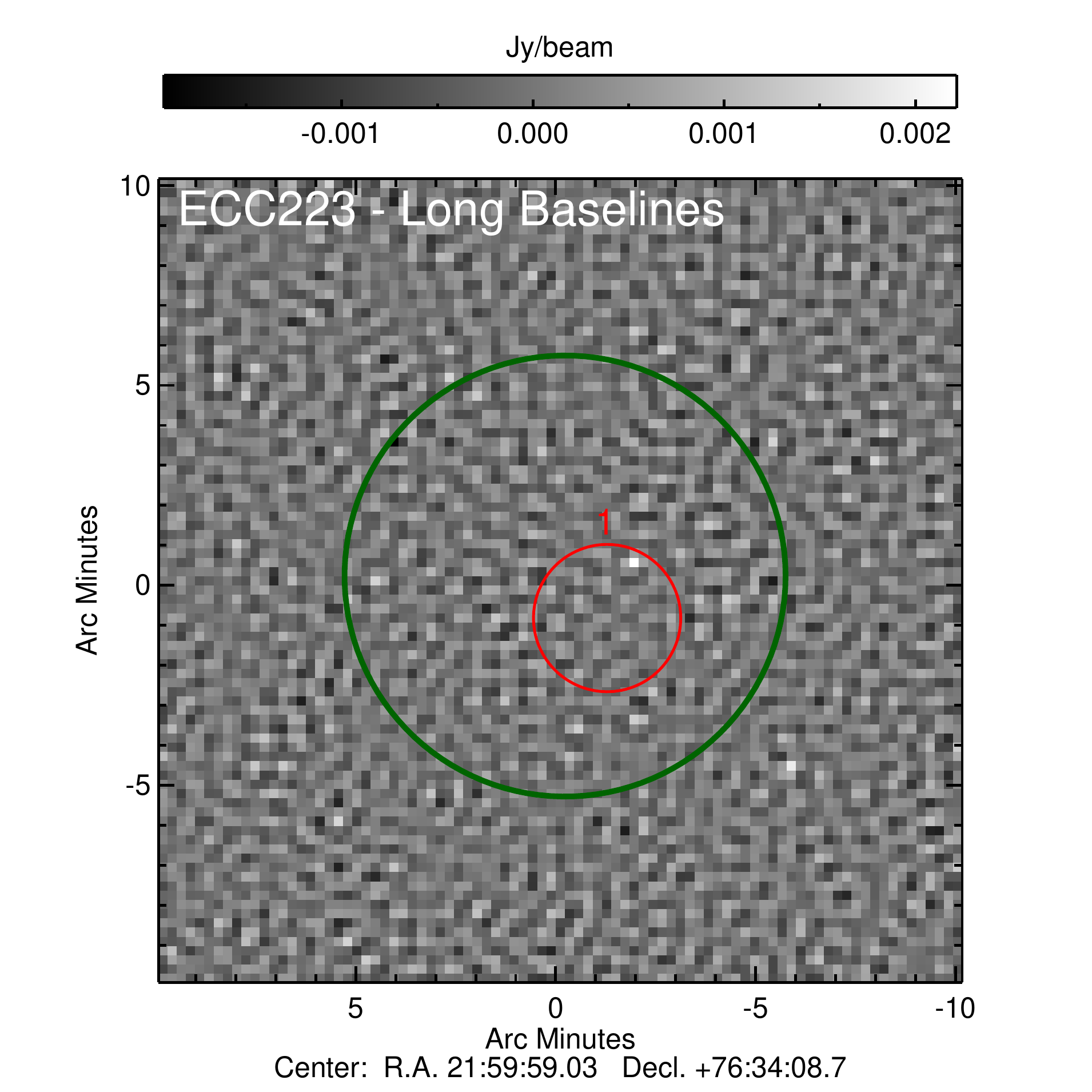} \\

\includegraphics[angle=0,scale=0.375]{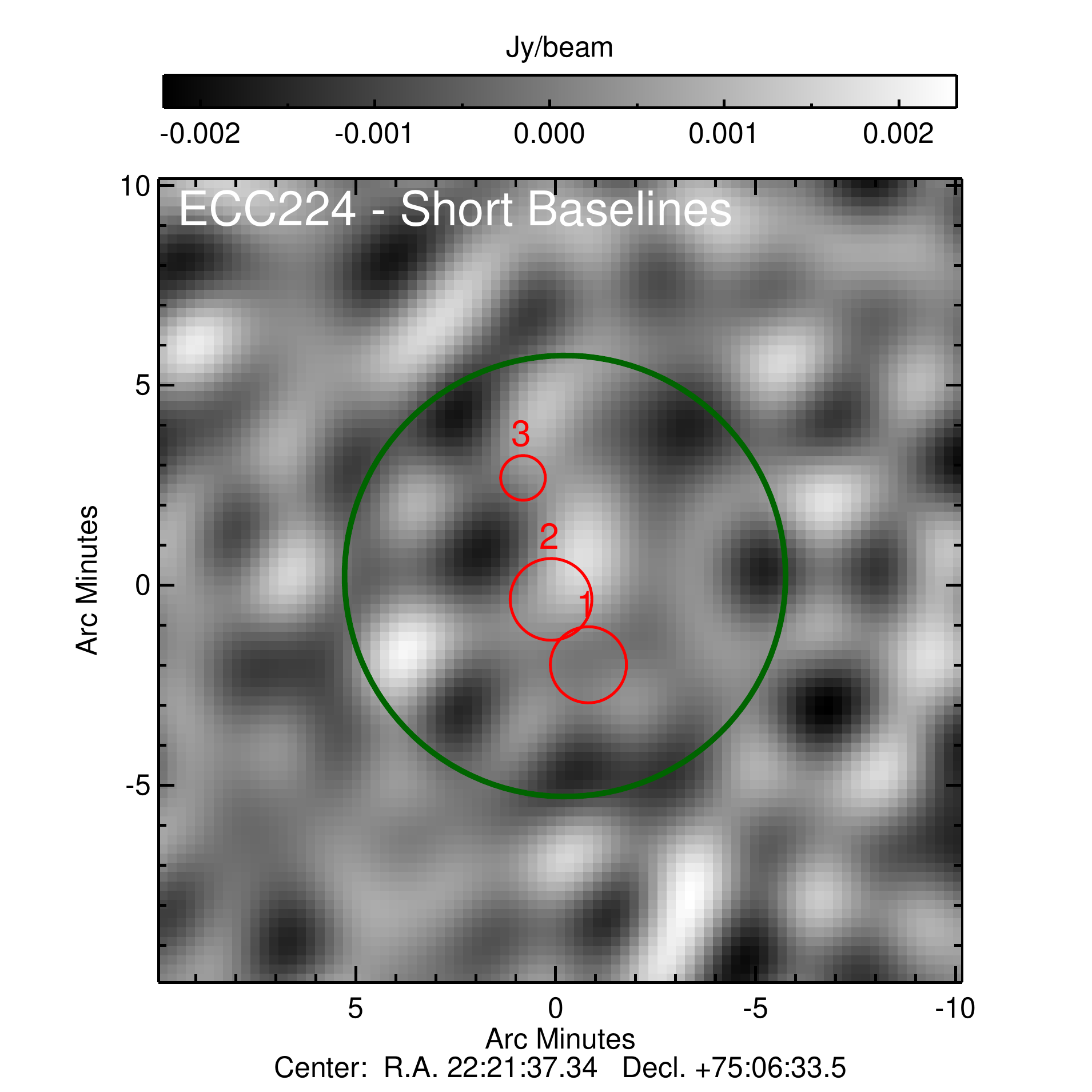}
\includegraphics[angle=0,scale=0.375]{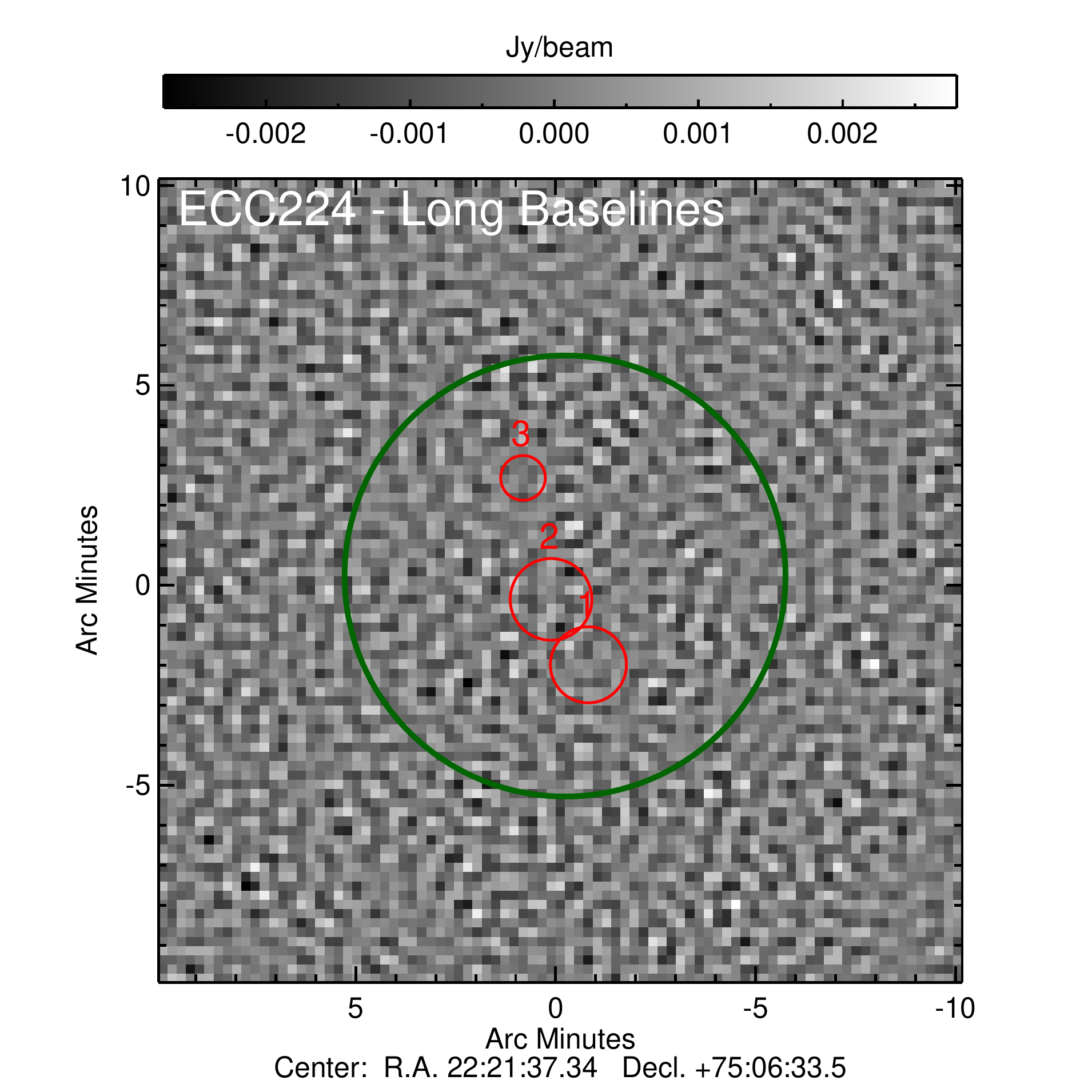} \\
\end{center}
\vspace{-0.4cm}
\caption{Continued}
\label{Fig:CARMA_Maps}
\end{figure*}

\begin{figure*}
\ContinuedFloat
\begin{center}
\includegraphics[angle=0,scale=0.375]{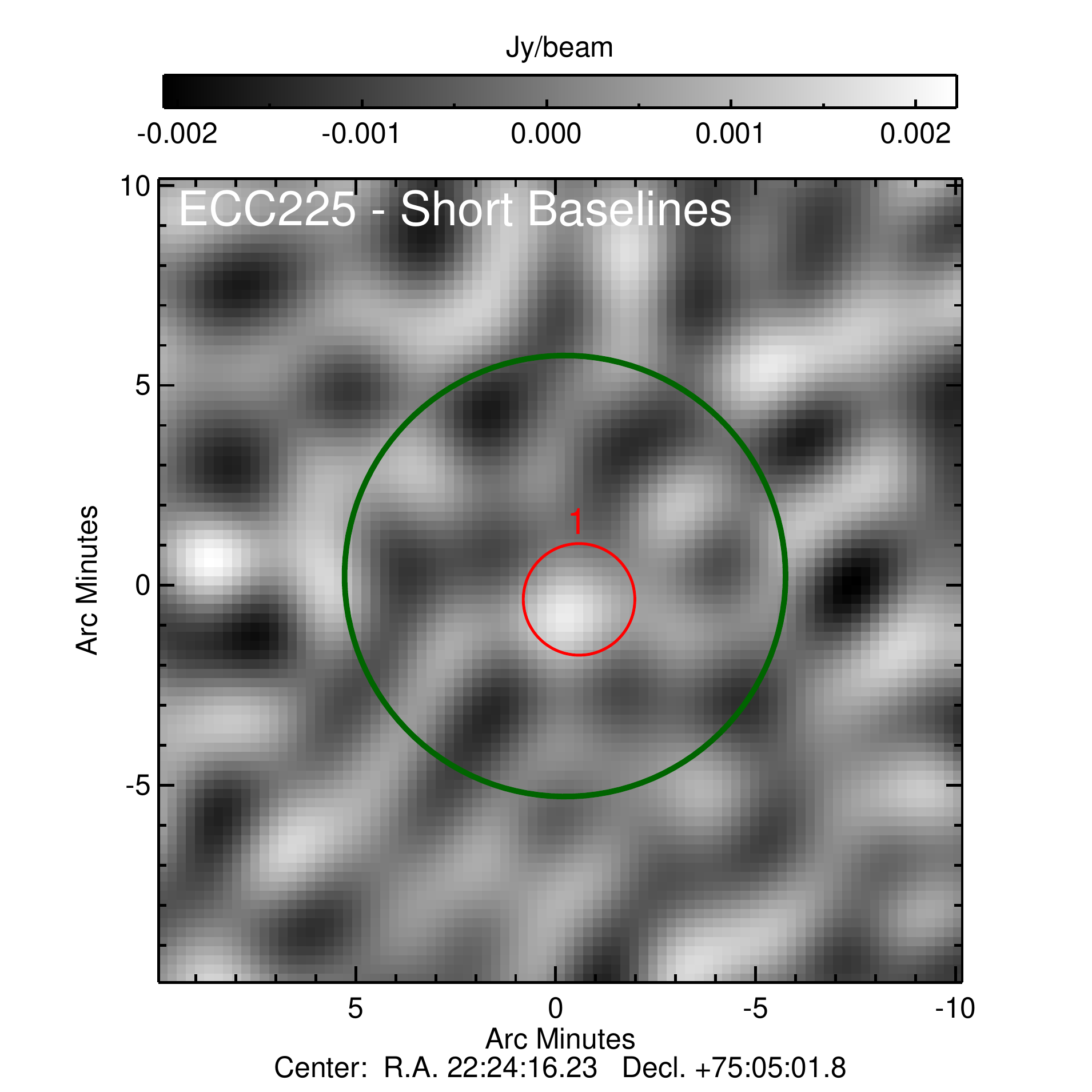}
\includegraphics[angle=0,scale=0.375]{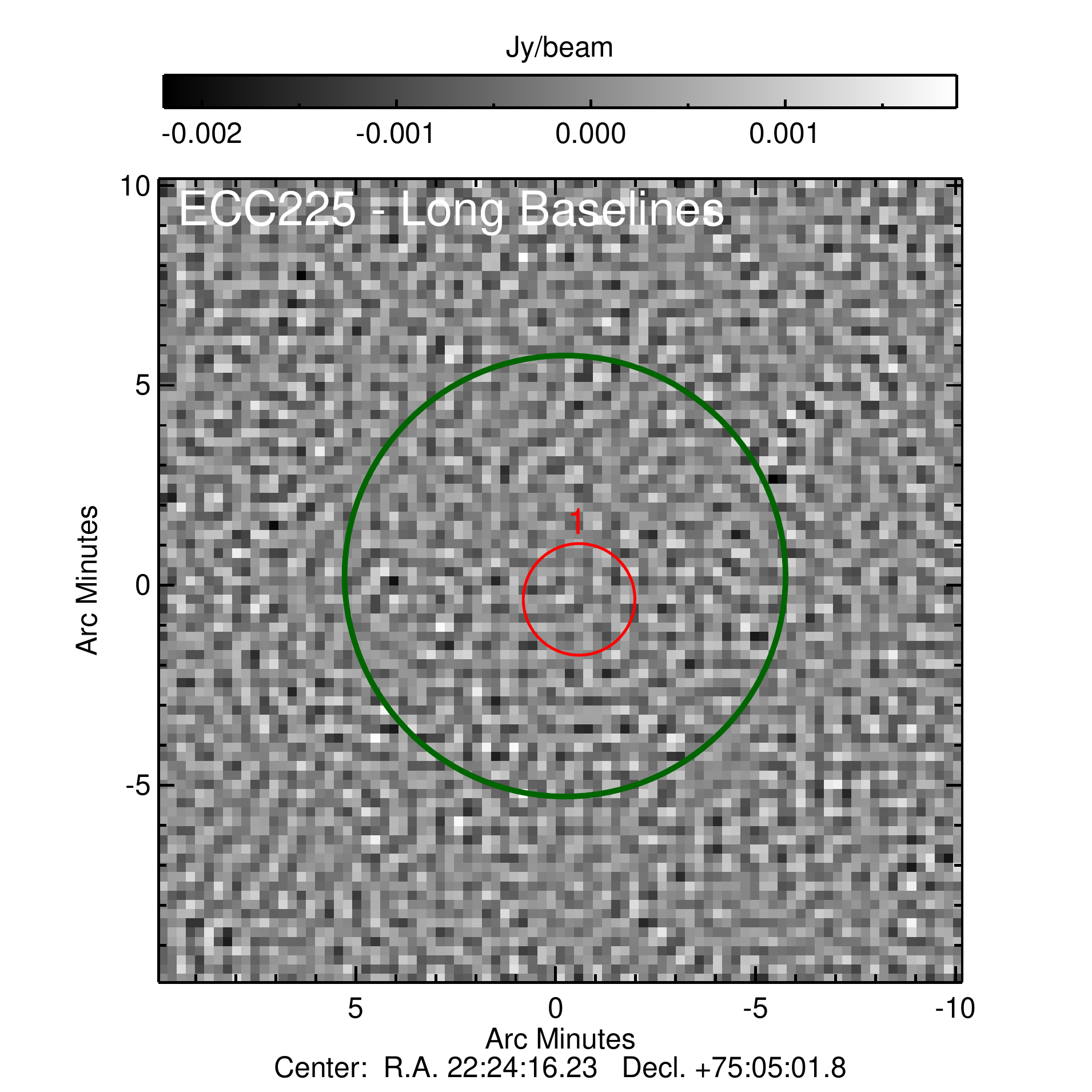} \\

\includegraphics[angle=0,scale=0.375]{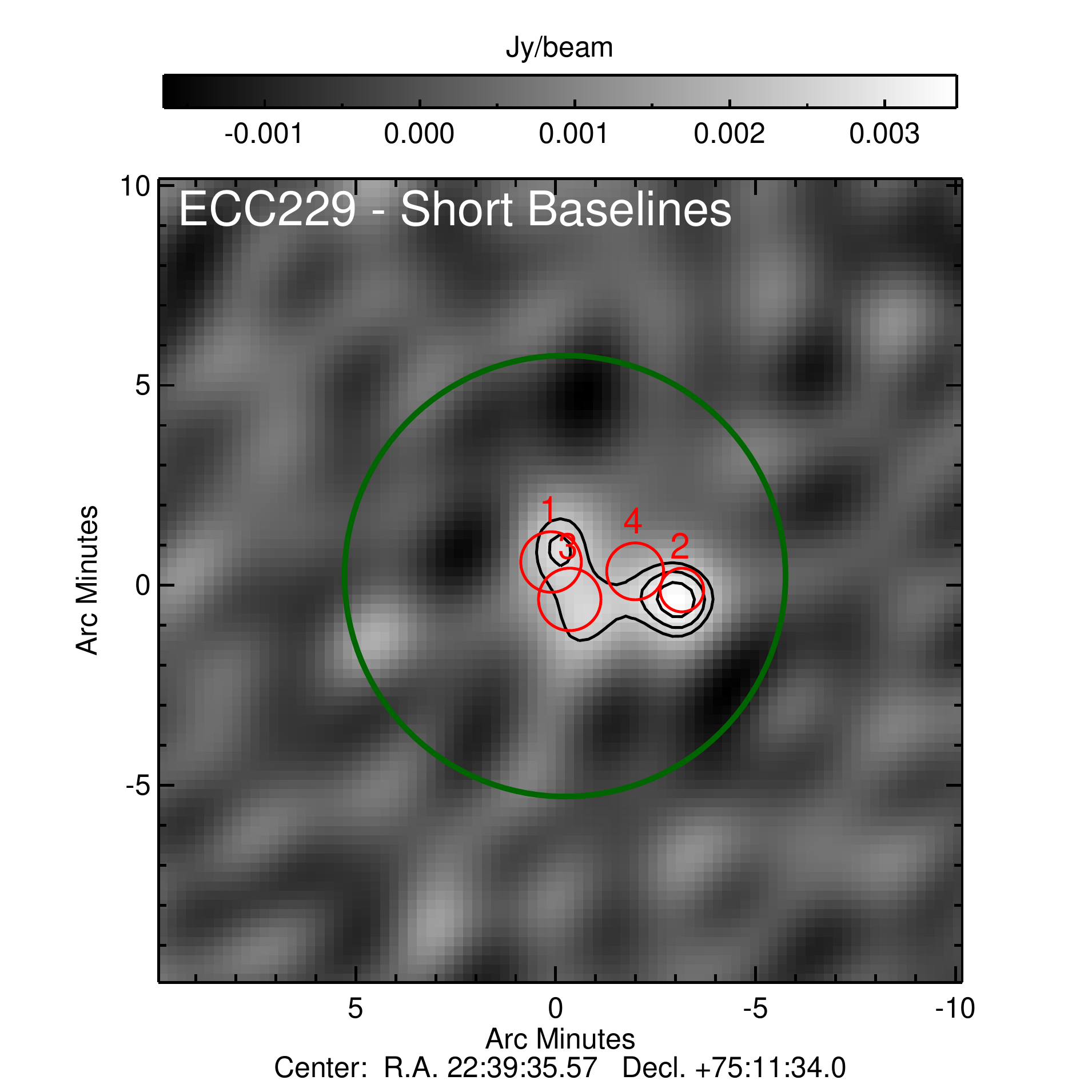}
\includegraphics[angle=0,scale=0.375]{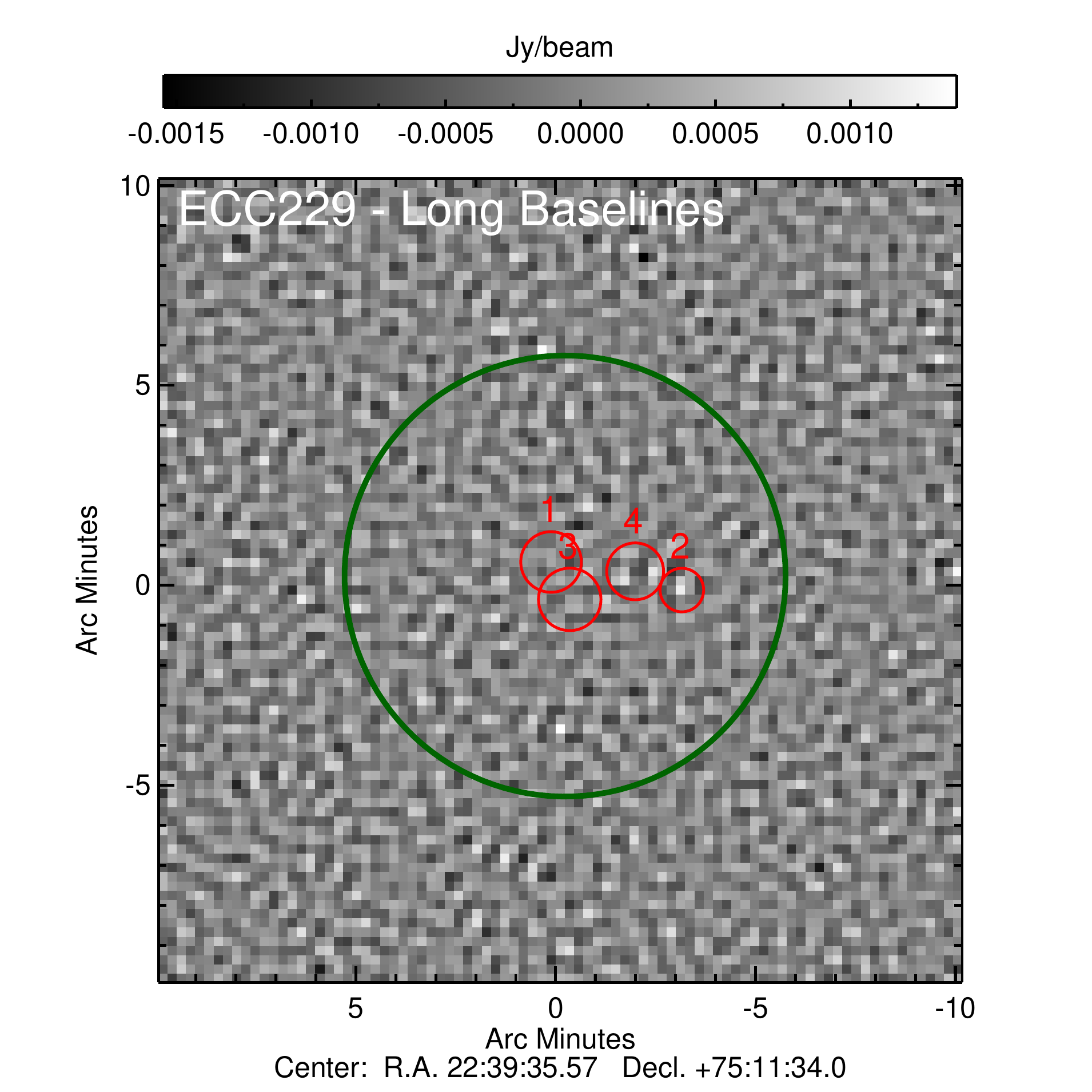} \\

\includegraphics[angle=0,scale=0.375]{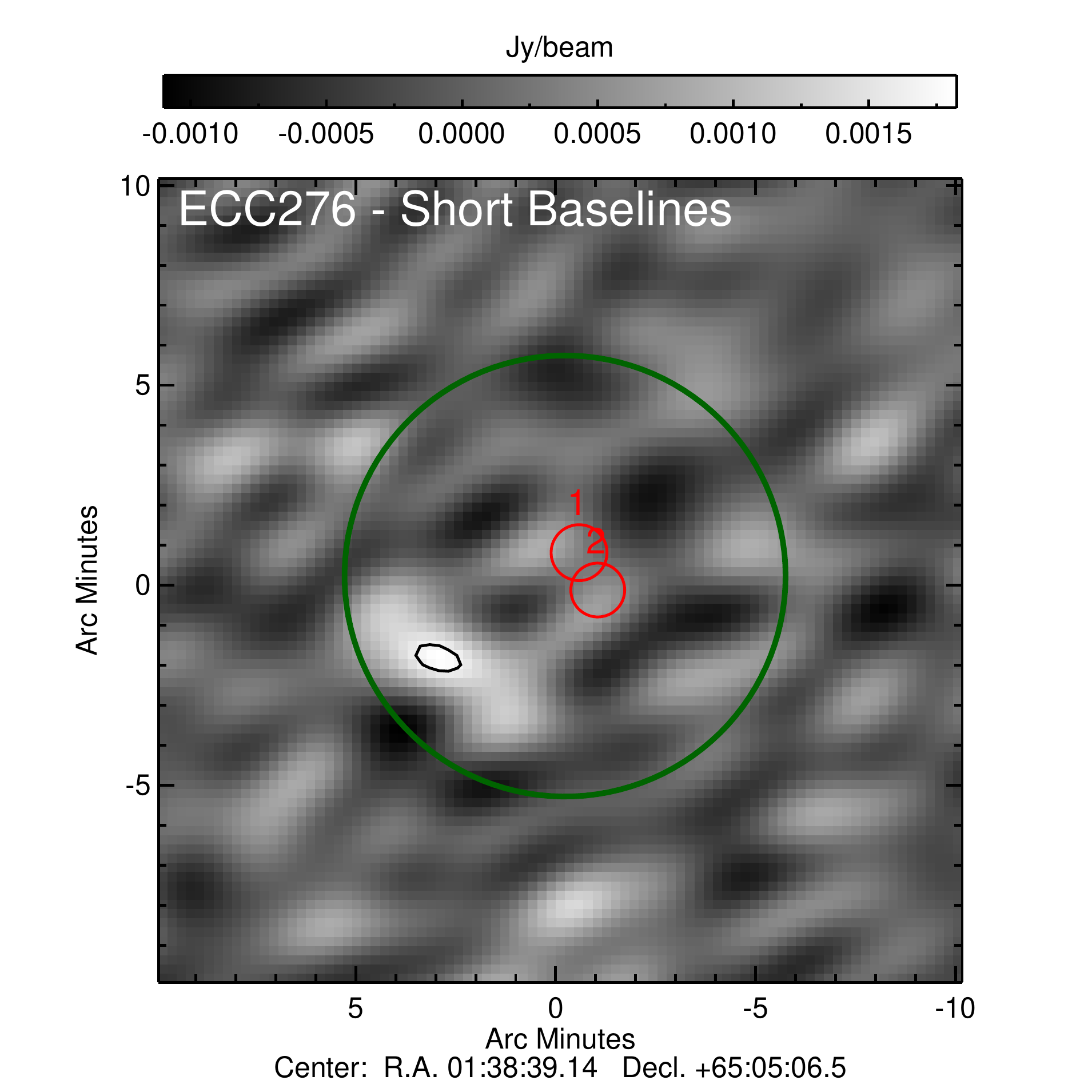}
\includegraphics[angle=0,scale=0.375]{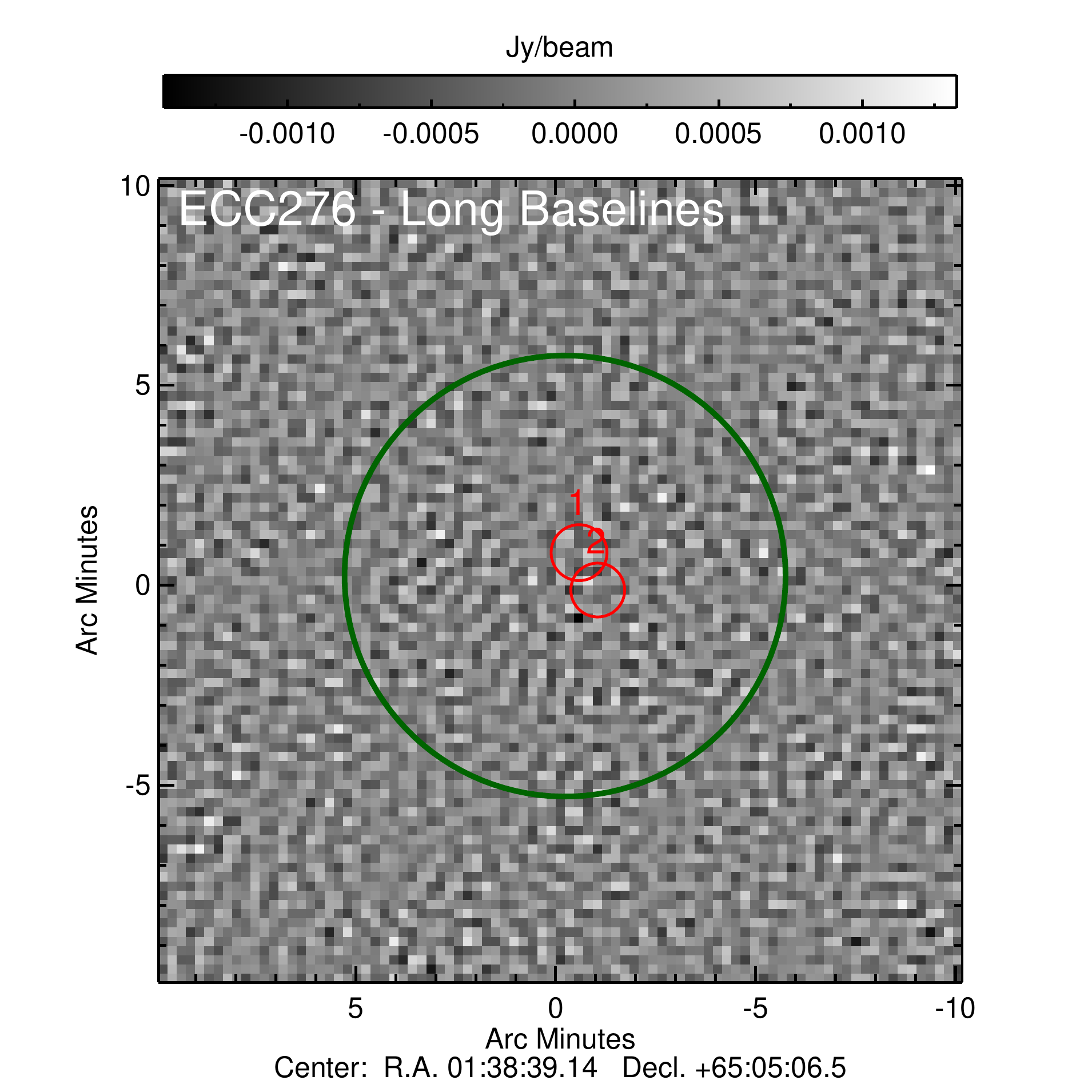} \\
\end{center}
\vspace{-0.4cm}
\caption{Continued}
\label{Fig:CARMA_Maps}
\end{figure*}

\begin{figure*}
\ContinuedFloat
\begin{center}
\includegraphics[angle=0,scale=0.375]{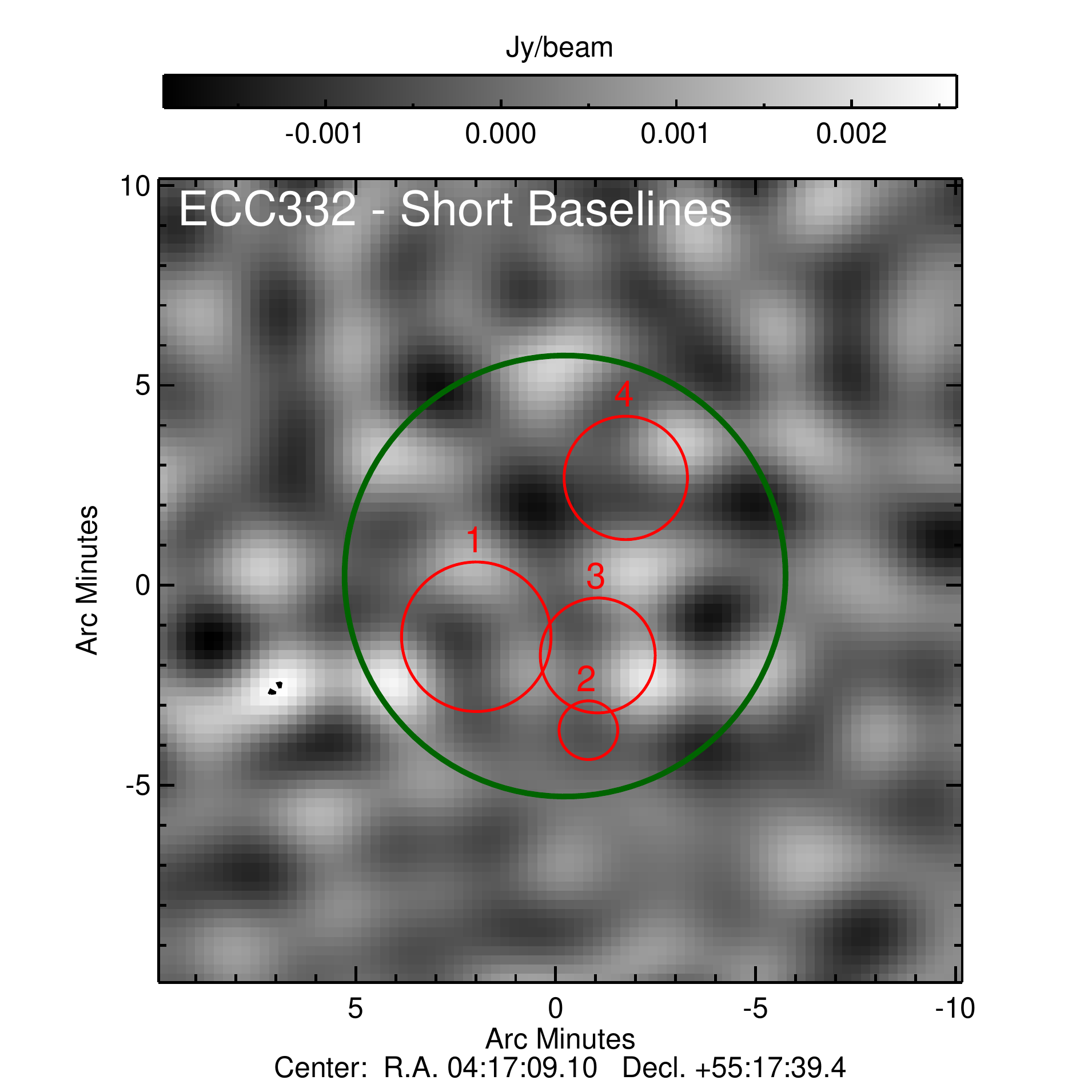}
\includegraphics[angle=0,scale=0.375]{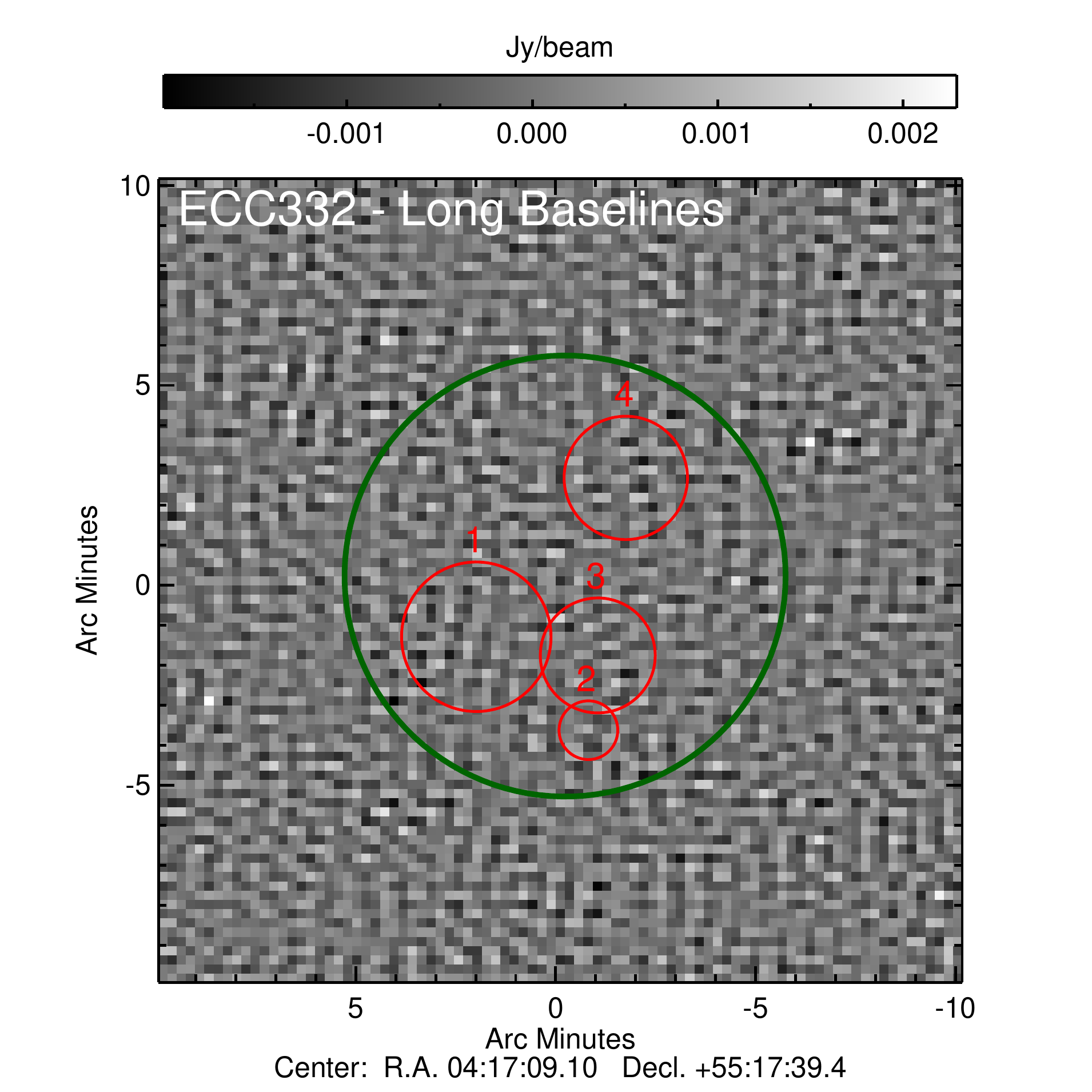} \\

\includegraphics[angle=0,scale=0.375]{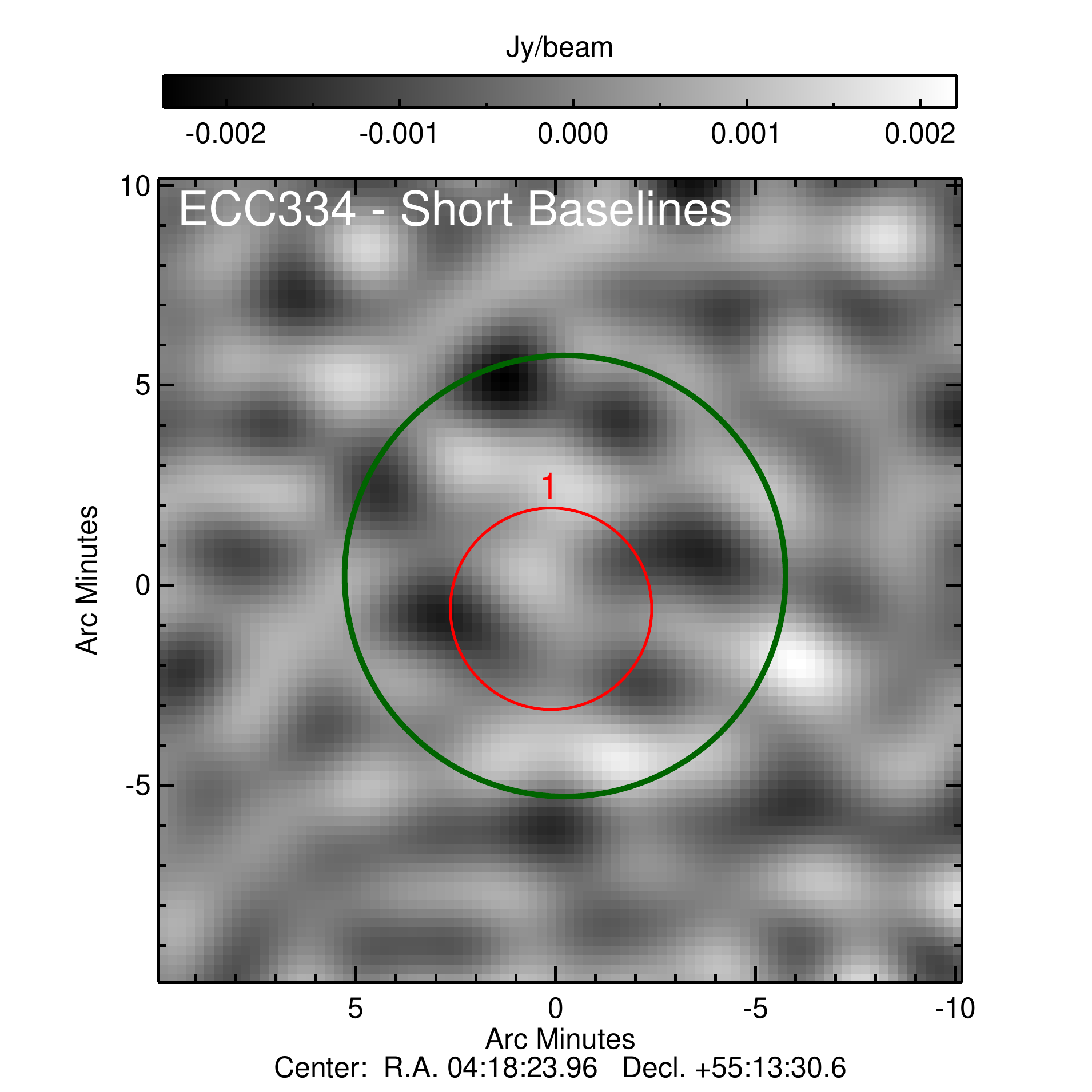}
\includegraphics[angle=0,scale=0.375]{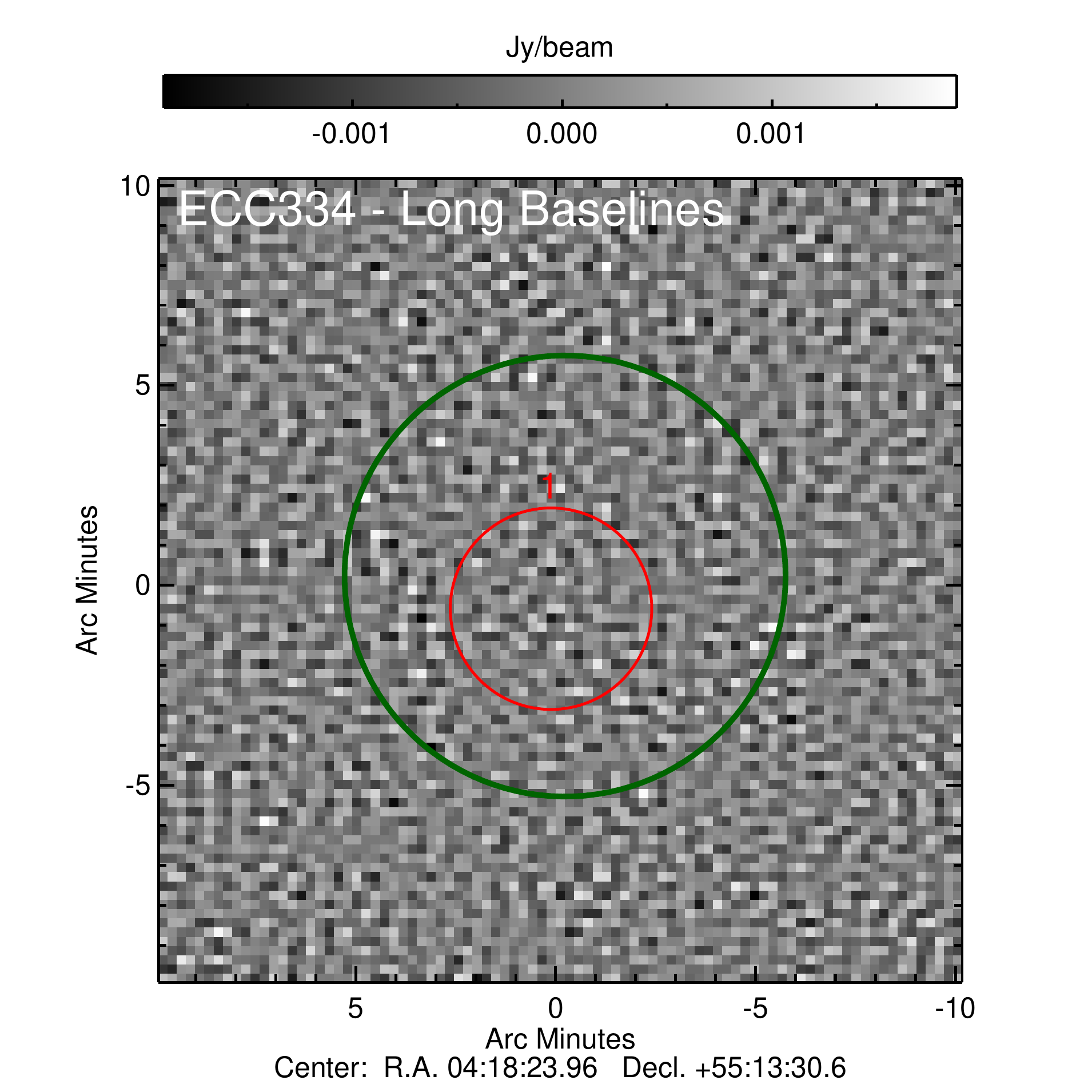} \\

\includegraphics[angle=0,scale=0.375]{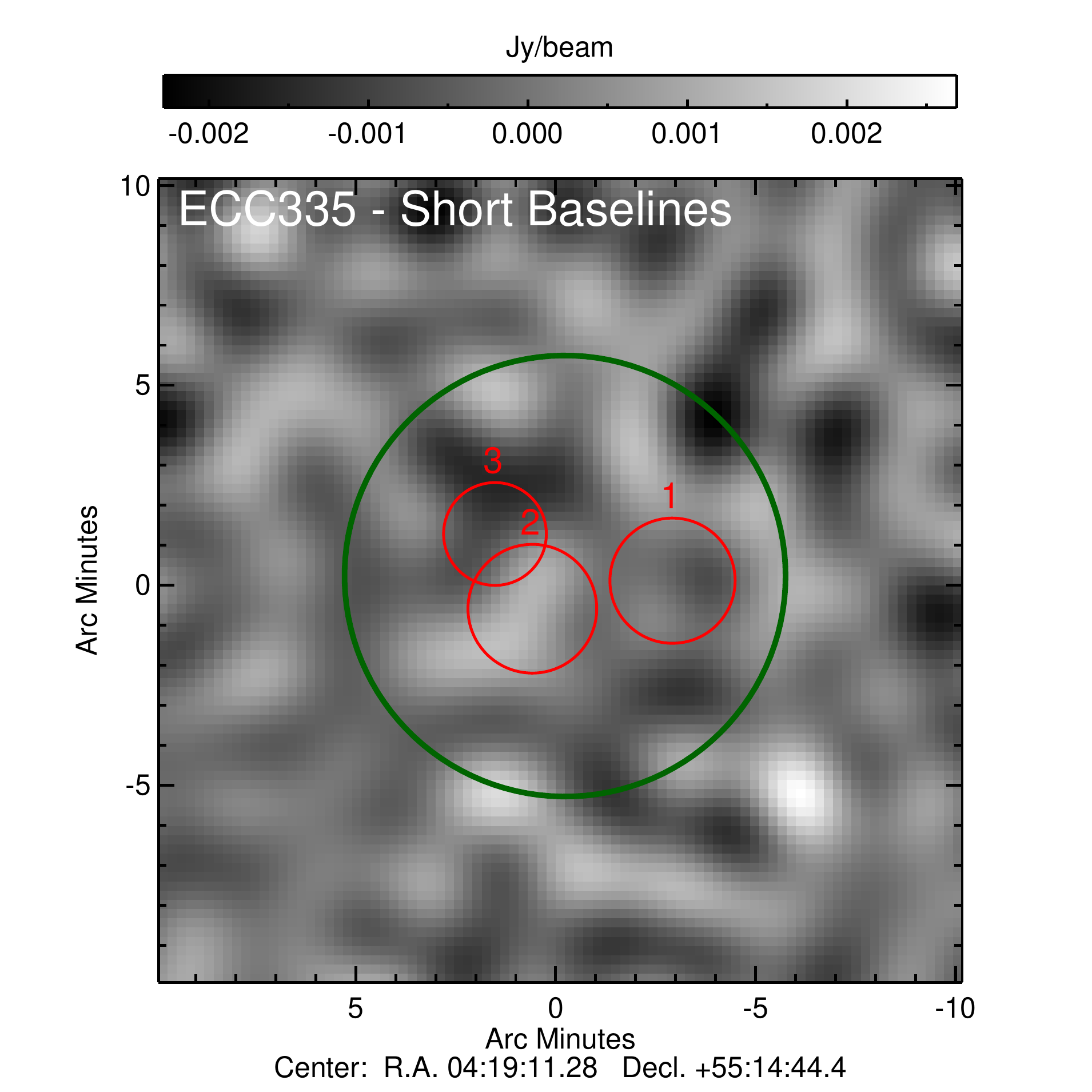}
\includegraphics[angle=0,scale=0.375]{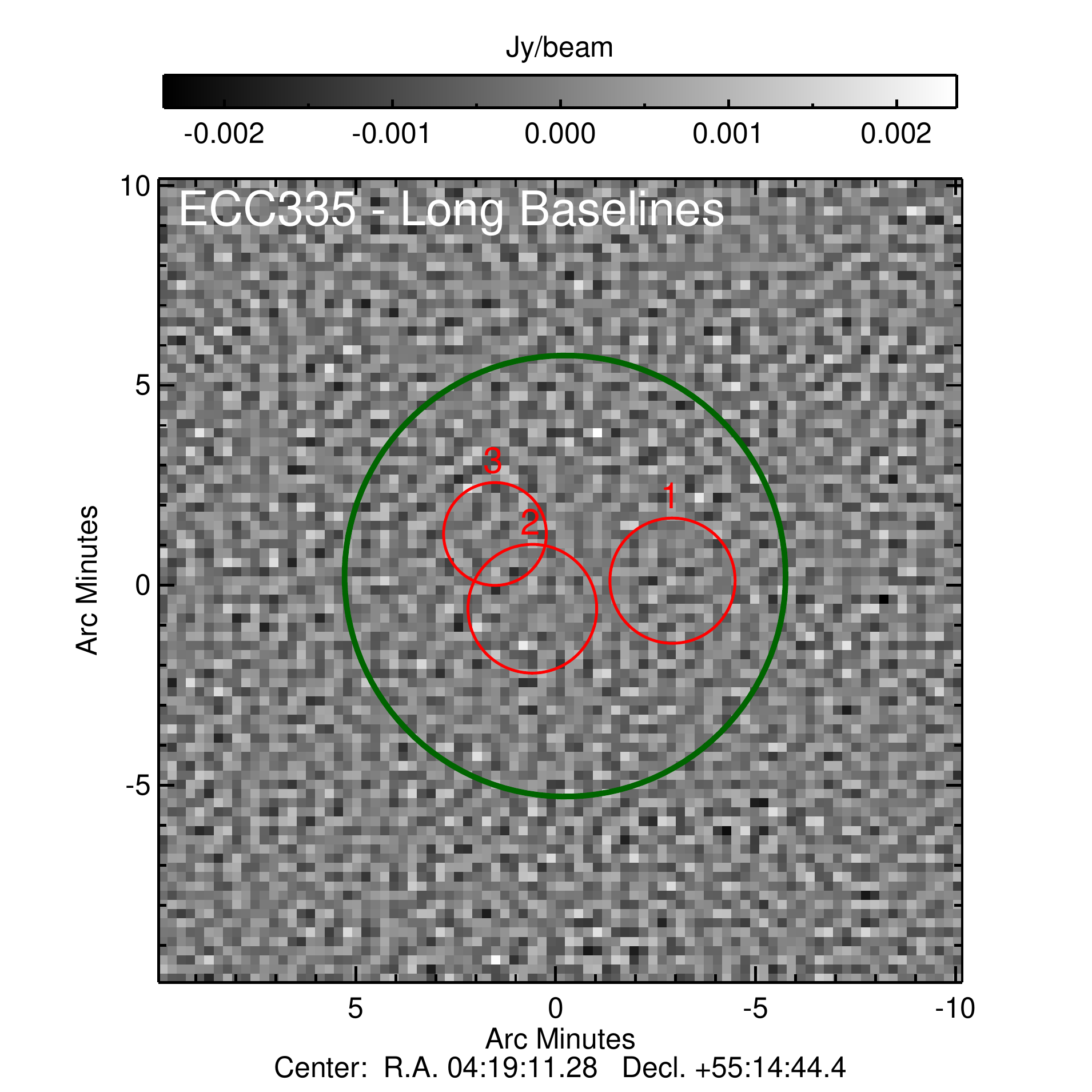} \\
\end{center}
\vspace{-0.4cm}
\caption{Continued}
\label{Fig:CARMA_Maps}
\end{figure*}

\begin{figure*}
\ContinuedFloat
\begin{center}
\includegraphics[angle=0,scale=0.375]{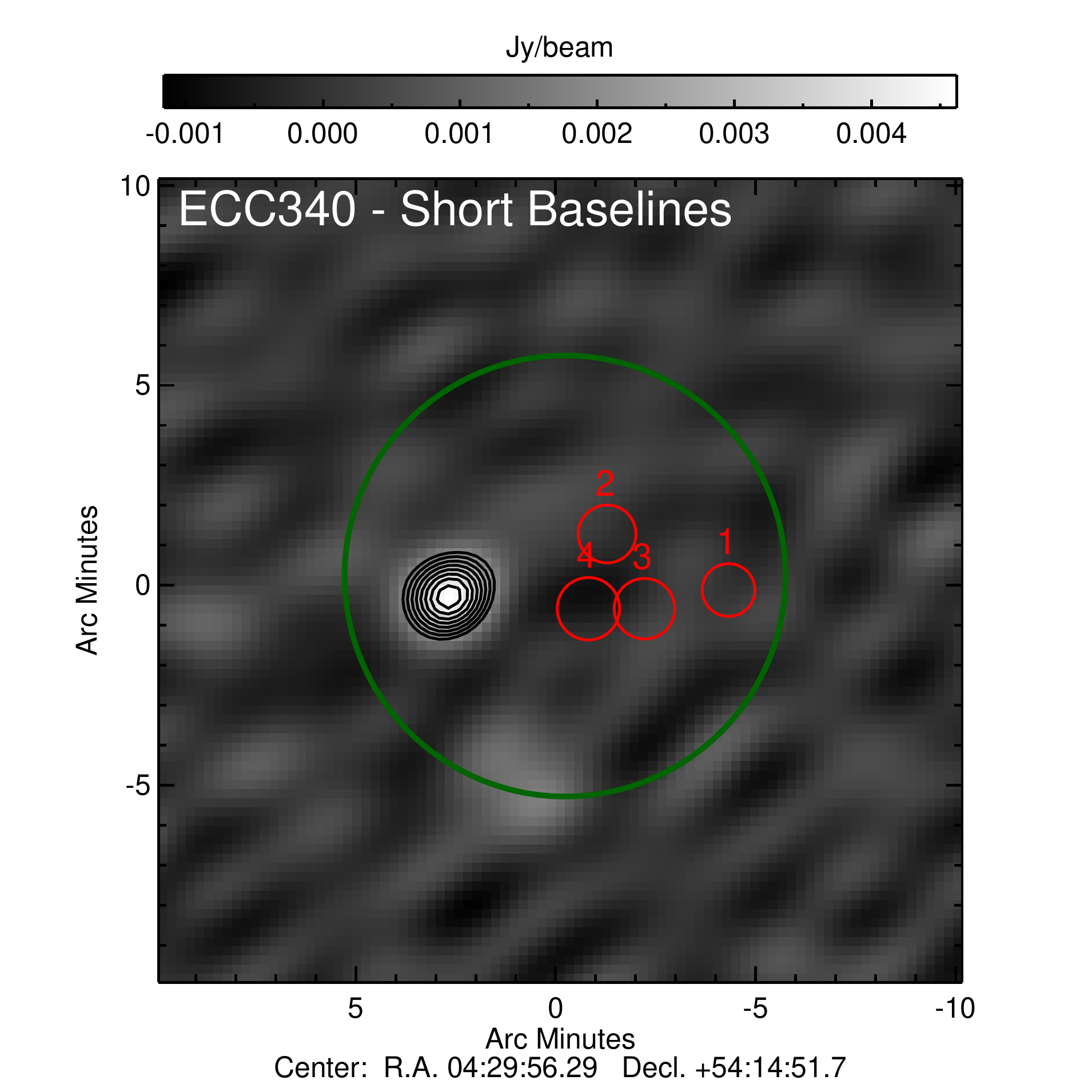}
\includegraphics[angle=0,scale=0.375]{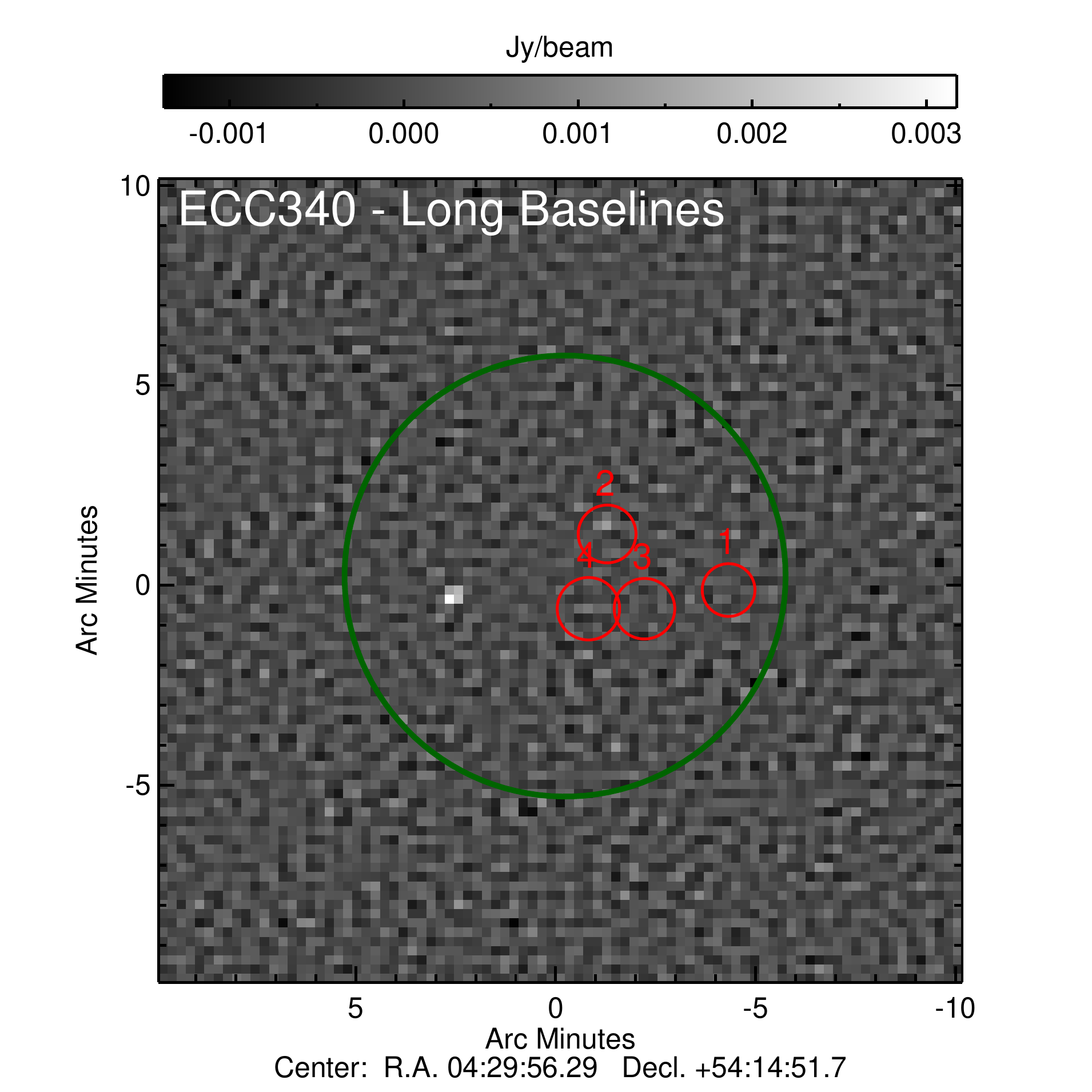} \\

\includegraphics[angle=0,scale=0.375]{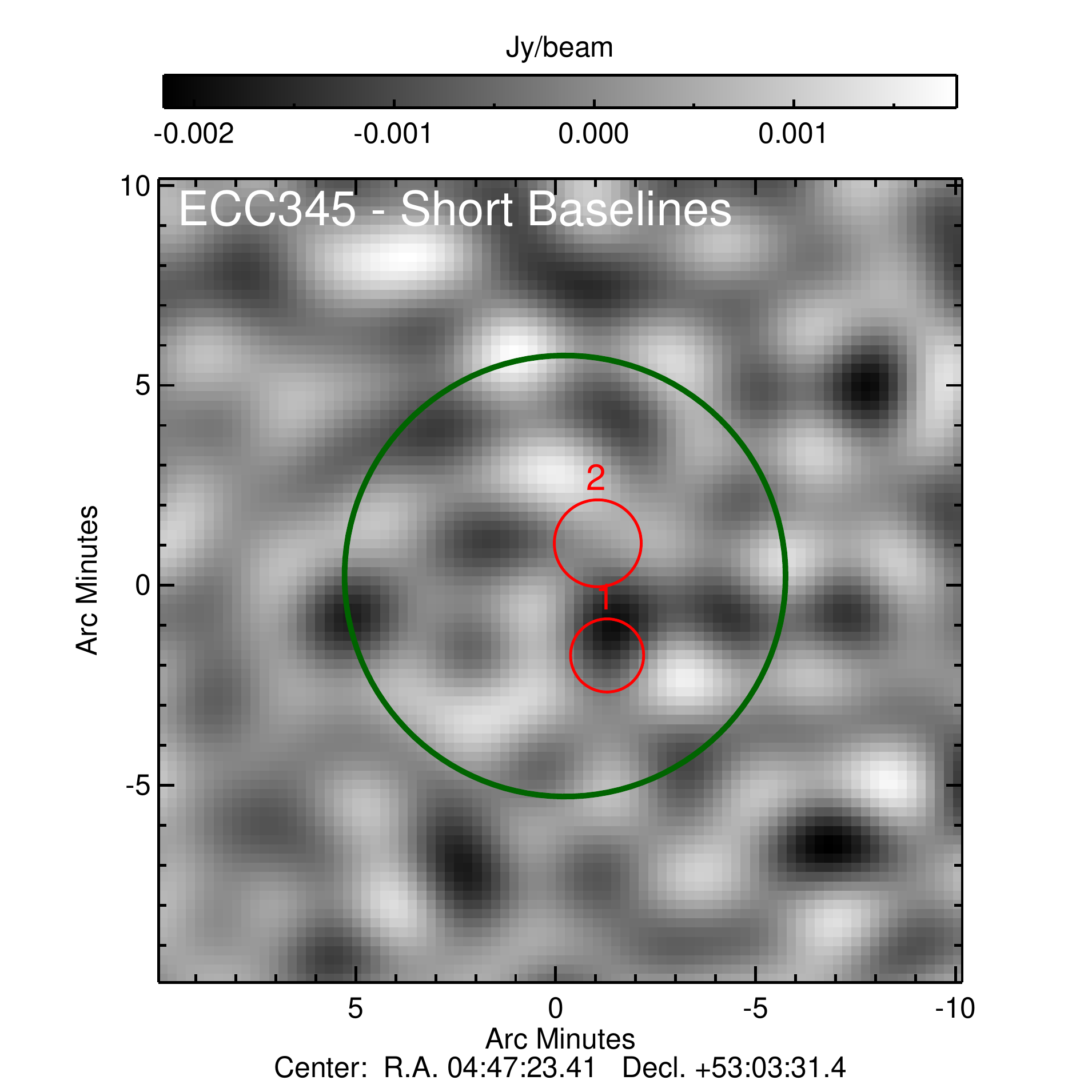}
\includegraphics[angle=0,scale=0.375]{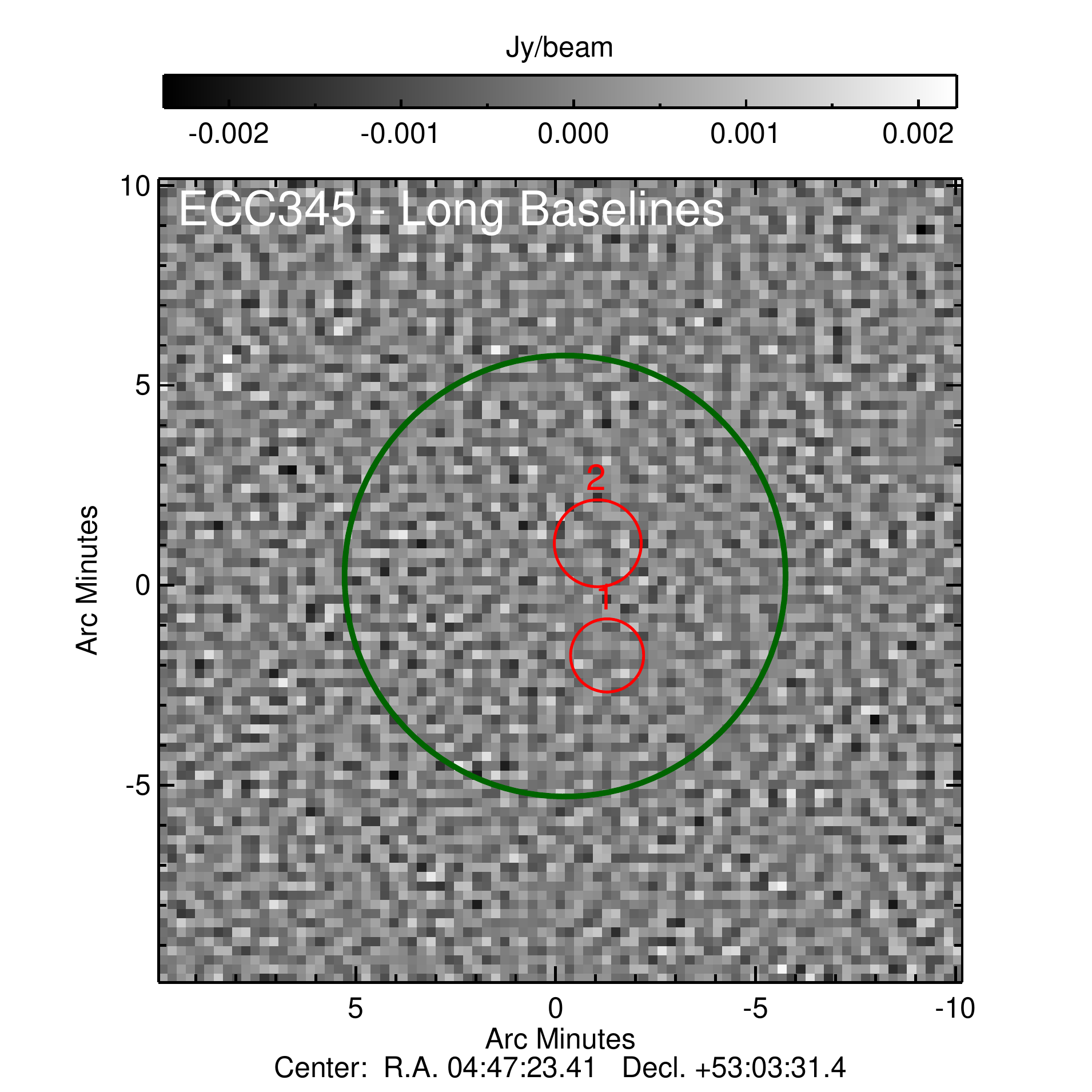} \\

\includegraphics[angle=0,scale=0.375]{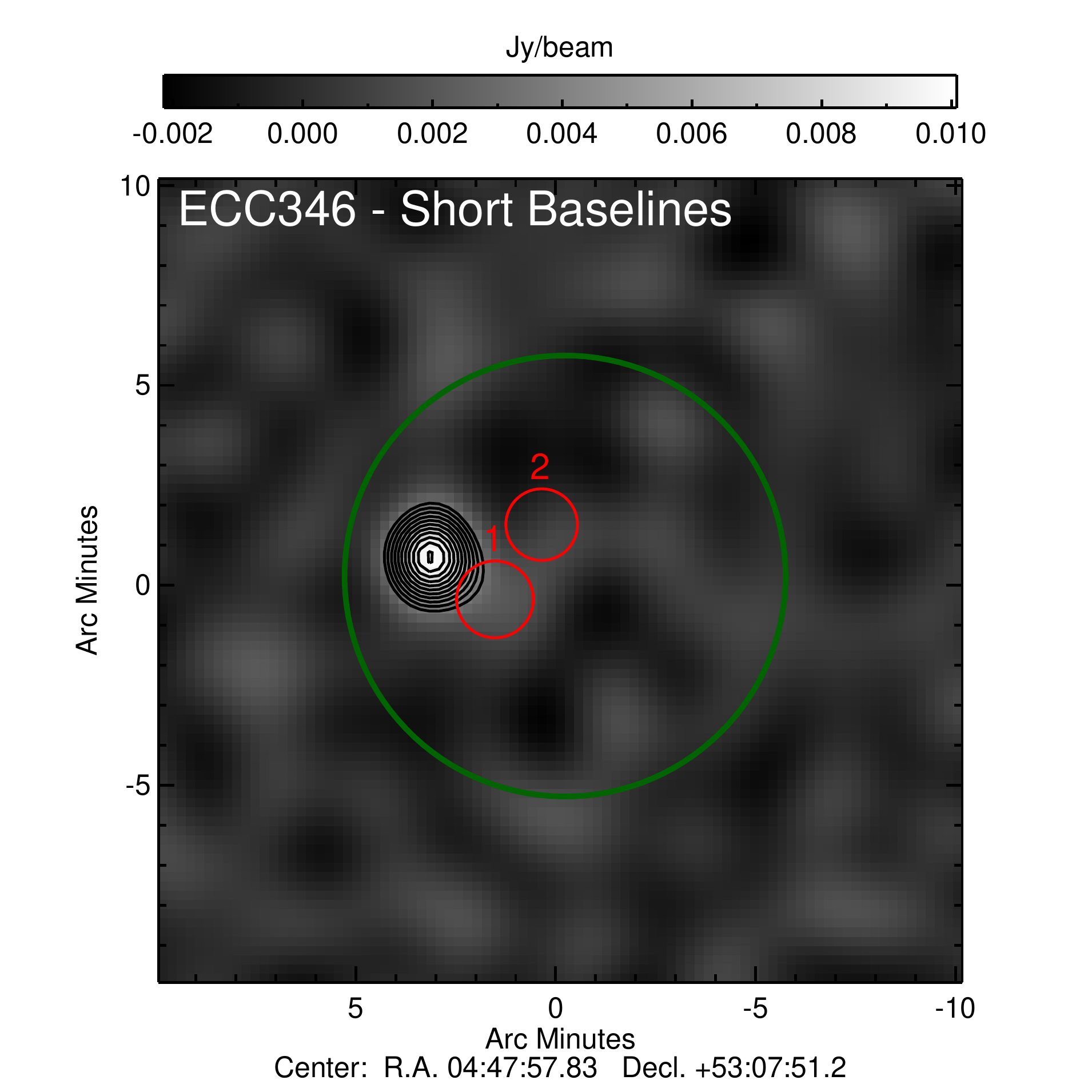}
\includegraphics[angle=0,scale=0.375]{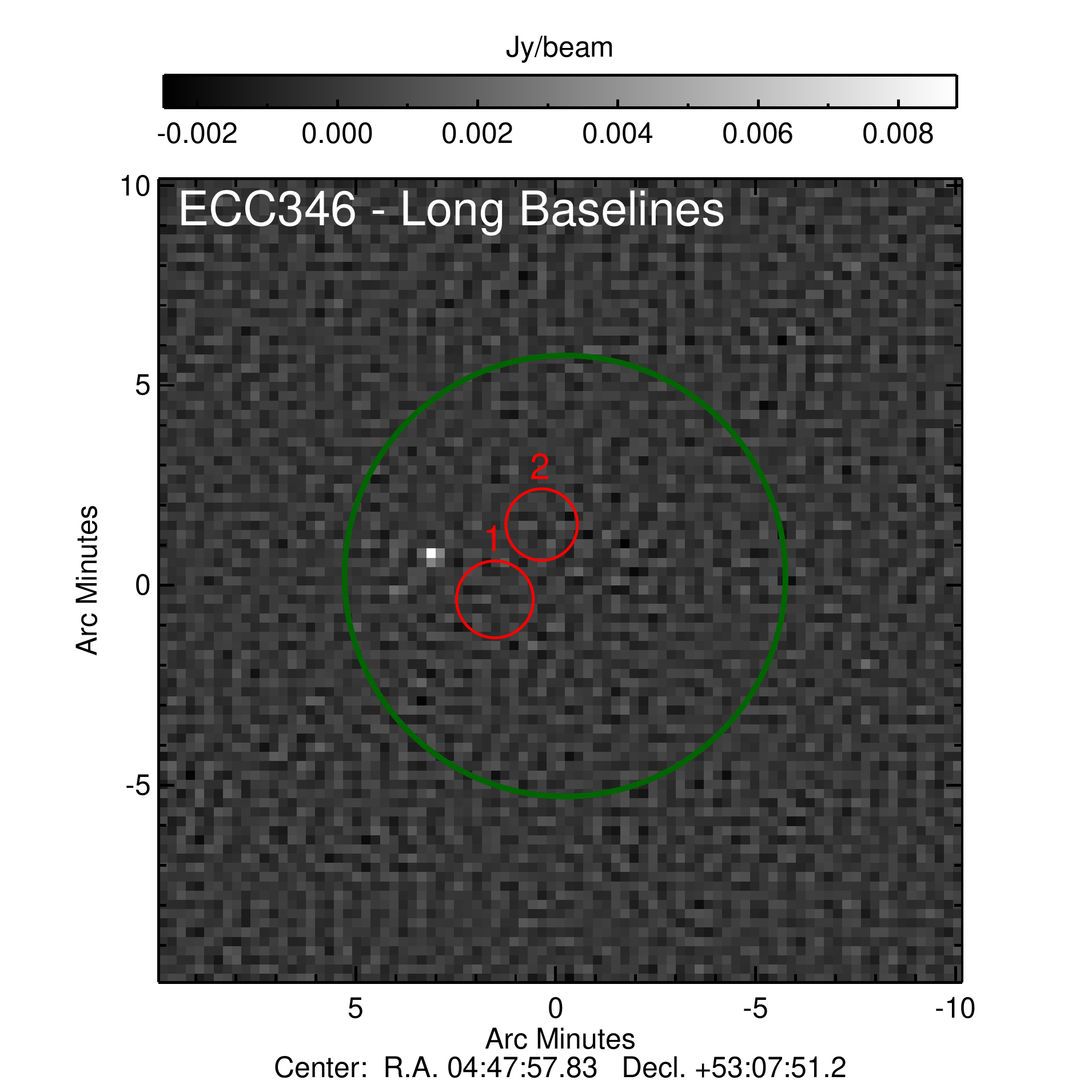} \\
\end{center}
\vspace{-0.4cm}
\caption{Continued}
\label{Fig:CARMA_Maps}
\end{figure*}

\clearpage
%\newpage
\section{$N_{\mathrm{H}}$ and $T_{\mathrm{d}}$ Maps}
\label{appendix2}

\vspace{-0.5cm}
\begin{figure*}
\begin{center}
\includegraphics[angle=0,scale=0.375]{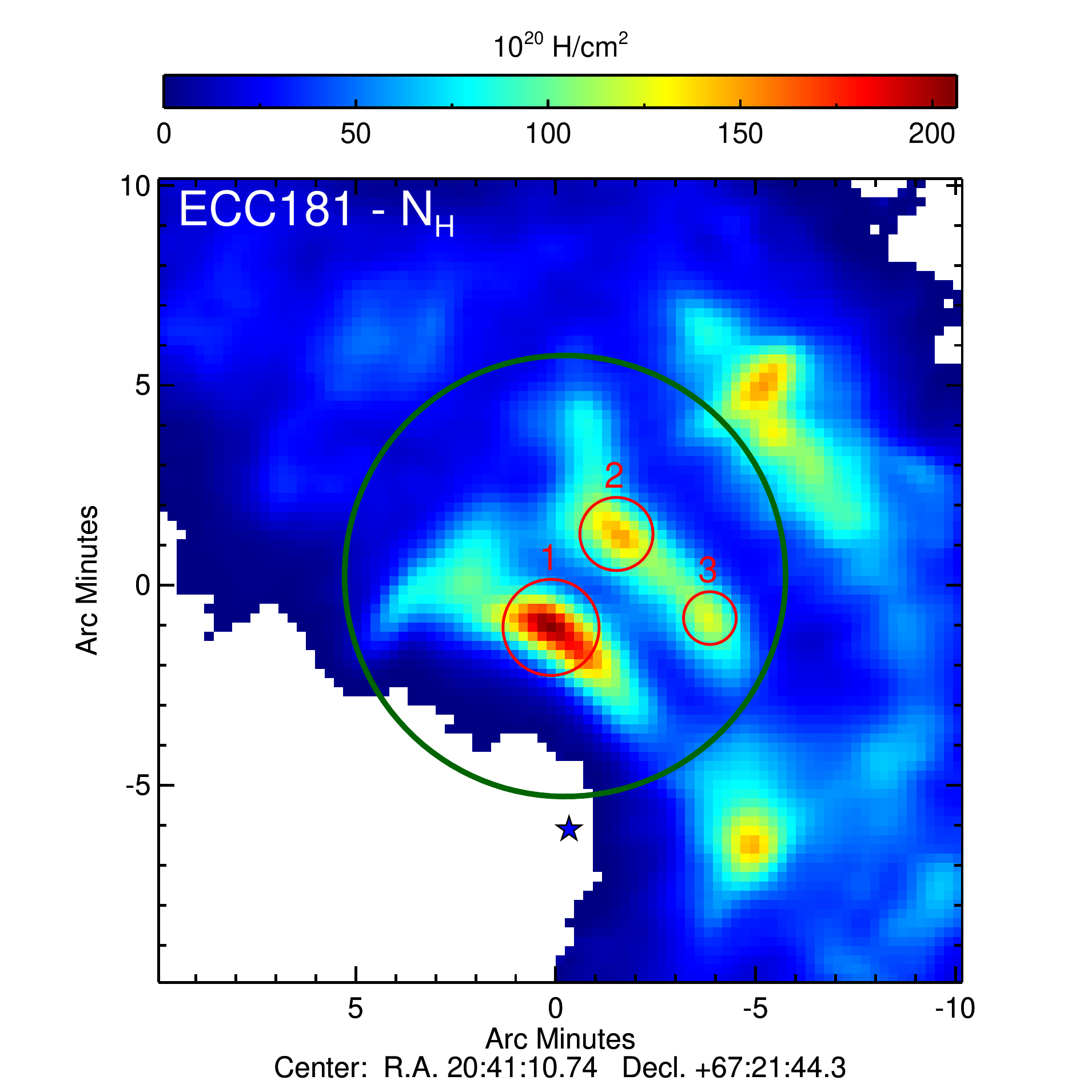}
\includegraphics[angle=0,scale=0.375]{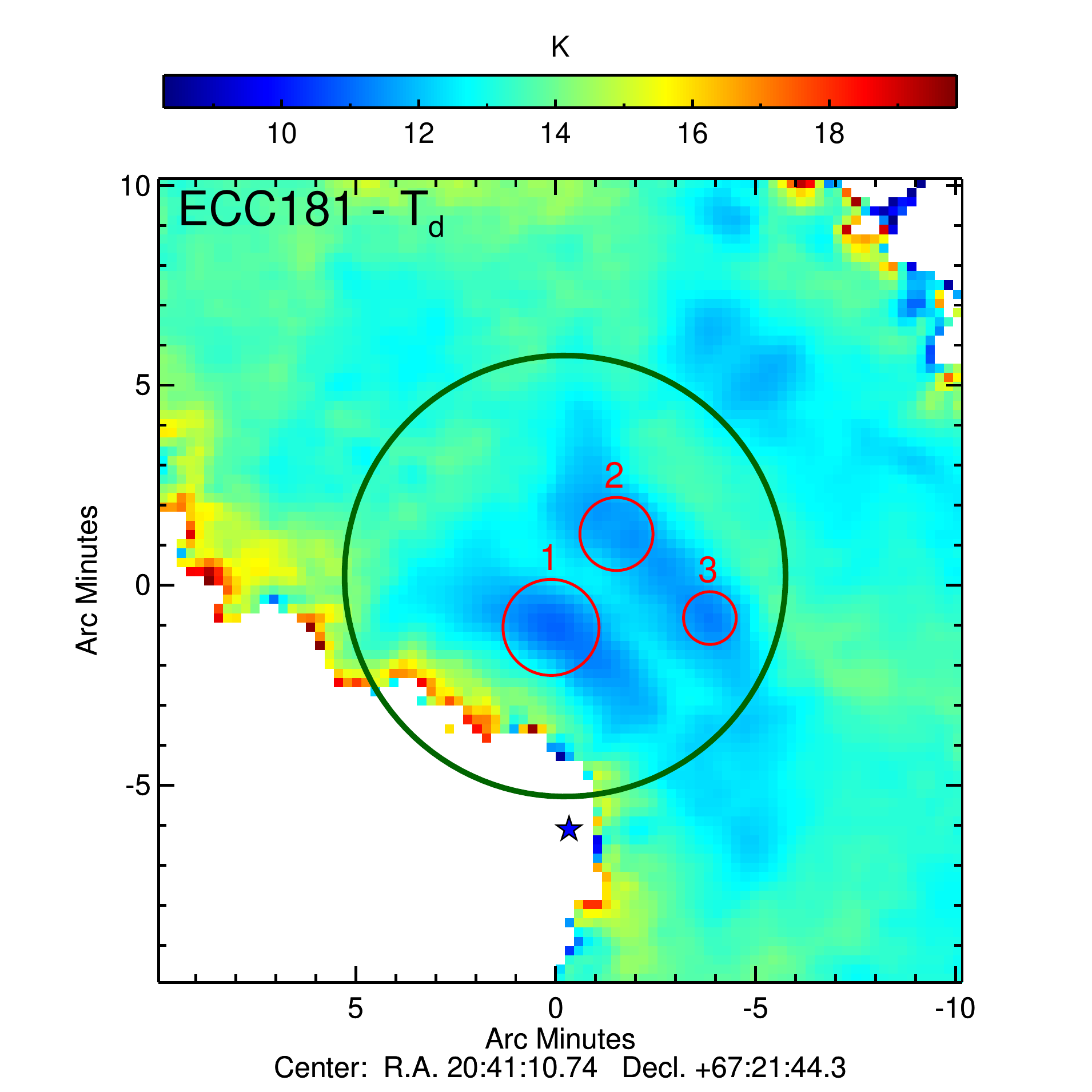} \\ 

\includegraphics[angle=0,scale=0.375]{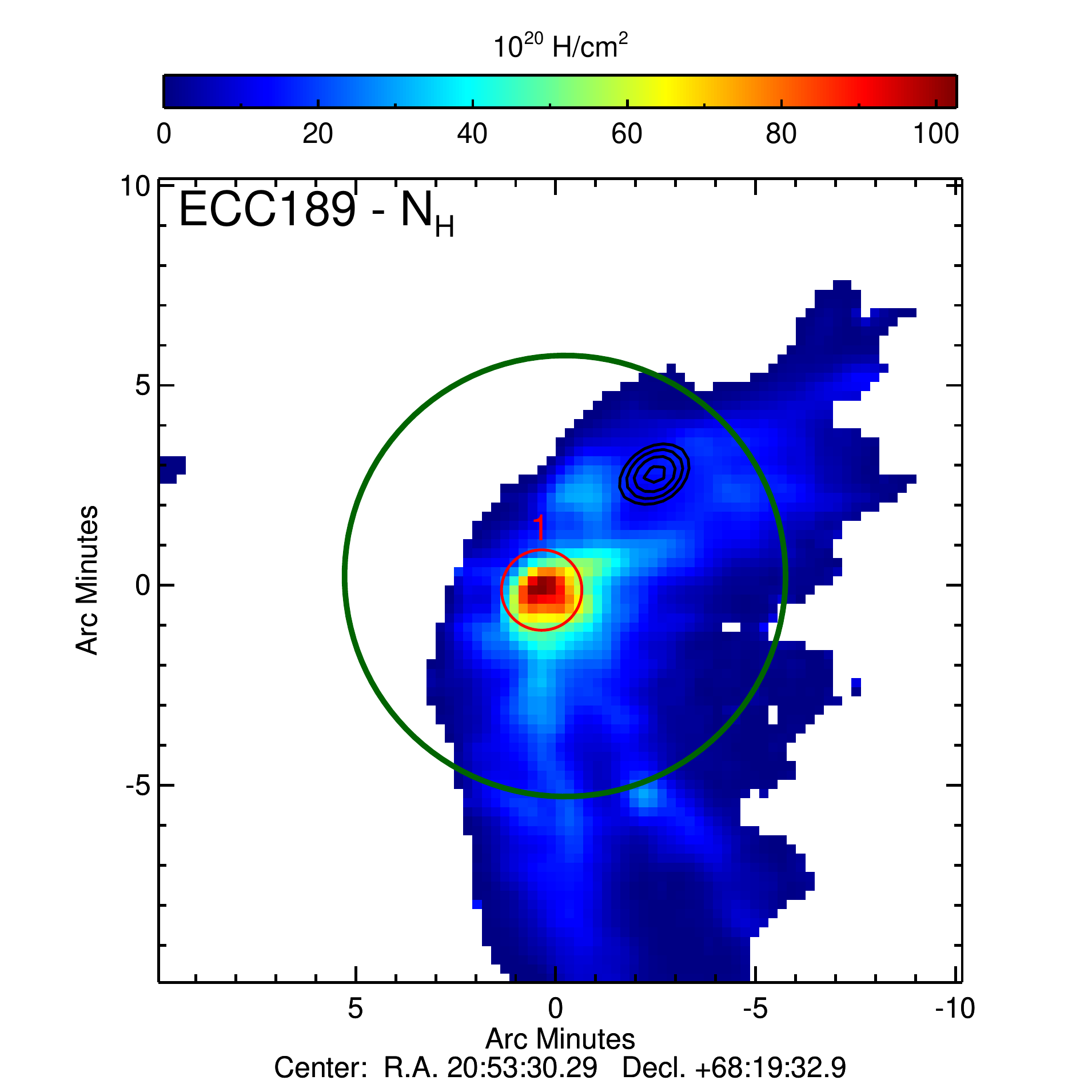}
\includegraphics[angle=0,scale=0.375]{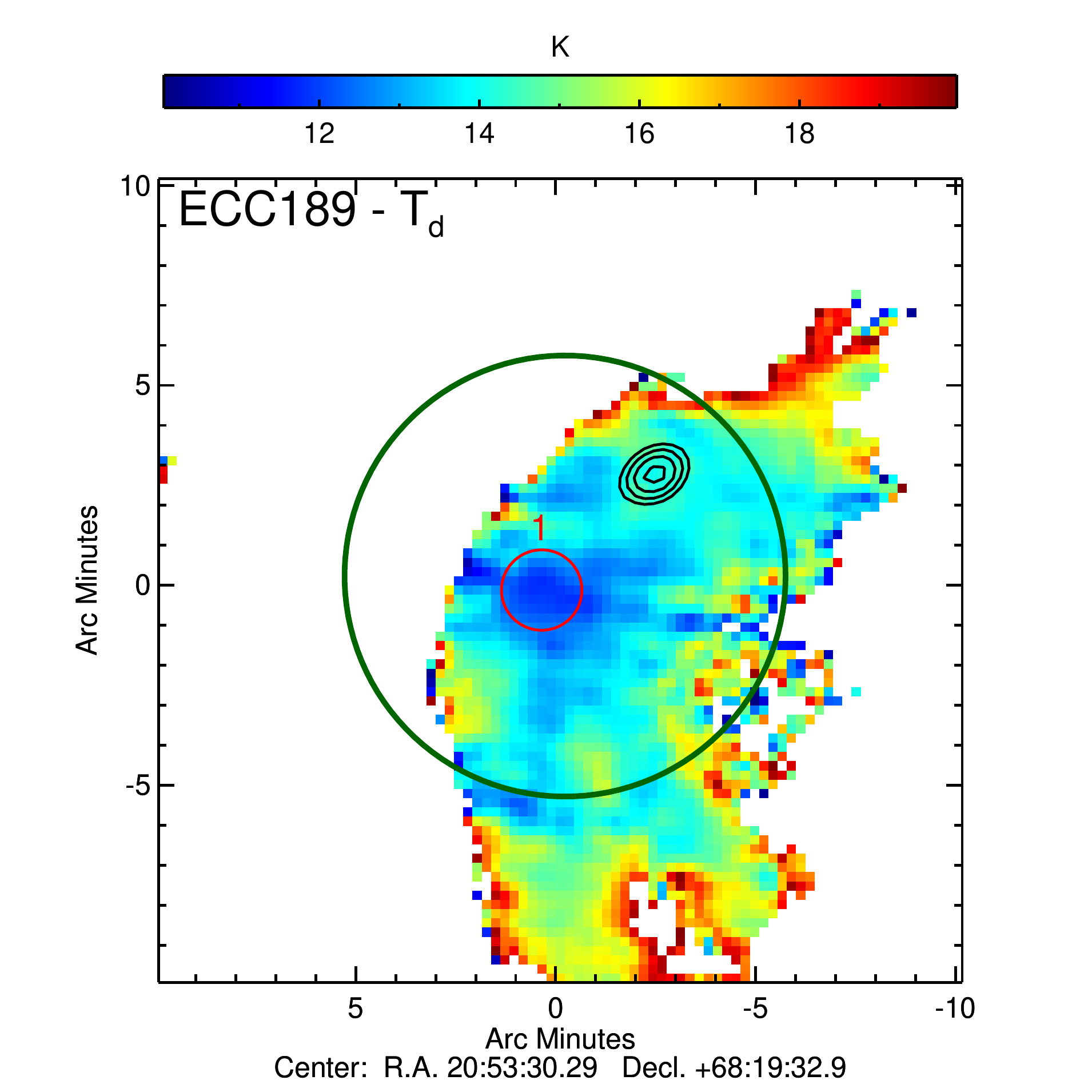} \\

\includegraphics[angle=0,scale=0.375]{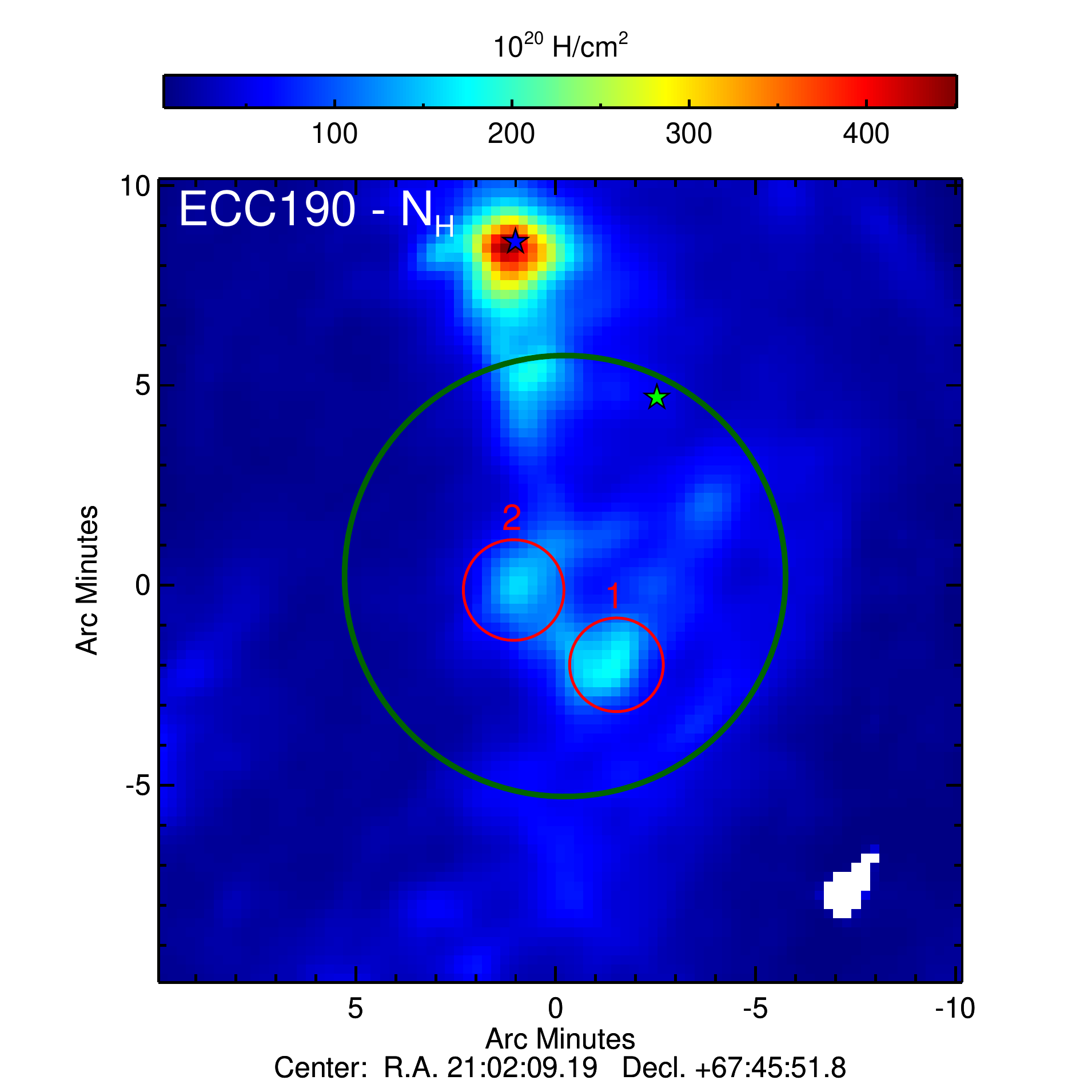}
\includegraphics[angle=0,scale=0.375]{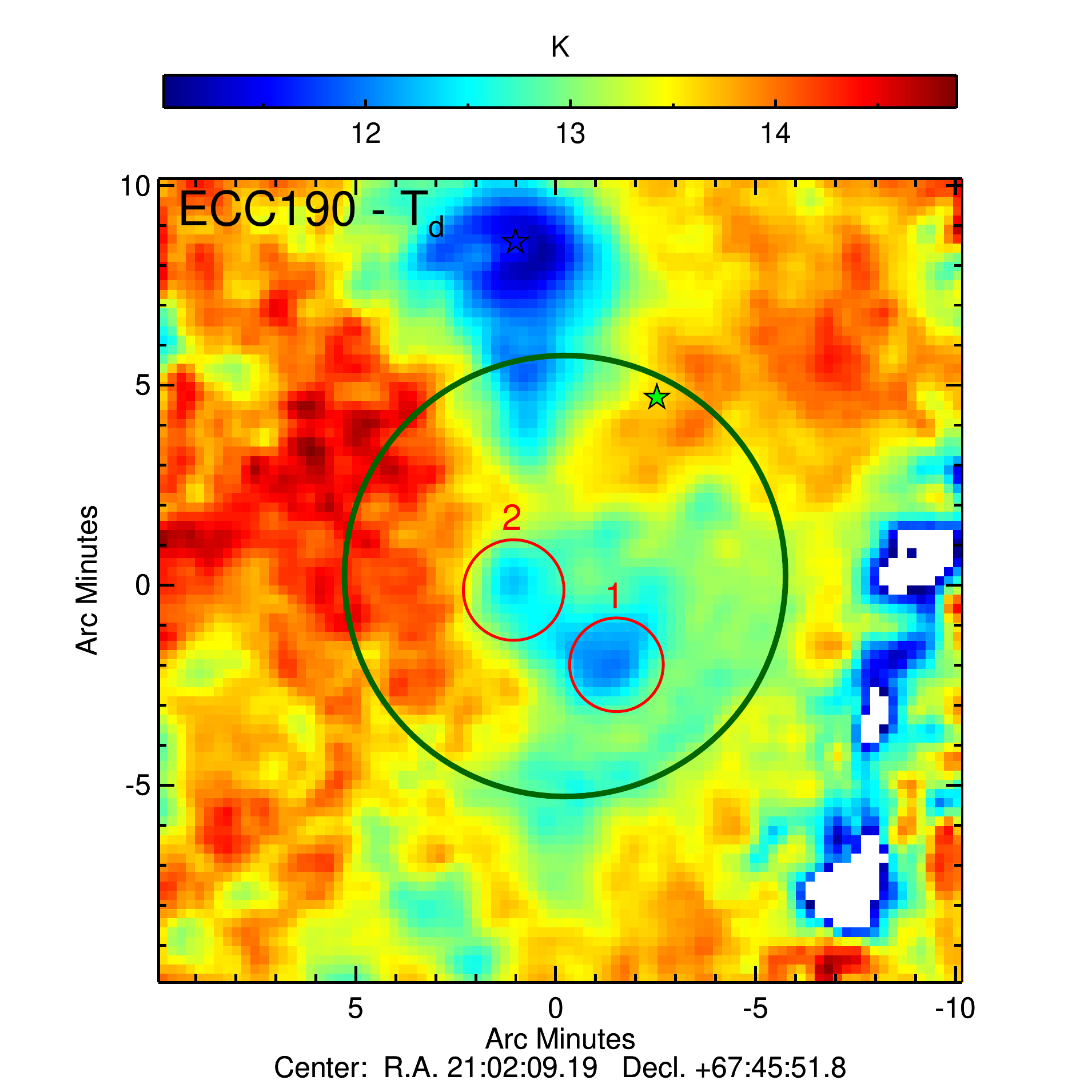} \\
\end{center}
\vspace{-0.4cm}
\caption{$N_{\mathrm{H}}$~(\textit{left}) and $T_{\mathrm{d}}$~(\textit{right}) maps for all fifteen cold clumps. The location of the cores are numbered and identified by red circles, while class I and class II YSO candidates are labelled as blue and green stars, respectively. CARMA short-baseline data contours, starting at 5$\sigma$ and increasing linearly, are overplotted (except for ECC191 where the contours increase in steps of 10$\sigma$) along with the CARMA 11~arcmin primary beam~(\textit{green circle}).}
\label{Fig:NH_Td_Maps}
\end{figure*}

\begin{figure*}
\ContinuedFloat
\begin{center}
\includegraphics[angle=0,scale=0.375]{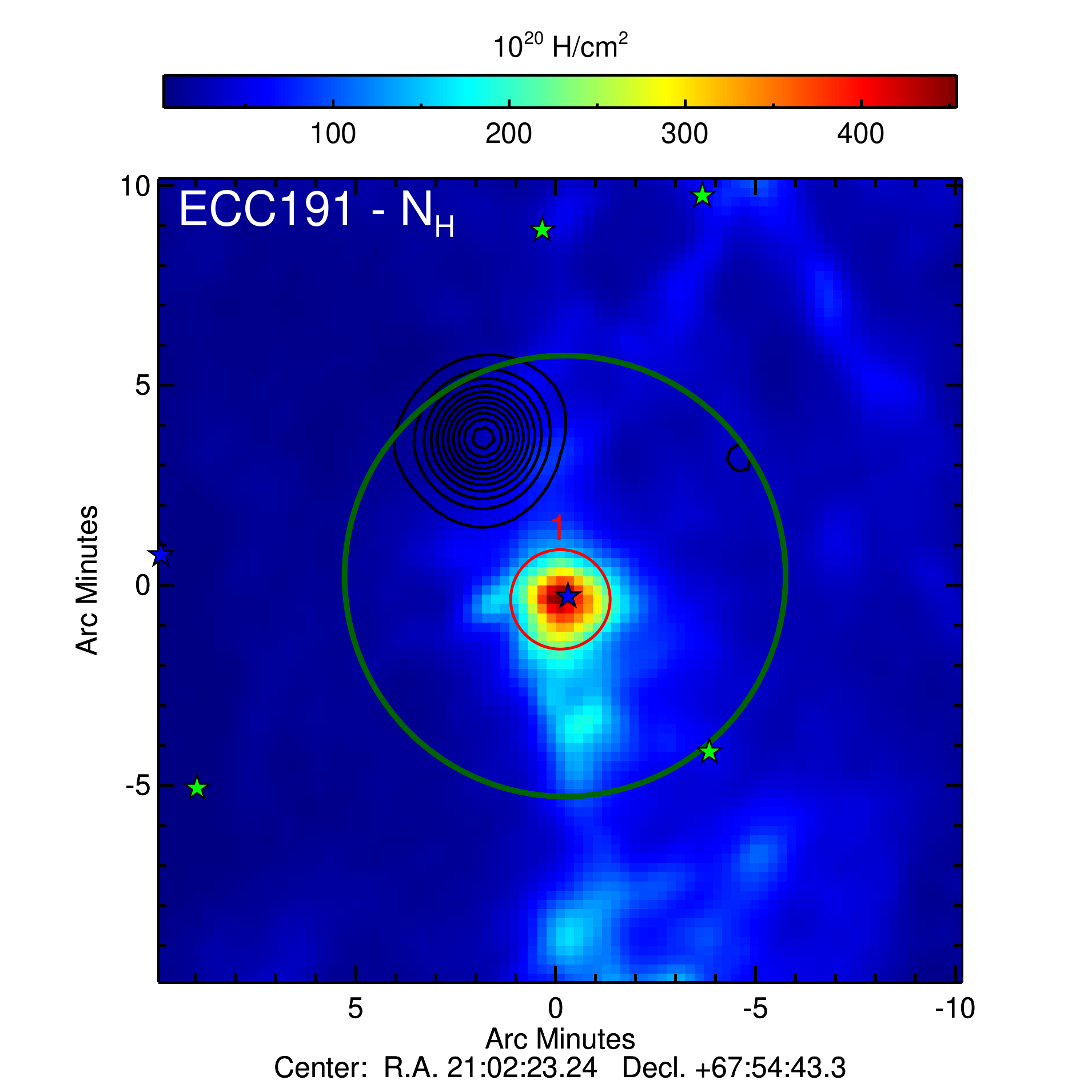}
\includegraphics[angle=0,scale=0.375]{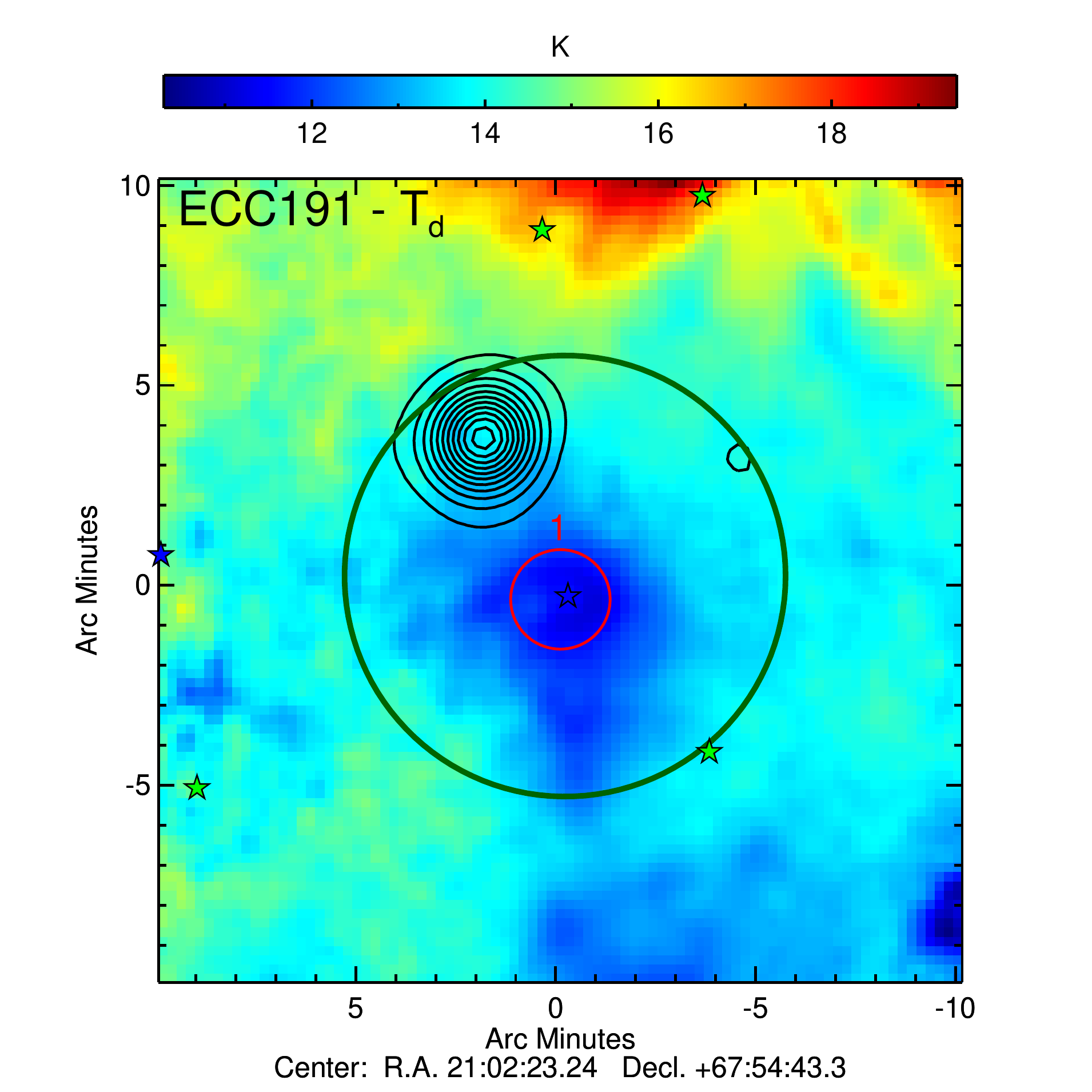} \\

\includegraphics[angle=0,scale=0.375]{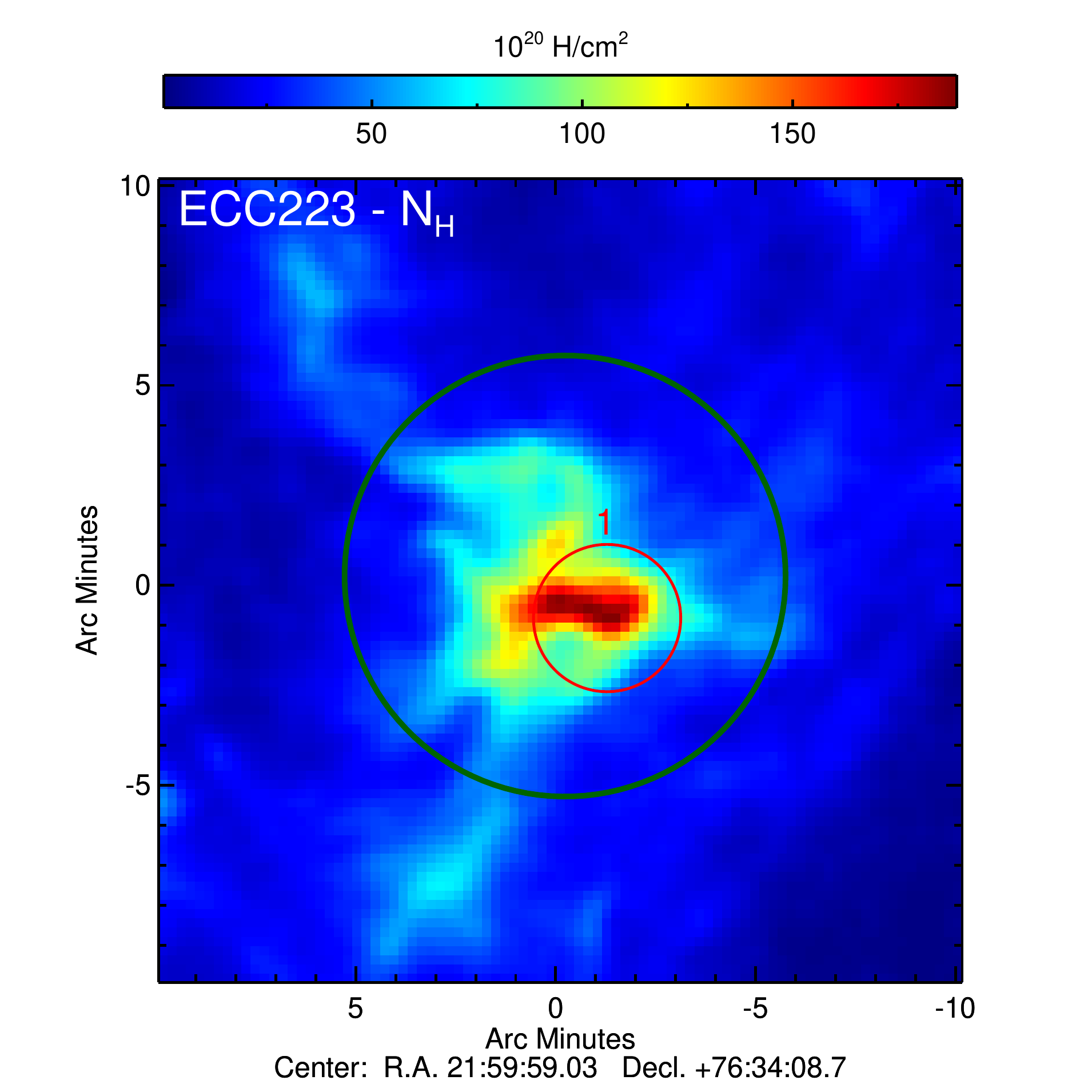}
\includegraphics[angle=0,scale=0.375]{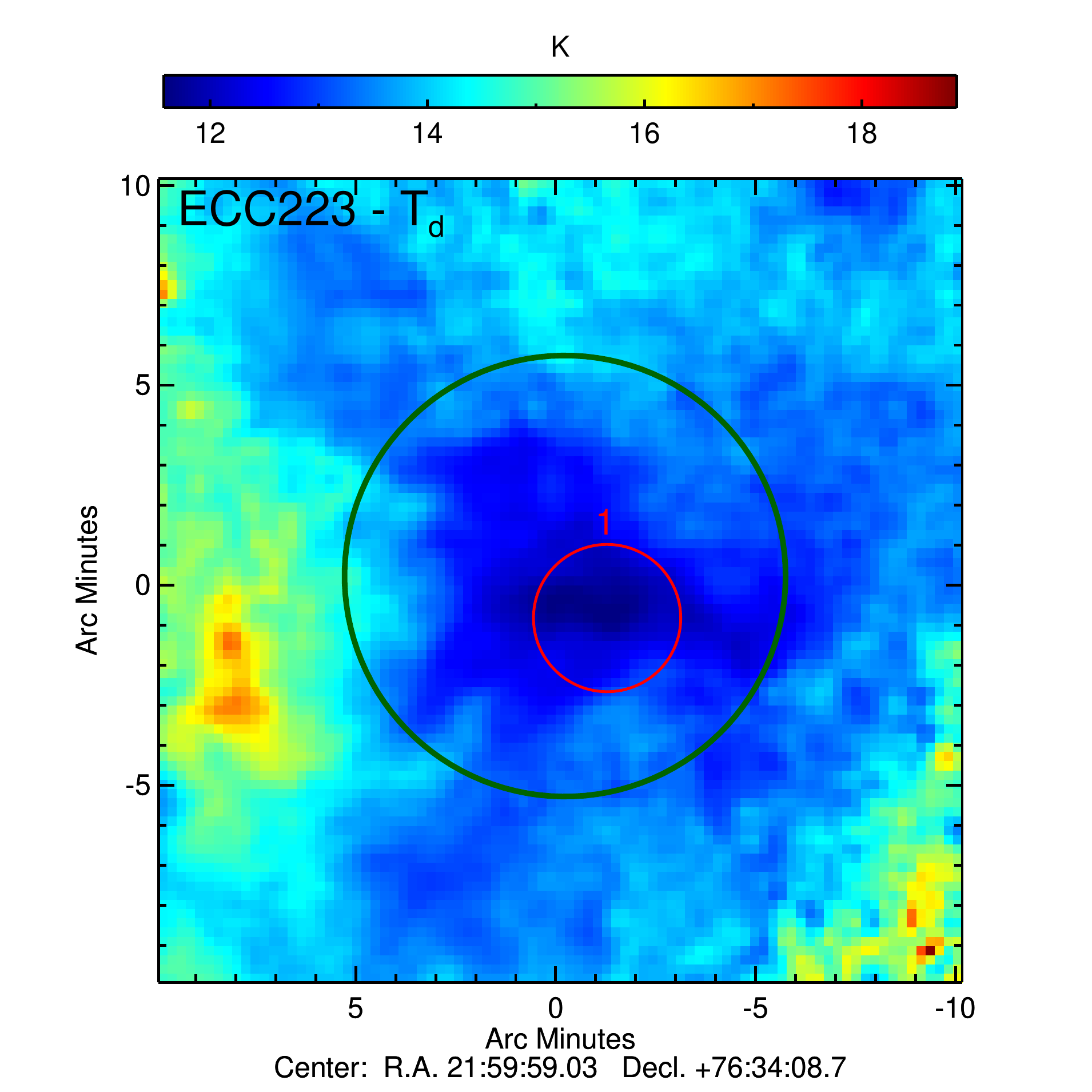} \\

\includegraphics[angle=0,scale=0.375]{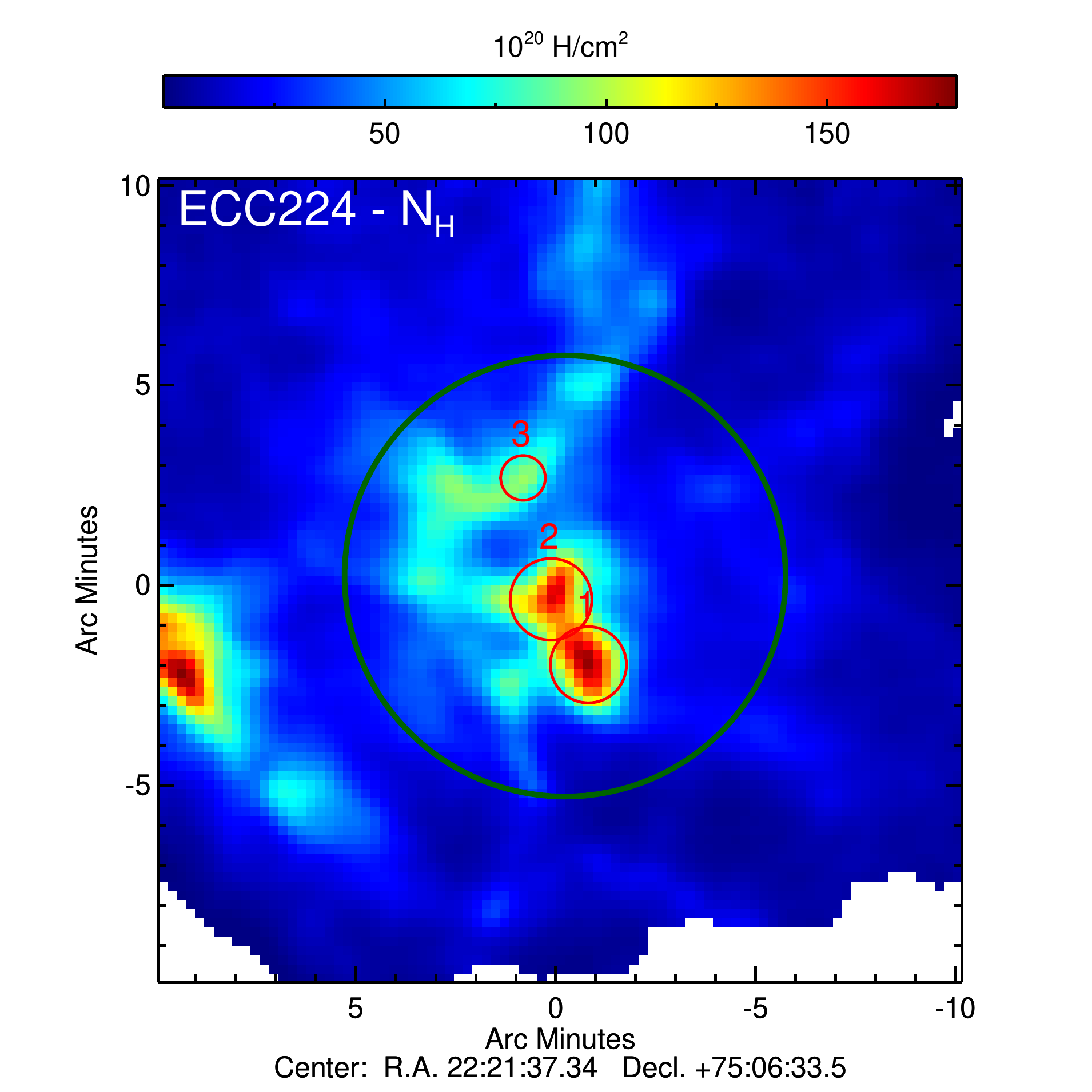}
\includegraphics[angle=0,scale=0.375]{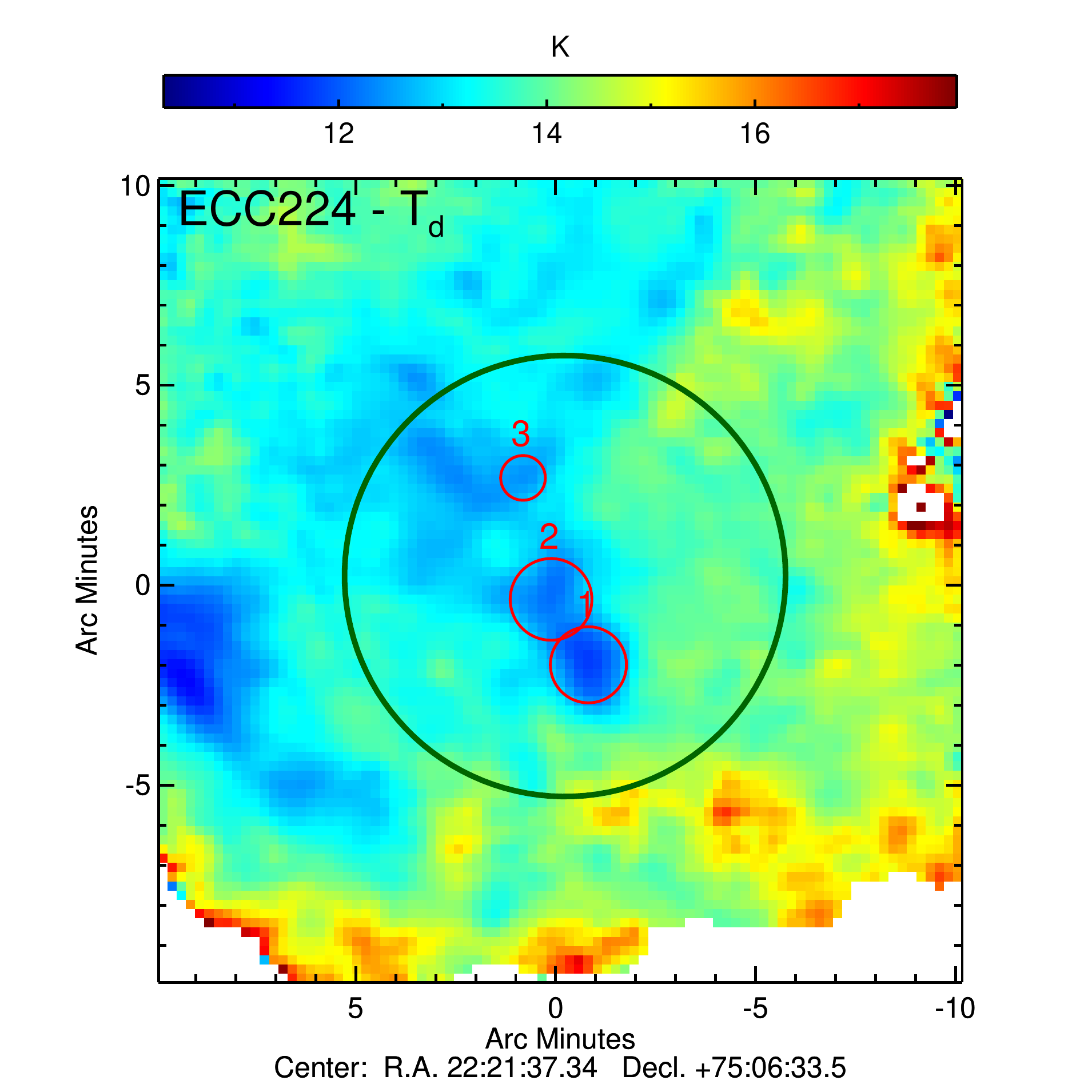} \\
\end{center}
\vspace{-0.4cm}
\caption{Continued}
\label{Fig:NH_Td_Maps}
\end{figure*}

\begin{figure*}
\ContinuedFloat
\begin{center}
\includegraphics[angle=0,scale=0.375]{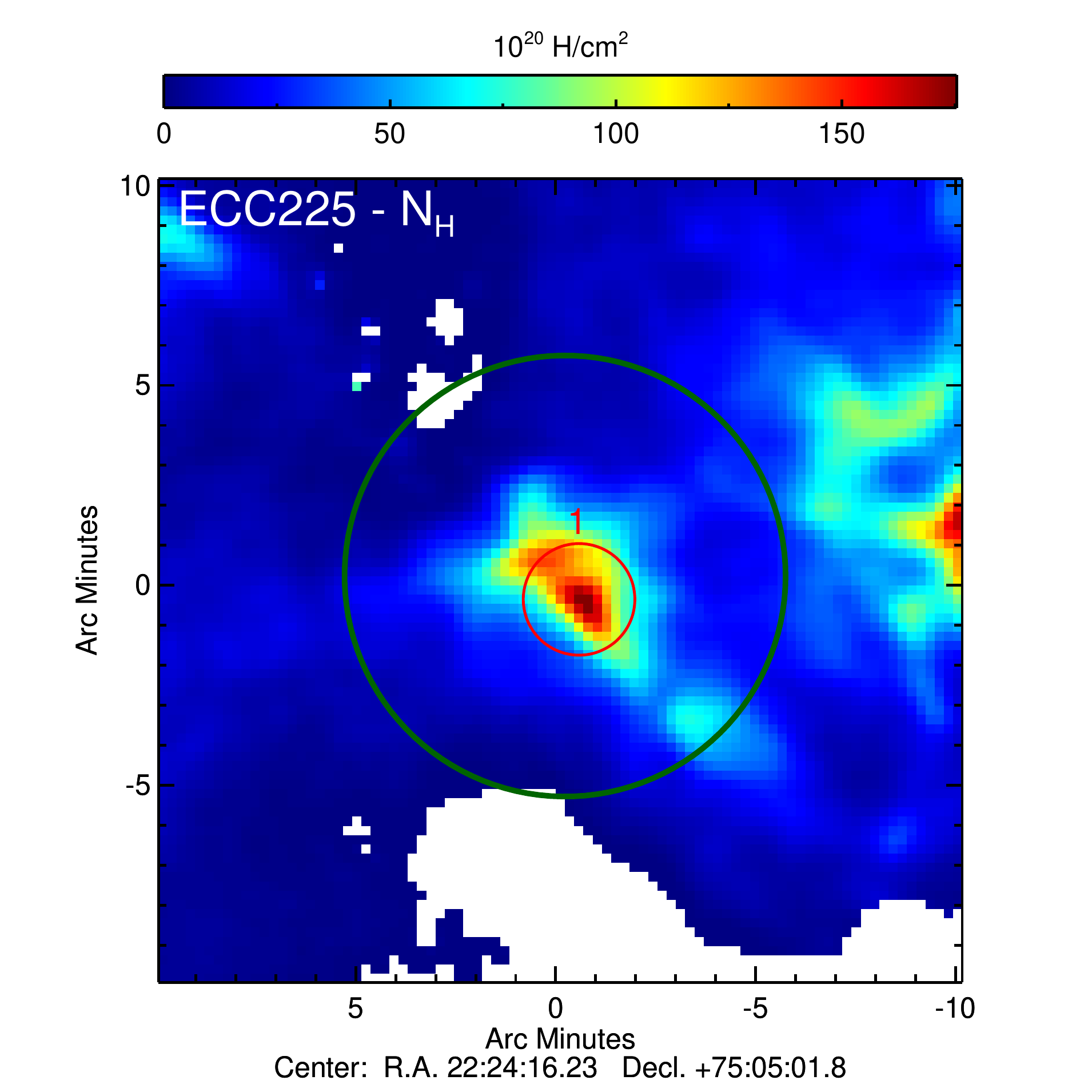}
\includegraphics[angle=0,scale=0.375]{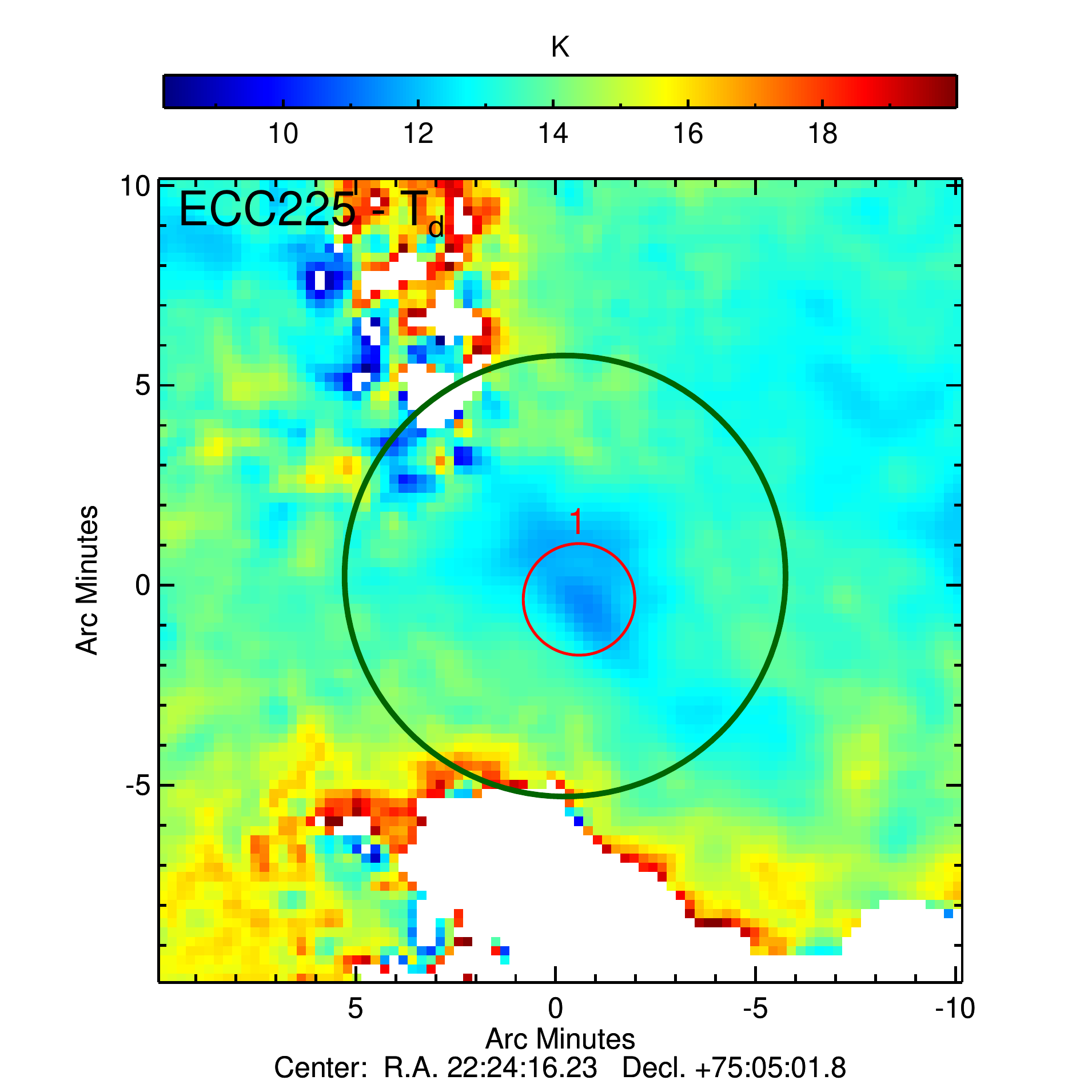} \\

\includegraphics[angle=0,scale=0.375]{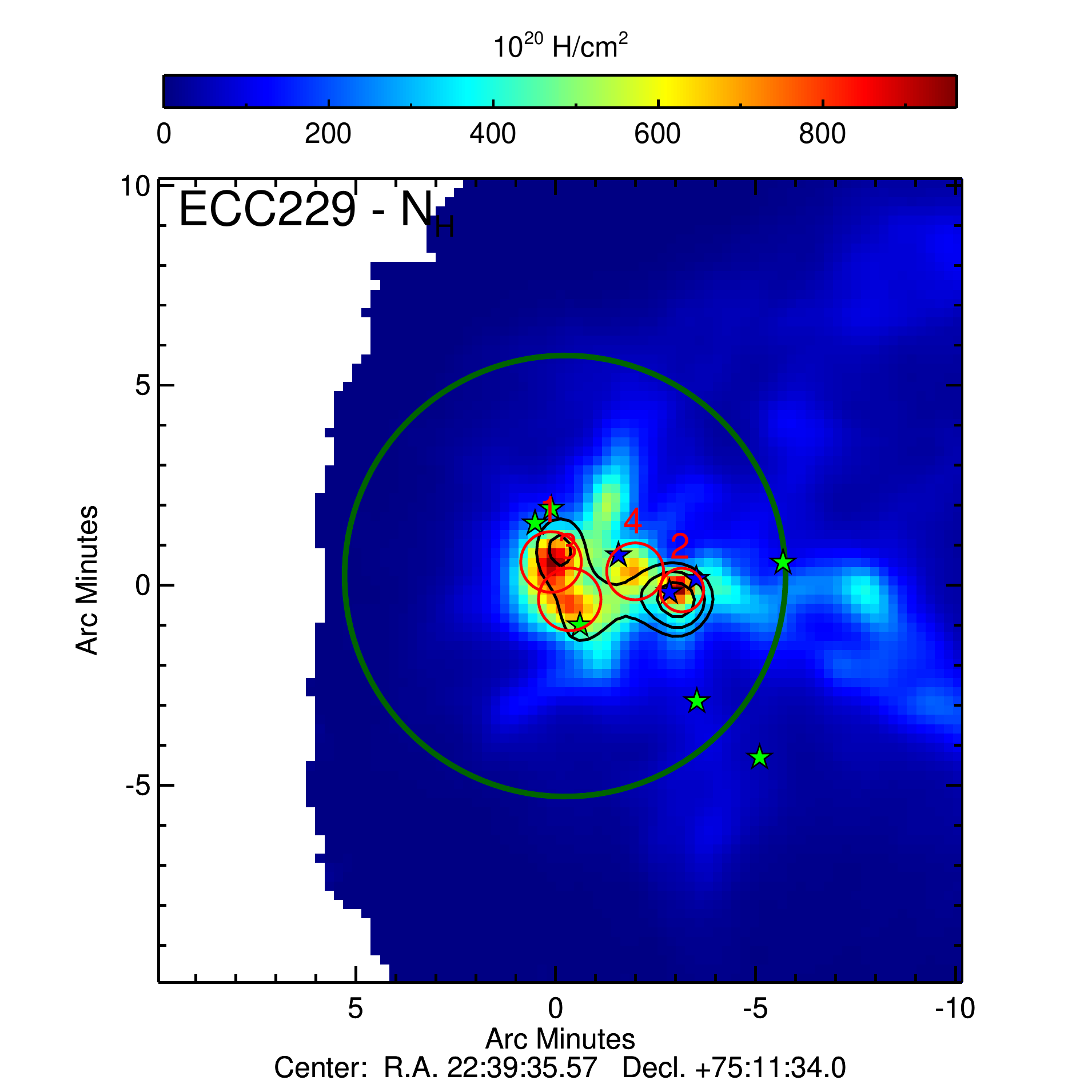}
\includegraphics[angle=0,scale=0.375]{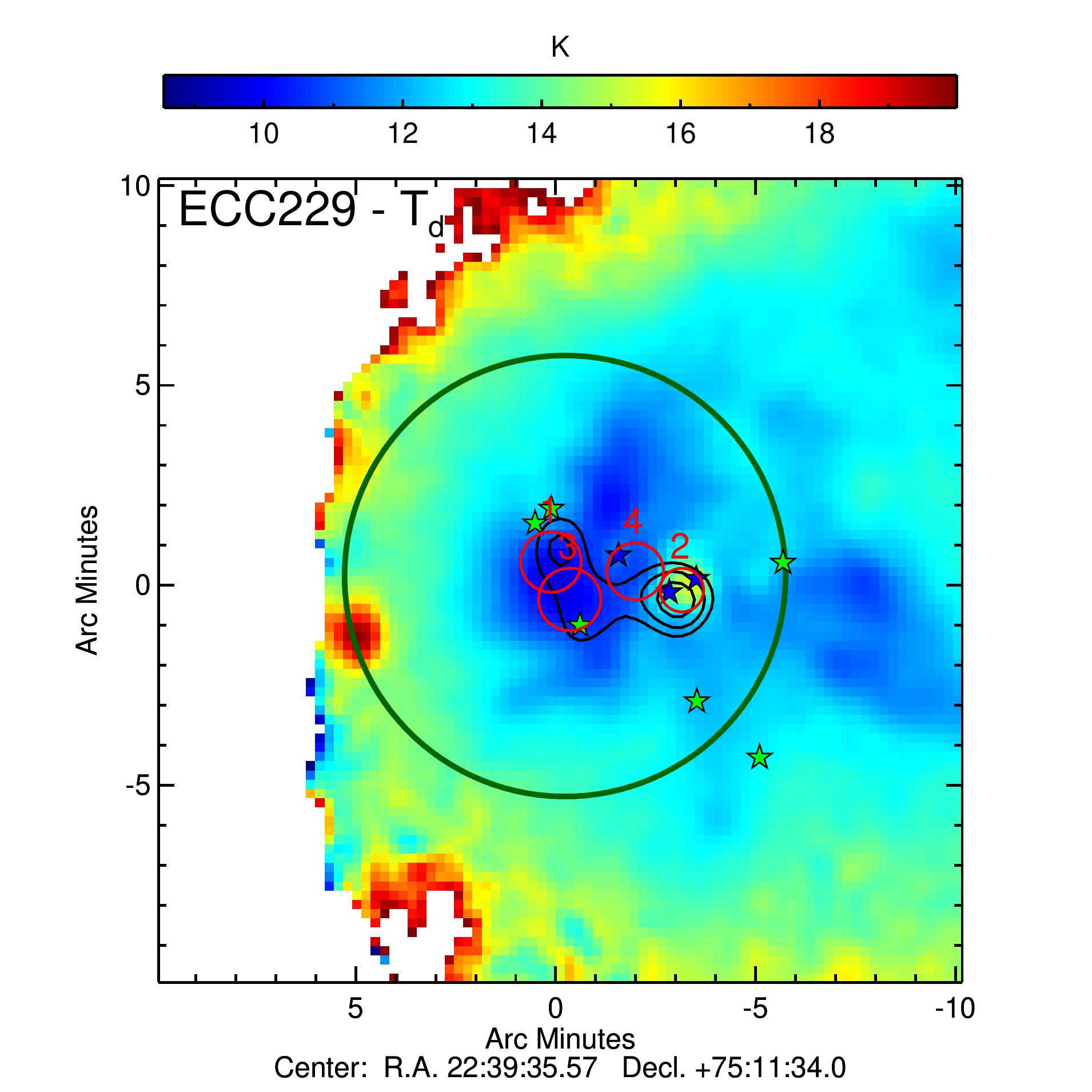} \\

\includegraphics[angle=0,scale=0.375]{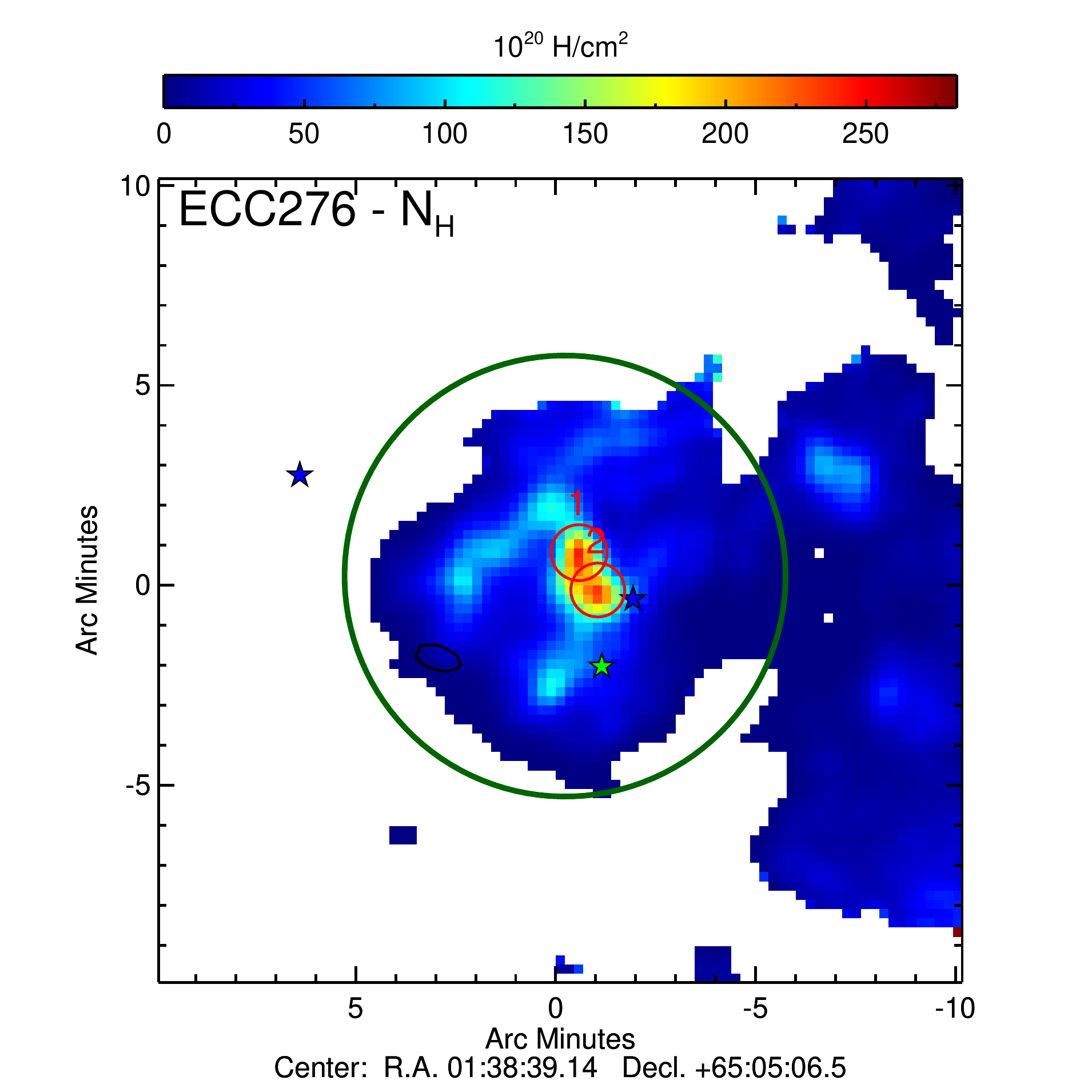}
\includegraphics[angle=0,scale=0.375]{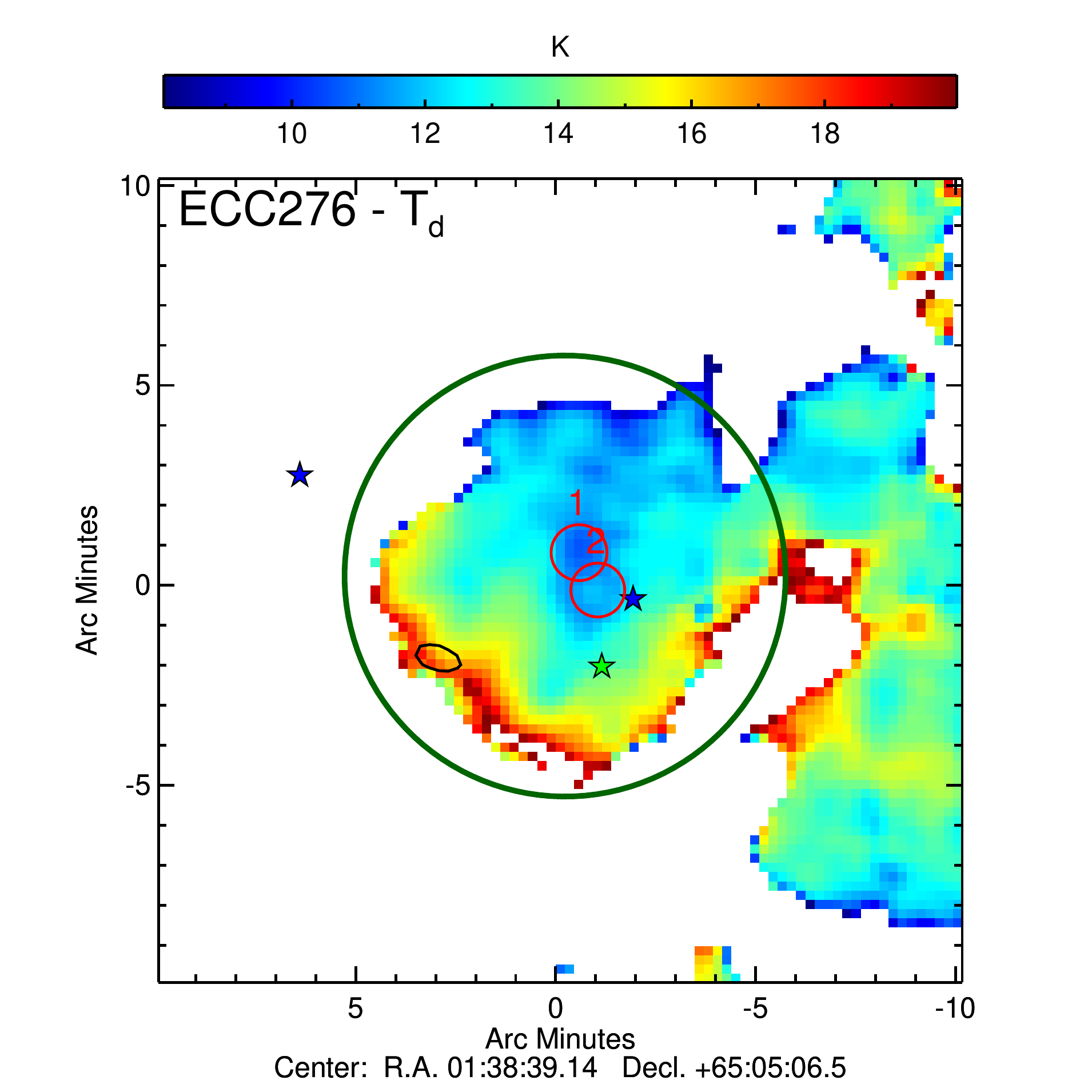} \\
\end{center}
\vspace{-0.4cm}
\caption{Continued}
\label{Fig:NH_Td_Maps}
\end{figure*}

\begin{figure*}
\ContinuedFloat
\begin{center}
\includegraphics[angle=0,scale=0.375]{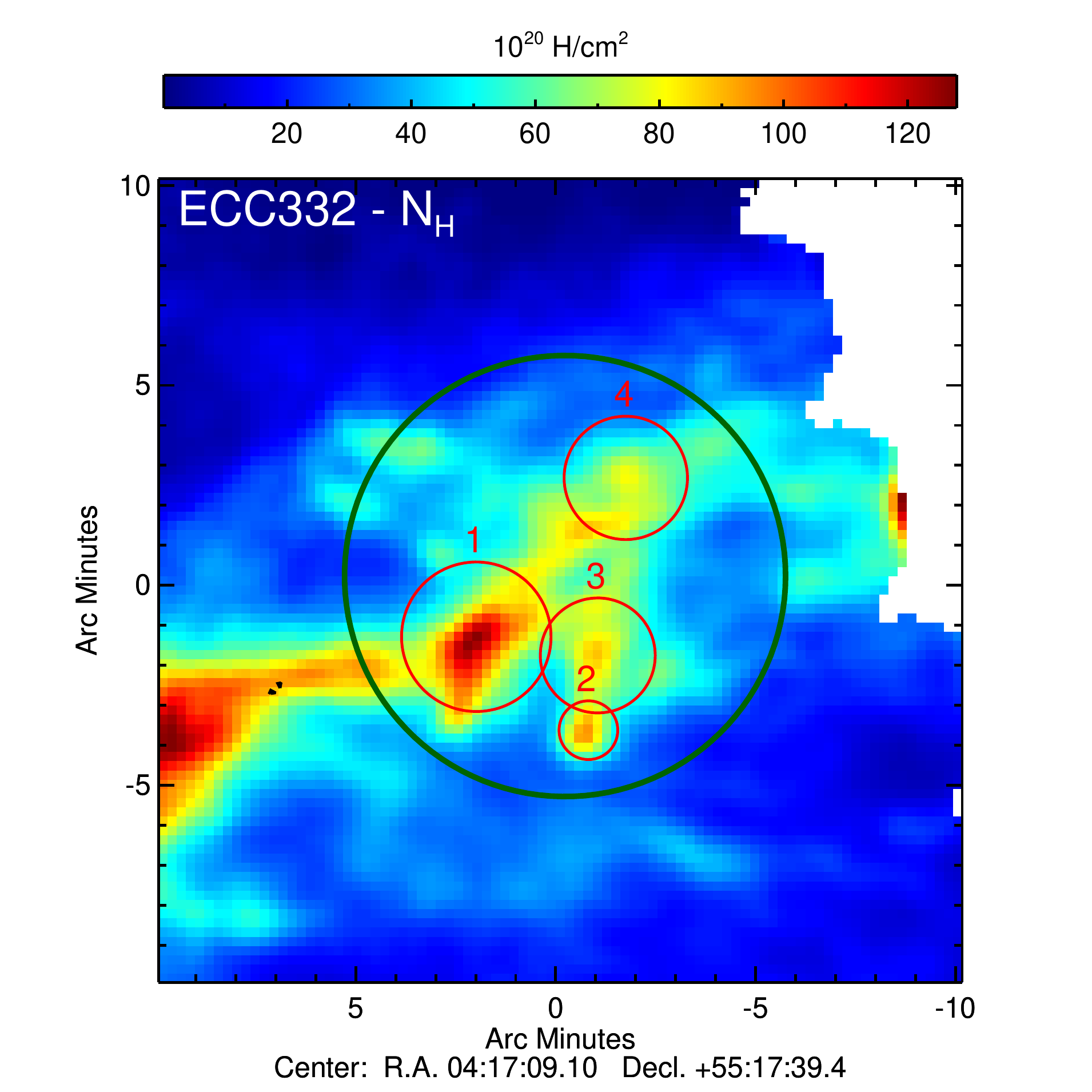}
\includegraphics[angle=0,scale=0.375]{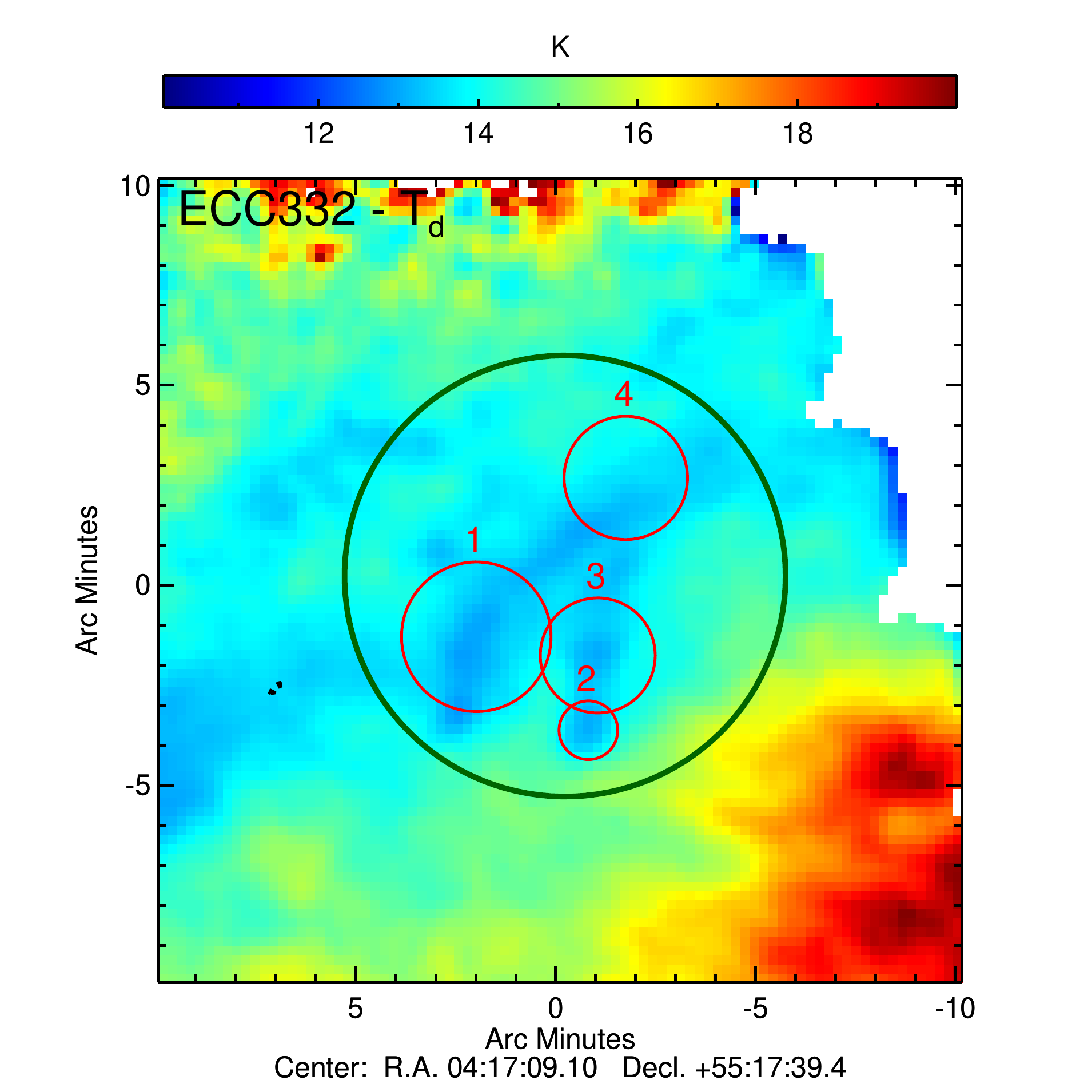} \\

\includegraphics[angle=0,scale=0.375]{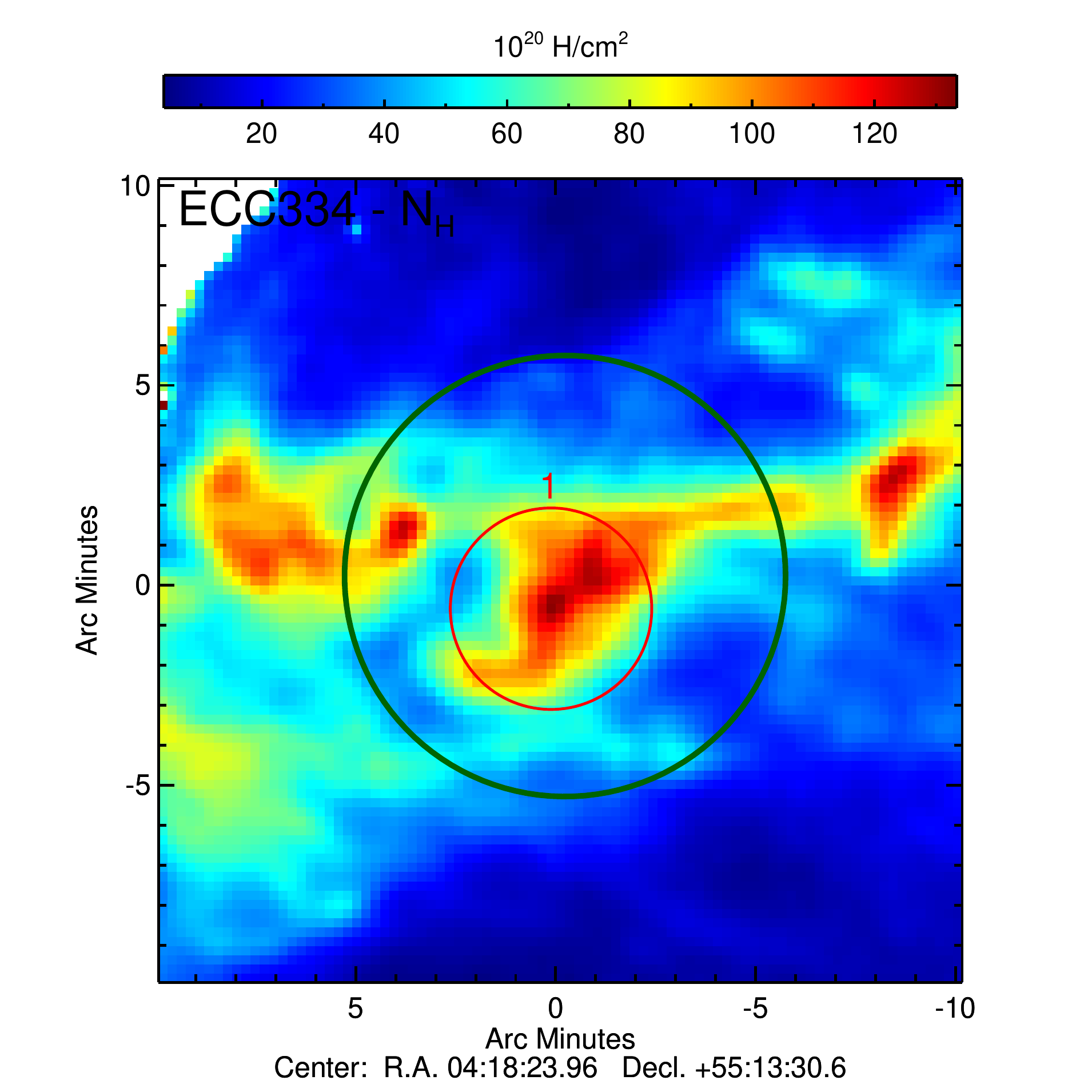}
\includegraphics[angle=0,scale=0.375]{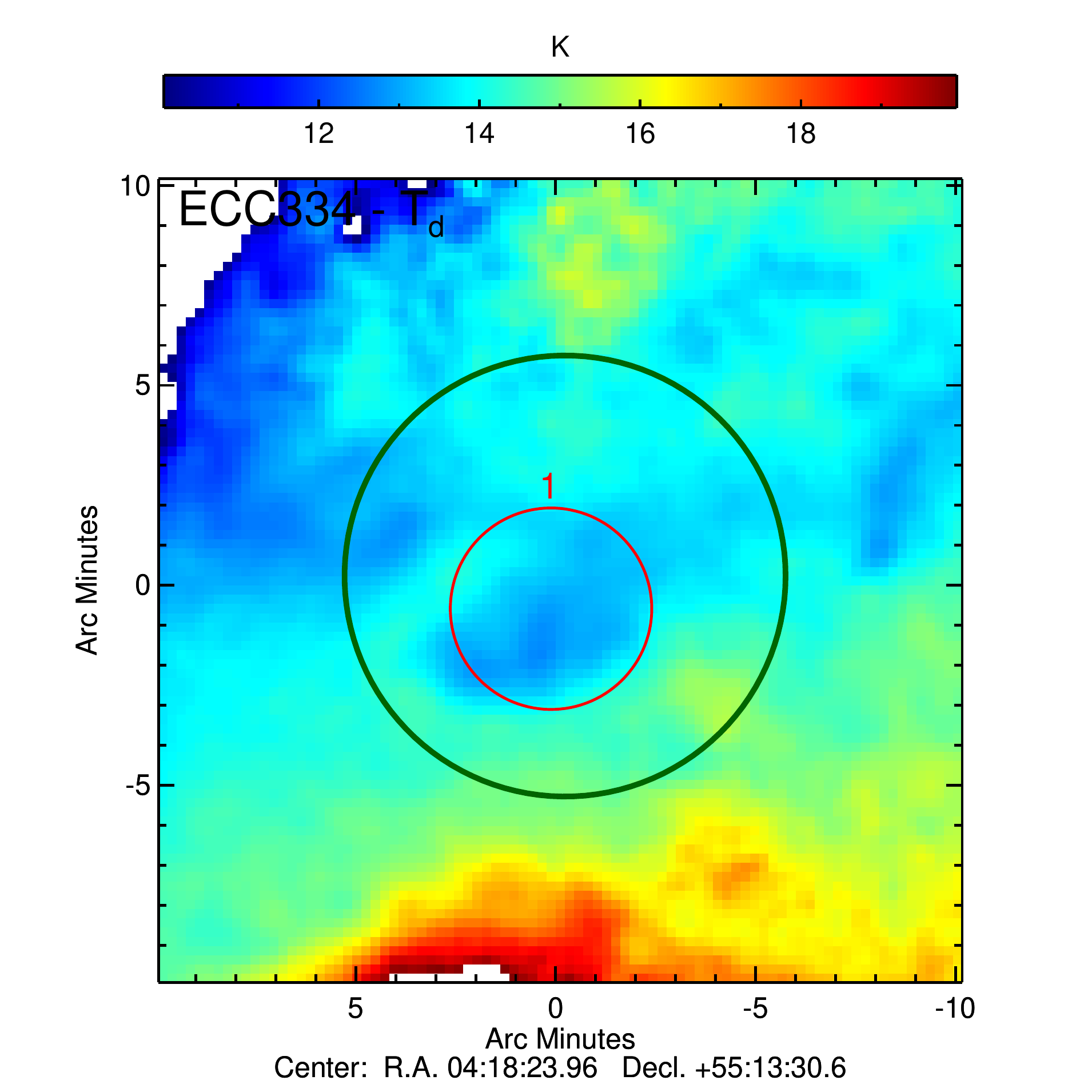} \\

\includegraphics[angle=0,scale=0.375]{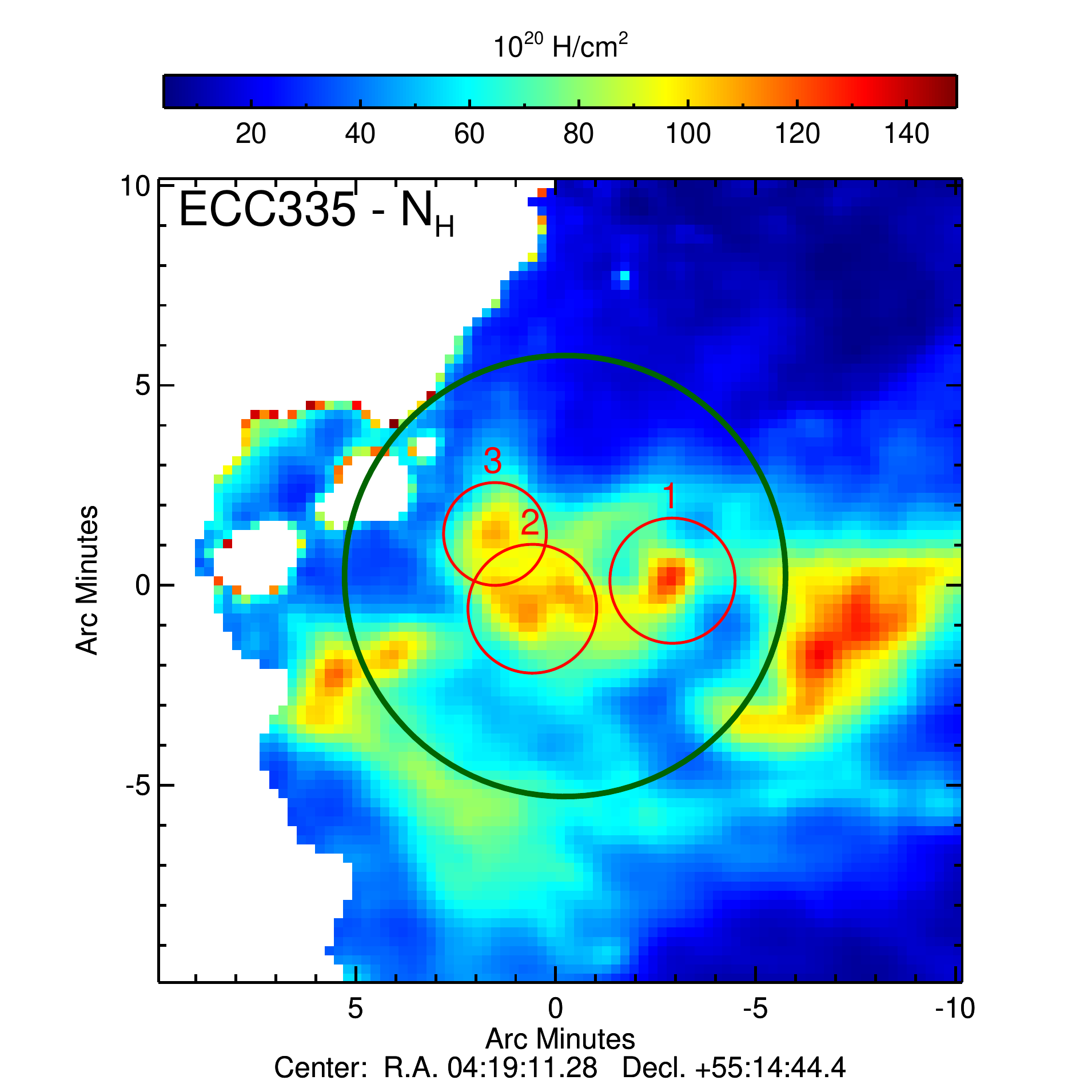}
\includegraphics[angle=0,scale=0.375]{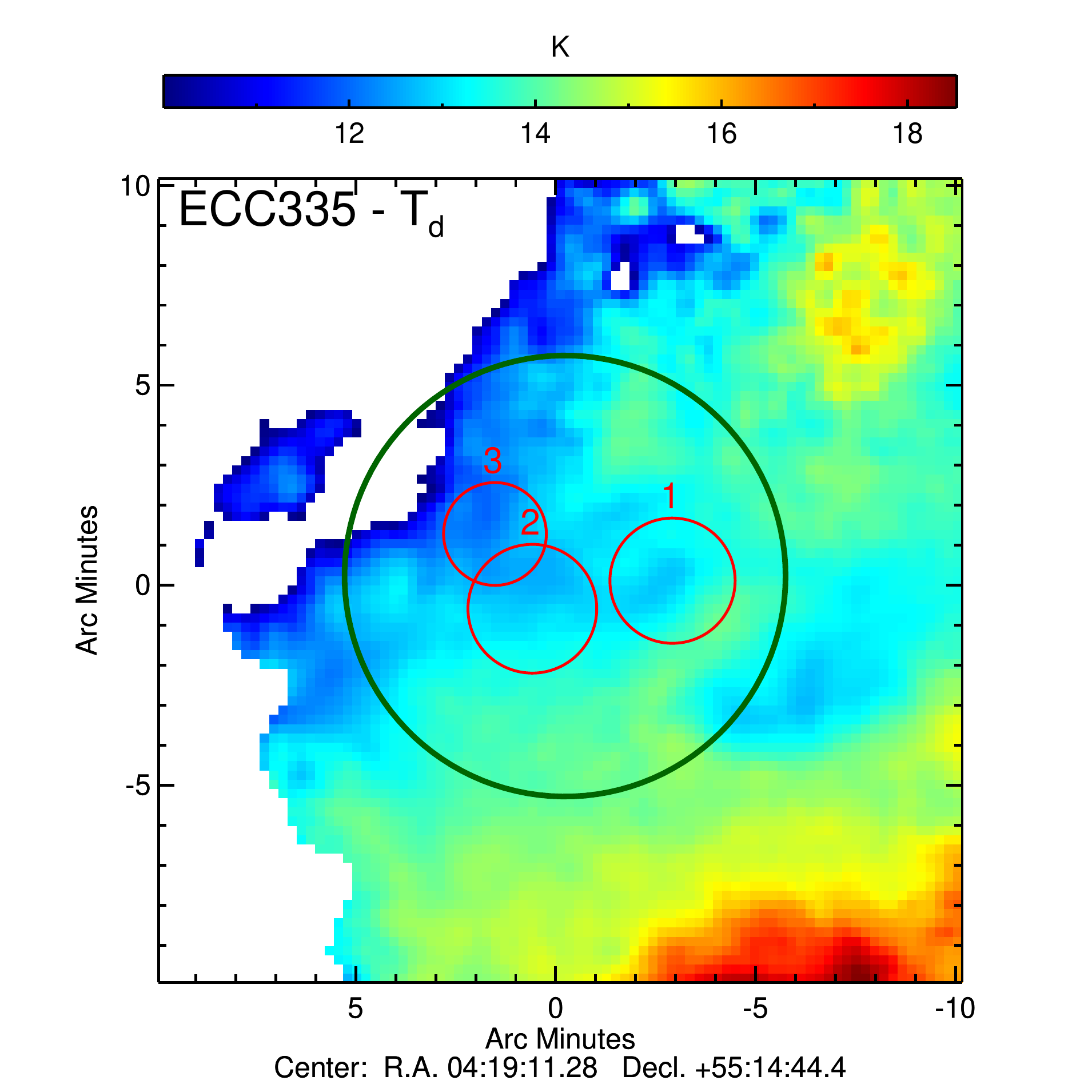} \\
\end{center}
\vspace{-0.4cm}
\caption{Continued}
\label{Fig:NH_Td_Maps}
\end{figure*}

\begin{figure*}
\ContinuedFloat
\begin{center}
\includegraphics[angle=0,scale=0.375]{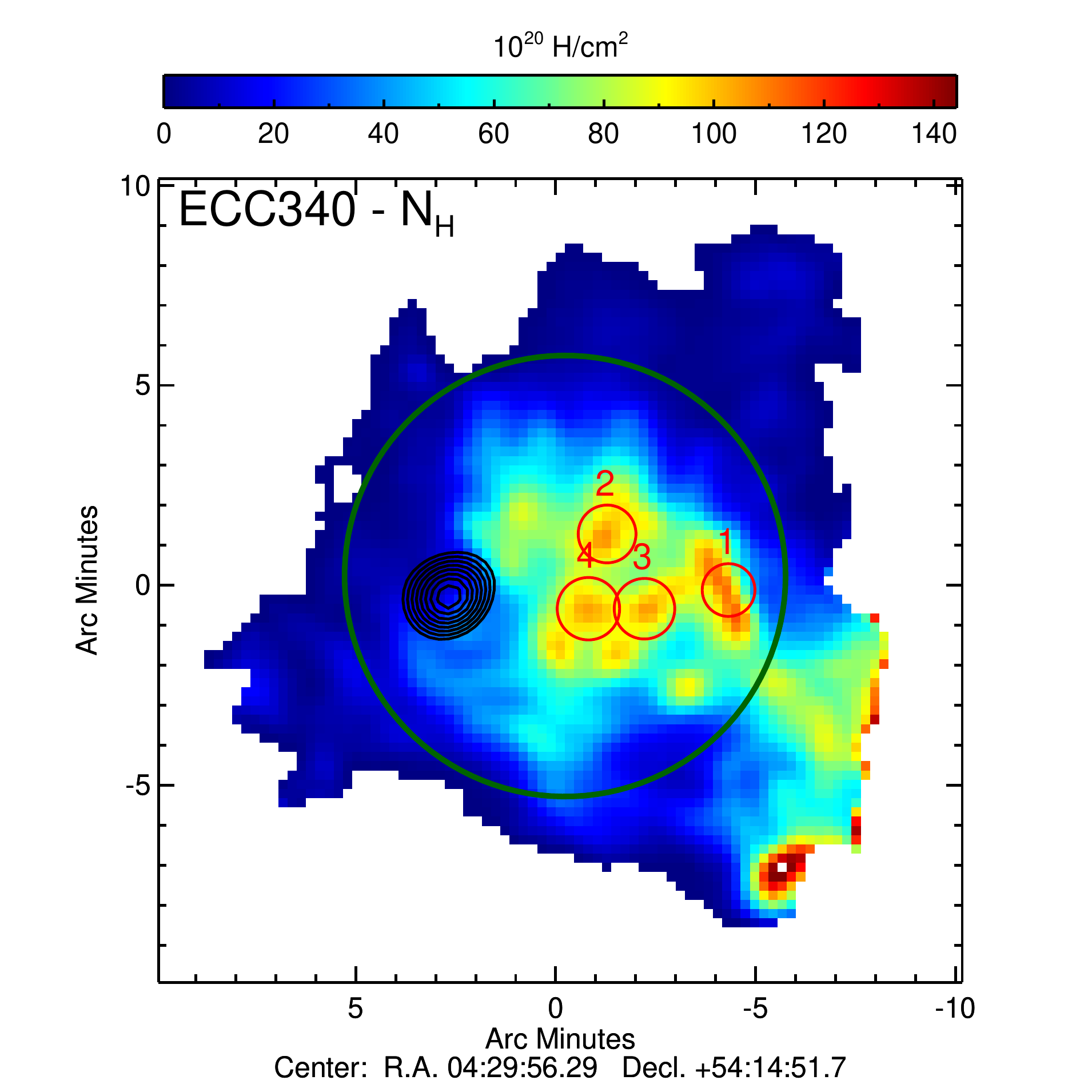}
\includegraphics[angle=0,scale=0.375]{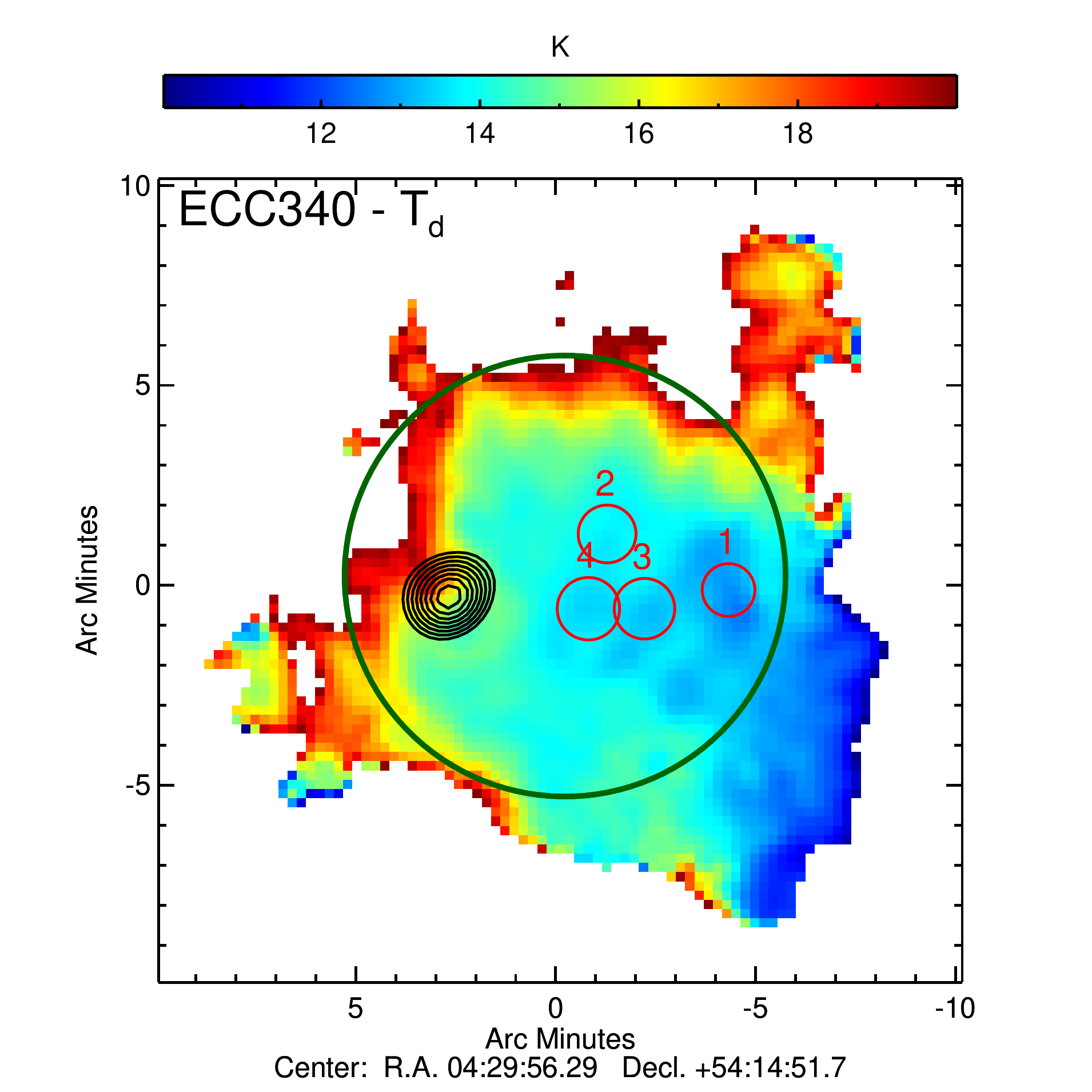} \\

\includegraphics[angle=0,scale=0.375]{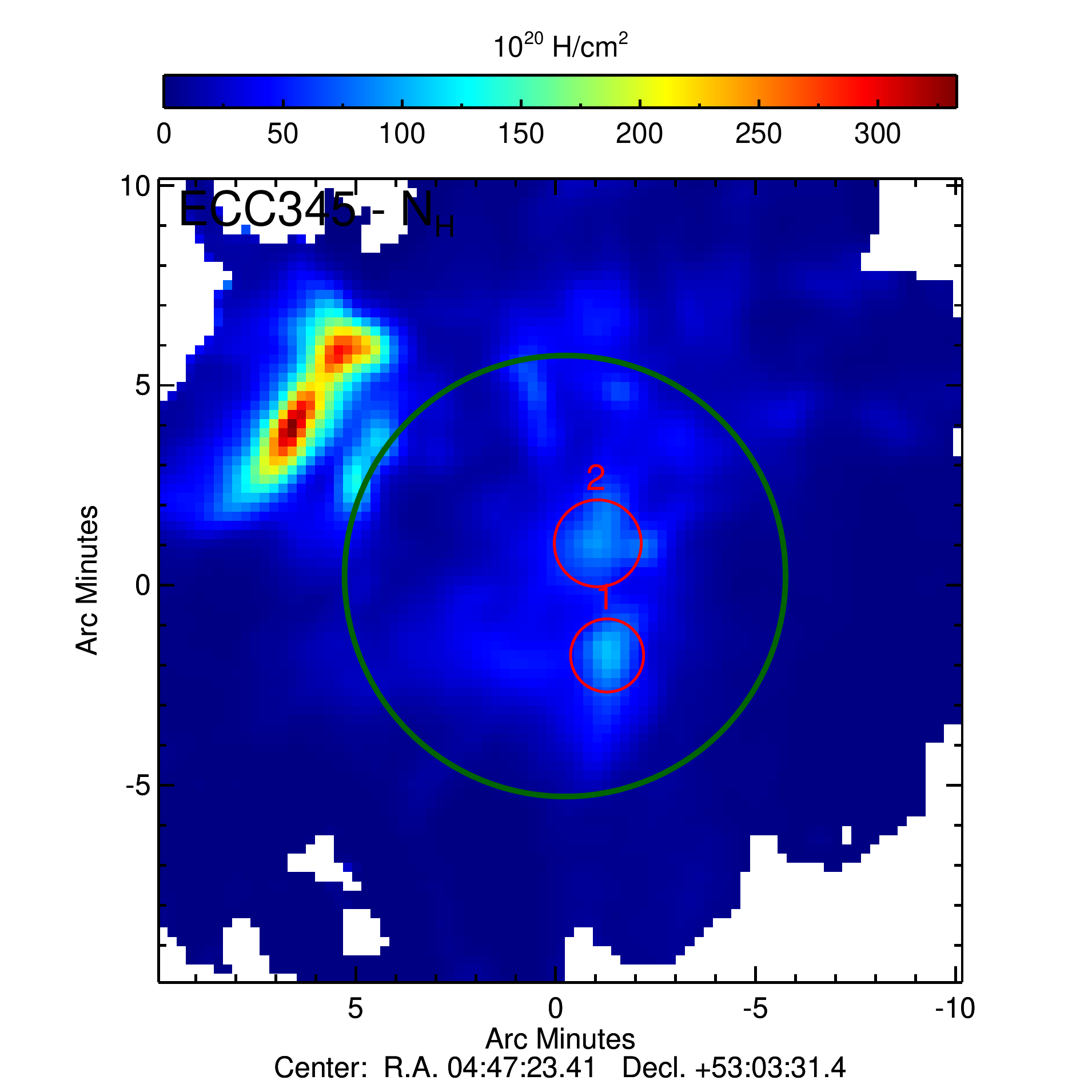}
\includegraphics[angle=0,scale=0.375]{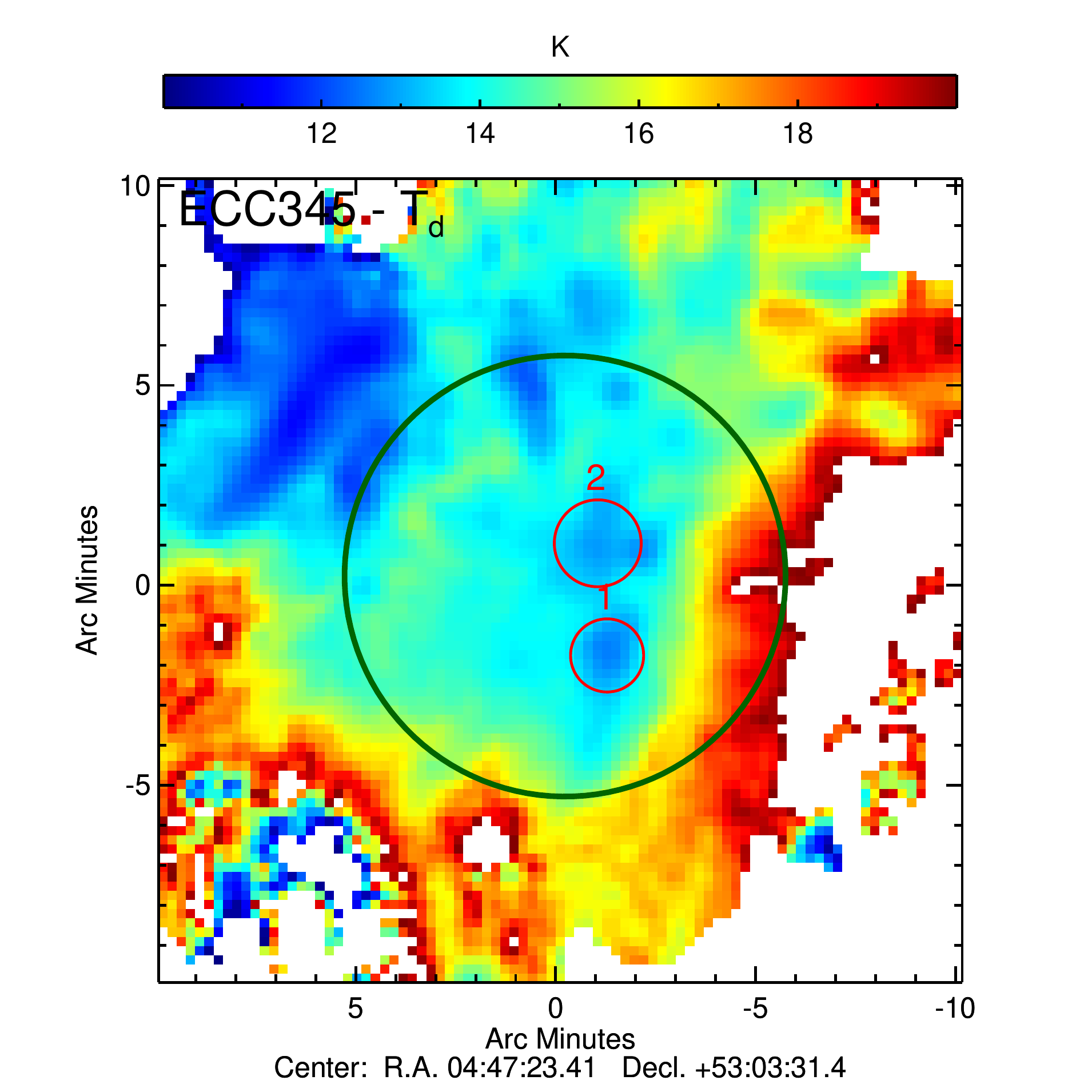} \\

\includegraphics[angle=0,scale=0.375]{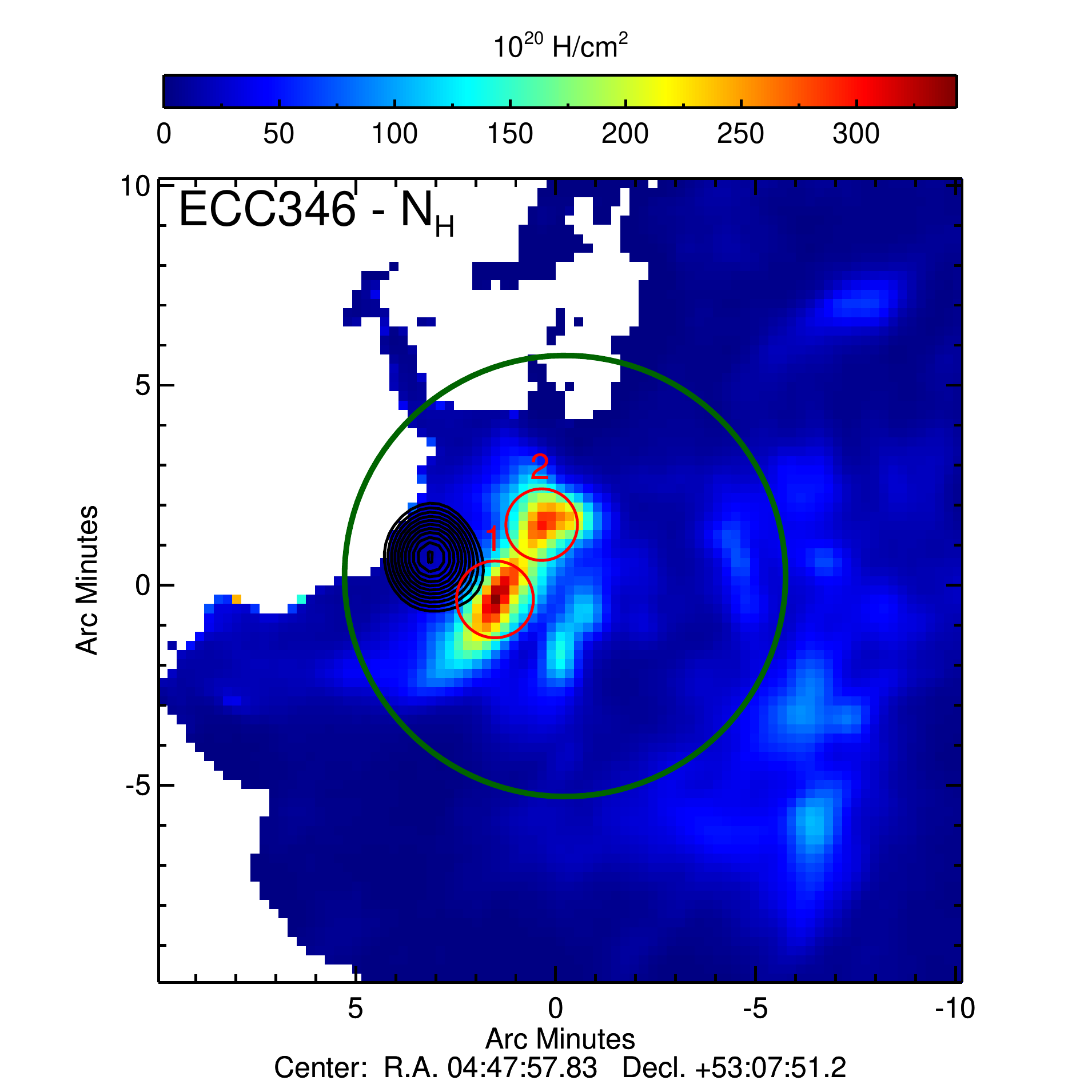}
\includegraphics[angle=0,scale=0.375]{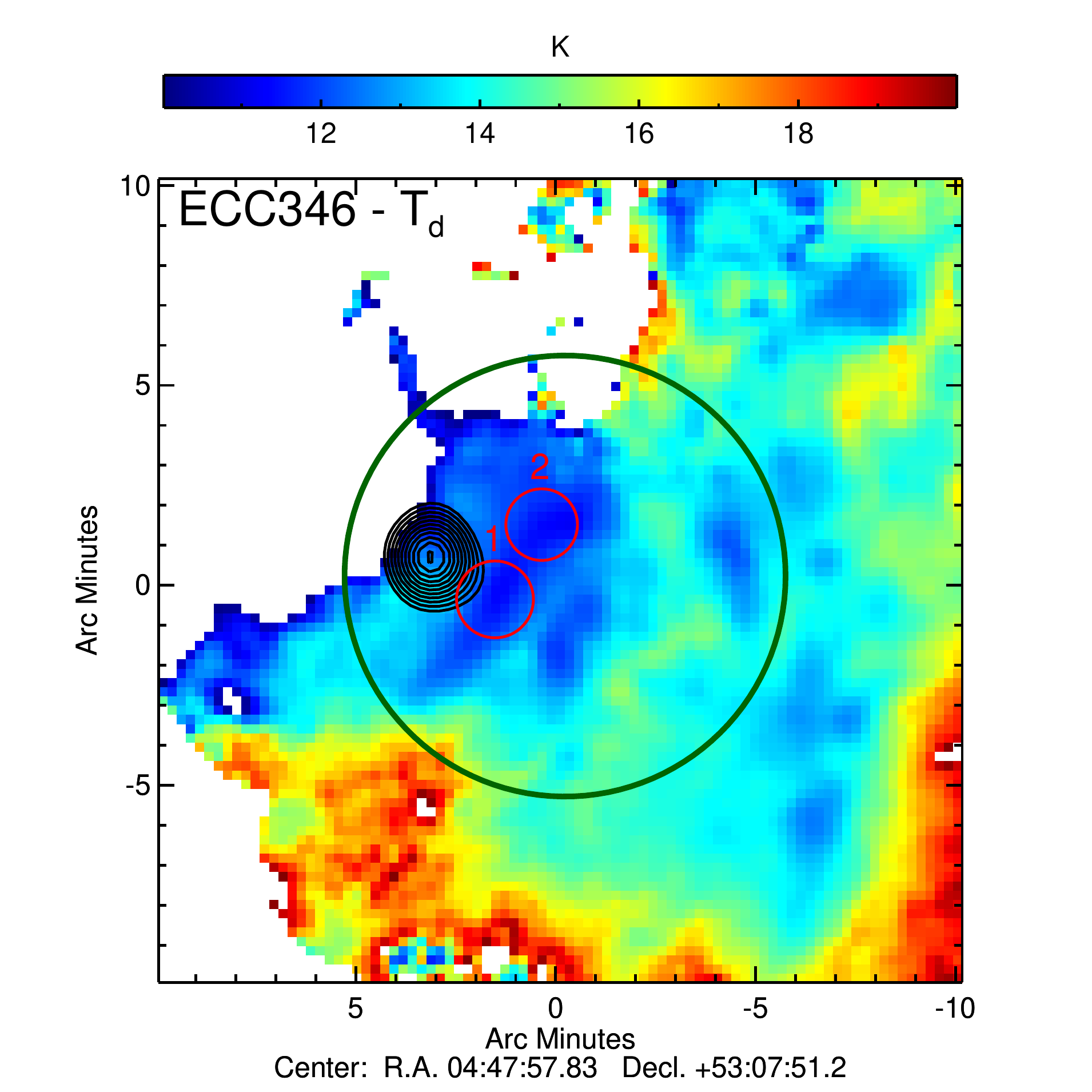} \\
\end{center}
\vspace{-0.4cm}
\caption{Continued}
\label{Fig:NH_Td_Maps}
\end{figure*}

\label{lastpage}

\end{document}